\documentclass[pdflatex,sn-basic]{sn-jnl}

\usepackage{caption}
\usepackage{subcaption}

\jyear{2021}%

\theoremstyle{thmstyleone}%
\theoremstyle{thmstyletwo}%

\theoremstyle{thmstylethree}%

\begin{document}

\title[Hedging Cryptocurrency options]{Hedging Cryptocurrency options}

\author*[1]{\fnm{Jovanka} \sur{Matic}}\email{jovanka.matic@hu-berlin.de}

\author[2]{\fnm{Natalie} \sur{Packham}}\email{natalie.packham@hwr-berlin.de}

\author[3]{\fnm{Wolfgang Karl} \sur{Härdle}} \email{haerdle@hu-berlin.de}

\affil*[1]{\small \orgdiv{BRC Blockchain Research Center, International Research Training Group
1792}, \orgname{Humboldt-Universität zu Berlin}, \orgaddress{\street{Dorotheenstr. 1}, \city{Berlin}, \postcode{10117}, \country{Germany}}
}

\affil[2]{\small \orgdiv{Department of Business and Economics}, \orgname{Berlin School of Economics and Law}, \orgaddress{\street{Badensche Str. 52}, \city{Berlin}, \postcode{10825}, \country{Germany}}}

\affil[3]{\small \orgdiv{BRC Blockchain Research Center, International Research Training Group
1792}, \orgname{Humboldt-Universität zu Berlin}, \orgaddress{\street{Dorotheenstr. 1}, \city{Berlin}, \postcode{10117}, \country{Germany;}}
\footnote[7]{\footnotesize{
\orgdiv{Sim Kee Boon Institute}, \orgname{Singapore Management University}, \orgaddress{\street{50 Stamford Rd}, \city{Singapore}, \postcode{178899}, \country{Singapore;}}\orgdiv{Asian Competitiveness Institute}, \orgname{National University of Singapore}\orgaddress{\street{469C Bukit Timah Road}, \city{Singapore}, \postcode{259772}, \country{Singapore;}} \orgdiv{Dept Information Science and Finance}, \orgname{National Yang-Ming Chiao Tung University}\orgaddress{\street{1001 University Road}, \city{Hsinchu}, \postcode{300093}, \country{Taiwan;}} \orgdiv{Dept Mathematics and Physics}, \orgname{Charles University}, \orgaddress{\street{Ke Karlovu 2027/3}, \city{Prague 2}, \postcode{121 16 }, \country{Czech Republic}}
}}
}

\abstract{The cryptocurrency market is volatile, non-stationary and non-continuous. Together with liquid derivatives markets, this poses a unique opportunity to study risk management, especially the hedging of options, in a turbulent market. We study the hedge behaviour and effectiveness for the class of affine jump diffusion models and infinite activity L\'evy processes. First, market data is calibrated to
stochastic volatility inspired (SVI)-implied volatility surfaces to price options. To cover a wide range of market dynamics, we generate Monte Carlo price paths using an SVCJ model (stochastic volatility with correlated jumps), a close-to-actual-market GARCH-filtered kernel density estimation as well as a historical backtest. In all three settings, options are dynamically hedged with Delta, Delta-Gamma, Delta-Vega and Minimum Variance strategies. Including a wide range of market models allows to understand the trade-off in the hedge performance between complete, but overly parsimonious models, and more complex, but incomplete models. The calibration results reveal a strong indication for stochastic volatility, low jump frequency and evidence of infinite activity. Short-dated options are less sensitive to volatility or Gamma hedges. For longer-dated options, tail risk is consistently reduced by multiple-instrument hedges, in particular by employing complete market models with stochastic volatility.}




\maketitle

\section{Introduction}\label{sec1}
Consider the problem of hedging contingent claims written on cryptocurrencies (CC). The dynamics of this new expanding market is characterized by high volatility, as is evident from the Cryptocurrency volatility index \href{https://thecrix.de/}{
\texttt{VCRIX}} (see \cite{Kim2019Crix}) and large price jumps \citep{Scaillet2017}. We approach hedging options written on \href{https://bitcoin.org/de/}{\texttt{Bitcoin}} (BTC) with models from the class of affine jump diffusion models and infinite activity L\'evy processes. Similarly to \cite{Branger2012Hedging}, we assess the hedge performance of implausible, yet complete as well as plausible, but incomplete asset pricing models. Since April 2019, contingent claims written on BTC and \href{https://ethereum.org/en/}{\texttt{Ethereum}} (ETH) have been actively traded on \href{https://www.deribit.com/}{\texttt{Deribit}} (\href{https://www.deribit.com/}{\texttt{www.deribit.com}}). The Chicago Merchantile Exchange (CME) introduced options on BTC futures in January 2020.  In contrast to traditional asset classes such as equity or fixed income, the market for CC options has only recently emerged and is still gaining liquidity, see e.g.\ \citep{TrimbornWKH2015} for an early description of the market. 
Cryptocurrency markets are known to exhibit high volatility and frequent jumps, see e.g.\ market crashes on 12 March 2020, 19 May 2021, 17 June 2022, posing challenges to valuation and risk management. This erratic price behaviour may be attributed to the lack of a fundamental value, see e.g. \cite{biais2022equilibrium}, \cite{Athey2016} and \cite{MAKAROV2020293}.


As the option market is still immature and illiquid, in the sense that quotes for many specific strikes or maturities are not directly observable or may be stale, we derive options prices by interpolating prices from stochastic volatility inspired (SVI) parametrized implied volatility (IV) surfaces \citep{Gatheral2004APA}. 

Aside from conducting a historical backtest, and in order to capture a variety of market dynamics, the BTC market is imitated with two different Monte Carlo simulation approaches. In a parametric price path generation approach, we assume that the data-generating process is described by a SVCJ model.  The second scenario generation method is based on GARCH-filtered Kernel-density estimation (GARCH-KDE), which can be thought of as a smooth historical simulation taking into account the historical volatility dynamics, and which is therefore close to actual market dynamics. 

Under each of the two different market simulation methods, options are hedged by a hedger employing models of different complexity. This deliberately includes models that are ``misspecified'' in the sense that relevant risk factors may be omitted \citep{Branger2012Hedging}. On the other hand, those models are possibly parsimonious enough to yield a complete market. It is known that, when comparing the hedge performance to a more realistic, albeit incomplete market model, the simpler model may outperform the complex model \citep{Detering2015}. In our context, a model is ``misspecified'' if it contains fewer or different parameters than the SVCJ model. 
Specifically, as models included in the class of SVCJ models, we consider the \cite{BlackScholes1973BS} (BS) model, the \cite{merton1976option} jump-diffusion model (JD), the \cite{Heston1993} stochastic volatility model (SV), the stochastic volatility with jumps model (SVJ) \citep{bates1996jumps} and the SVCJ model itself. Infinite activity L\'evy hedge models under consideration are the Variance-Gamma $(VG)$ model \citep{MadanVC1998} and the CGMY model \citep{Carr2002}. Options are hedged dynamically with the following hedge strategies:  Delta ($\Delta$), Delta-Gamma ($\Delta$-$\Gamma$), Delta-Vega ($\Delta$-$\mathcal V$) and minimum variance strategies (MV). 

To gain further insights, we separate the full time period, ranging from April 2019 to June 2020, into 3 different market scenarios with a bullish market behavior, calm circumstances with low volatility and a stressed scenario during the SARS-COV-2 crisis. In addition to evaluating the hedge performance, we aim to identify BTC risk-drivers such as jumps. This contributes to the understanding of what actually drives fluctuations on this market. 
A historical backtest of the hedge performance, in the spirit of \cite{Detering2016} and \cite{TingEwald2013}, completes and confirms the findings of the SVCJ- and GARCH-KDE approaches. 

A number of papers investigate the still young market of CC options. \cite{TrimbornWKH2015} describe the CC market dynamics via the cryptocurrency index \href{https://thecrix.de/}{\texttt{CRIX}}. \cite{Schoutens2019} price BTC options and calibrate parameters for a number of option pricing models, including the Black-Scholes, stochastic volatility and infinite activity  models. \cite{crixpricing2019} price 
\href{https://thecrix.de/}{\texttt{CRIX}} options under the assumption that the dynamics of the underlying are driven by the  (SVCJ) model introduced in \cite{duffie2000transform} and \cite{eraker2003impact}. The literature on the aspects of risk management in CC markets is scarce but growing. \cite{DYHRBERG2016139}, \cite{BOURI2017192} and \cite{Selmi2018} investigate the role of BTC as a hedge instrument on traditional markets. \cite{SEBASTIAO2020101230} and \cite{Alexander2021} investigate the hedge effectiveness of BTC futures, while \cite{NEKHILI2021} hedge BTC risk with conventional assets.  To the best of our knowledge, hedging of CC options has not yet been investigated in this depth and detail. The aspect of risk management and the understanding of the dynamics of CCs is therefore a central contribution of this study. 

The remainder of the paper is structured as follows: Section \ref{sec:methodology} describes the methodology, decomposed into market scenario generation, option valuation and hedge routine. The hedge routine presents the hedge models and explains the model parameter calibration and hedge strategy choices. In Section \ref{sec:results}, we present and evaluate the results of the hedge routine and in Section \ref{sec:conclusion}, we conclude. The code is available as quantlets, accessible through \href{https://www.quantlet.com/}{\texttt{Quantlet}} under the name \href{https://github.com/QuantLet/hedging_cc}{\includegraphics[scale=0.2]{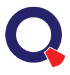}}\href{https://github.com/QuantLet/hedging_cc}{ hedging\_cc}.

\section{Methodology}\label{sec:methodology}
In this section, we introduce the methodology, comprising market scenario generation, option valuation and hedging. 
We take an option seller's perspective and sell 1- and 3-month (M) contingent claims. The choice is justified by the total trading volume of BTC options. Nearly $80 \%$ of the trading volume consists of options expiring in at most 1 month. Almost all remaining options expire in 3 months or less \citep{Carol2022}.

\subsection{Synthetic market data generation}
\begin{figure}[ht!]
\begin{center}
\includegraphics[scale=0.37]{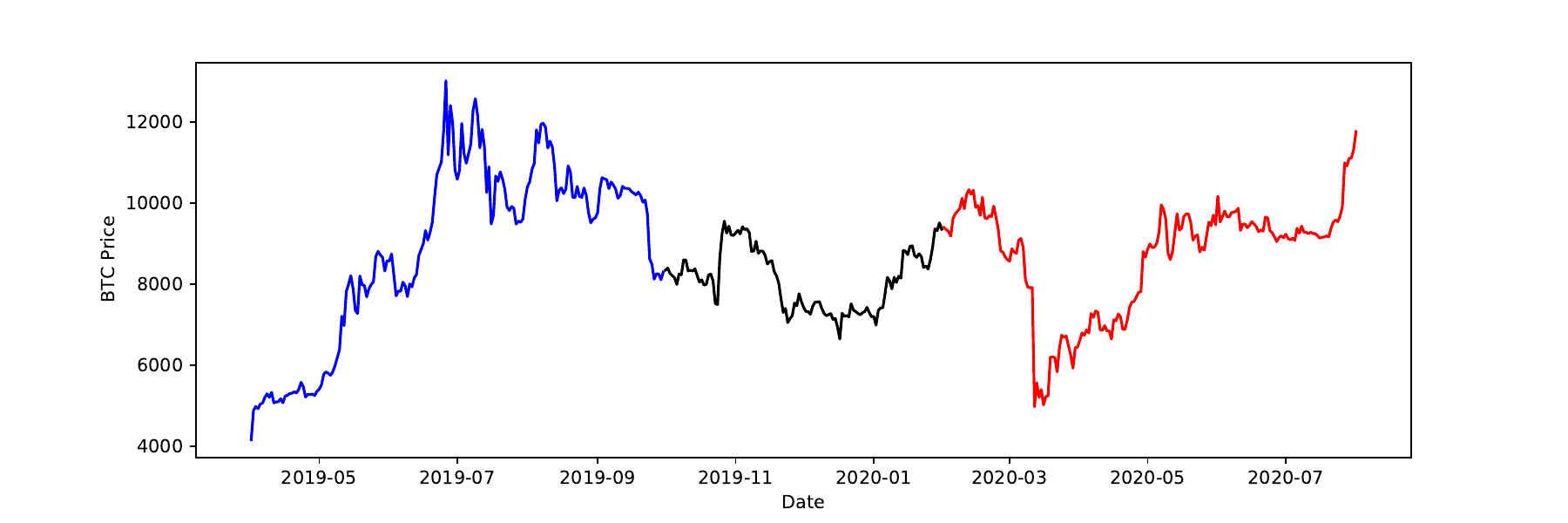}
\caption[Data]{BTC closing price from 1st April 2019 to 30th June 2020, where the \textcolor{blue}{blue trajectory} represents the \textcolor{blue}{bullish market behavior}, the \textcolor{black}{black path} the \textcolor{black}{calm period} and \textcolor{red}{red path} the stressed scenario during \textcolor{red}{ the Corona crisis}. \\
\includegraphics[scale=0.2]{quanlet.png}{ LoadBTC}}\label{btc_path}
\end{center}
\end{figure}

We describe how to generate synthetic market data, which serves as the input for the main analysis. The principal goal of synthetic scenario generation is to imitate the BTC market behavior, especially retaining its statistical properties, with the added flexibility of Monte Carlo simulation  to create a large amount of plausible scenarios. 
In addition, we consider two simulation methods capturing different statistical properties. They represent a trade-off between a parametric model with valuable and traceable risk-factor information and a flexible non-parametric closer-to-actual-market approach. The parametric model is simulated under the risk neutral measure $\mathbb{Q}$ with a forward looking perspective. The non-parametric simulation relates to the past market behavior performed under the physical measure $\mathbb{P}$. The time frame under consideration is from $1^{st}$ April 2019 to $30^{th}$ June 2020. The BTC market behavior in this time period is time-varying. This makes it convenient to segregate the time frame into three disjoint market segments from April to September 2019 (\textit{bullish}), October 2019 to February 2020 (\textit{calm}) and March to June 2020 (\textit{covid}), respectively. Bearing in mind that we are going to hedge  1-month and 3-month options, the minimal segment length is chosen to exceed three months. A graphical representation of the BTC closing price trajectory is illustrated in Figure \ref{btc_path} with the corresponding summary statistics in 
\mbox{Table \ref{tab:btc_behavior}}. The first interval is labeled as the \textit{bullish} segment, because, to a great extent, the market behaves upward-trending. The second period is labeled as the \textit{calm period}. With an overall standard deviation of $\hat{\sigma}=756.55$, price movements are more stagnant compared to the bullish segment. 
\begin{table}[t]
\centering
\scalebox{0.7}{
\begin{tabular}{lrrrrrrr}
\hline
\hline 
behavior     &     $\hat{\mu}$  &      $\hat{\sigma}$ &      min &      $q^{25}$ &      $q^{50}$  &      $q^{75}$ &       max \\
\hline
bullish &   0.0038 &  0.0428 & -0.1518 & -0.0157 &  0.0050 &  0.0227 &  0.1600 \\
calm    &  0.0009 &  0.0290 & -0.0723 & -0.0162 & -0.0015 &  0.0098 &  0.1448 \\
covid   &  0.0012 &  0.0490 & -0.4647 & -0.0107 &  0.0009 &  0.0162 &  0.1671 \\
\hline
\hline
\end{tabular}}
\caption{
Summary statistics of the bullish, calm and covid market log returns $R_t$. \includegraphics[scale=0.2]{quanlet.png}\href{https://github.com/QuantLet/hedging_cc}{ hedging\_cc}}
\label{tab:btc_behavior}
\end{table}
The last segment is the \textit{Corona crisis} or \textit{stressed scenario}, where financial markets, especially CC markets, experienced high volatility. A notable mention is the behavior of  BTC on $12^{th}$ March 2020, where its price dropped by nearly $50 \%$.

We now turn to a formal mathematical framework. Let the BTC market be a continuous-time, frictionless financial market. Borrowing and short-selling are permitted. The constant risk-free interest rate $r \geq 0$ and the time horizon $T < \infty $ are fixed. On a filtered probability space $\left( \Omega , \mathcal { F } , \left( \mathcal { F } _ { t } \right) _ { t \in [ 0 , T ] } ,\mathbb P  \right)$, the asset price process and the risk-free asset are defined by adapted semimartingales $(S_{t})_{t \geq 0}$ and $\left( B_{t} \right)_{t \geq 0}$, where $B_{0}=1$ and $B_{t} = e^{rt}, \ t \geq 0$, respectively. The filtration is assumed to satisfy the usual conditions (e.g.\ \citep{protter2005stochastic}). To ensure the absence of arbitrage, we assume the existence of a risk-neutral measure $\mathbb{Q}$. We consider an option writer's perspective and short a European call option. The price of the option with strike $K$ and time-to-maturity (TTM) $\tau = T-t$ at time $t < T$ is $C(t, \tau, K)$. For multiple-instrument hedges, we further assume the existence of a liquidly traded call option $C_2 (t,\tau,K_2)$, $K_2\not=K$, suitable for hedging. The dynamic, self-finance hedging strategy $\xi=( \xi^{0}, \xi^{1} )=(\xi^{0}_{t}, \xi^{1}_{t} ))_{0 \leq t \leq T}$ is an $\mathcal{F}$-predictable process, where $\xi^{0}_{t}$ and $\xi^{1}_{t}$ denote the amounts in the risk-free security and the asset, respectively. The resulting portfolio process $\Pi=(\Pi_t)_{t\geq 0}$ is admissible and self-financing. The evolution of the value process $\Pi$ is reviewed in detail in Appendices \ref{sec:hedge_routine}, \ref{sec:appendix_dynamic_delta} and  \ref{sec:appendix_dynamic_delta_vega}.

For the Monte Carlo simulation, the finite time horizon is partitioned into daily steps of size $\delta t = \frac{1}{365}$. The number of trajectories of the asset price process is set to $n=100,000$. 
\subsubsection{SVCJ model}\label{sec:svcj}
The parametric scenario generation approach assumes that the dynamics of the asset price process $(S_t)_{t\geq 0}$ and the volatility process $(V_t)_{t\geq 0}$ are described by the SVCJ model introduced in \cite{duffie2000transform}. This particular choice is motivated by the methodology in \cite{crixpricing2019}, where the model is applied to pricing options on the \href{https://thecrix.de/}{\texttt{CRIX}} . A high degree of free parameters enables to model various market dynamics. More specifically, the risk-neutral model dynamics
are \citep{Broadie2005}

\begin{equation}\label{eq:svcjsqe}
\begin{aligned}
    d S_t &=(r-\lambda \bar{\mu}) S_t d t+\sqrt{V_t} S_t\left[\rho d W_t^{(1)}+\sqrt{1-\rho^2} d W_t^{(2)}\right] + Z^{s}_{t} d N_t \\
    d V_t &=\kappa\left(\theta-V_t\right) d t+\sigma^{v} \sqrt{V_t} d W_t^{(1)}+Z^{v
}_{t} d N_t
\end{aligned}
\end{equation}
where $W_{t}^{(1)}, W_{t}^{(2)}$ are two independent standard Wiener processes. The scale of $V_{t}$ is given by $\sigma^{v}$, the mean reversion speed is denoted by $\kappa$ and $\theta$ is the mean reversion level. The model allows for simultaneous arrivals of jumps in returns and jumps in volatility governed by the Poisson process $(N_t)_{t\geq 0}$  with constant intensity $\lambda$. The jump sizes in volatility $Z^{v}_{t}$ are exponentially distributed $Z^{v}_{t} \sim \varepsilon \left( \mu^{v}\right)$ and the jumps sizes of the asset price are conditionally normally distributed
\begin{equation}\label{eq:svcj_jumps}
\Xi \stackrel{def}{=} Z^{s}_{t} \vert Z^{v}_{t} 
\sim N\left(\overline{\mu}^{s} + \rho^{j} Z^{v}_{t}, \left(\sigma^{s}\right)^2 \right)
\end{equation}
where $\overline{\mu}^{s}$ is the conditional mean jump size in the asset price given by 
\begin{equation*}
\overline{\mu}^{s}=\frac{\exp \left\{\mu^{s}+\frac{\left(\sigma^{s}\right)^{2}}{2}\right\}}{1-\rho^{j} \mu^{v}}-1
\end{equation*}
In detail, $\sigma^{s}$ denotes the jump size standard deviation. The unconditional mean is denoted by $\mu^{s}$, which is related to the jump compensator $\lambda {\mu}^{*}$ by $\mu^s=\log \left[(1+{\mu}^{*})\left(1-\rho^{j} \mu^{v}\right)\right]-\frac{1}{2} \left(\sigma^{s}\right)^2$. The correlation parameter $\rho^j$ governs the correlations between jump sizes. From an empirical point of view, in most markets, jumps occur seldomly and are difficult to detect, which, as a consequence, makes the calibration of $\rho^{j}$ unreliable \citep{broadie2007model}. 
\cite{Chernov2003Paper},
\cite{broadie2007model},  \cite{eraker2003impact}, \cite{eraker2004stock} and \cite{branger2010svcj} therefore recommend to set $\rho^{j}=0$. In fact, this finding extends to the BTC market, see \cite{crixpricing2019}, who find that "the jump correlation $\rho^j$ is negative but statistically insignificant...". Our main results are therefore calculated assuming $\rho ^j=0$. Nonetheless, we add some insights into calibrating $\rho^j$ and hedging with the calibrated parameter in Section \ref{sec:rhojsens}. Note that despite a correlation of zero, the SVCJ model does not reduce to an SVJ model, as it still features jumps in the volatility.

The resulting paths are simulated according to the Euler-Maruyama discretization of \eqref{eq:svcjsqe} suggested in \cite{belay2005paper}. The corresponding model parameters are re-calibrated on a daily basis according to the methodology described in Section \ref{sec:calibration}. 

\subsubsection{GARCH-KDE approach}
Compared to the empirical price process, the SVCJ may appear quite restrictive: aside from being an incomplete market model, the price dynamics are limited by the specification of the stochastic volatility component as well as the jump intensity and size. The semi-parametric method loosens the assumptions by generating scenarios using GARCH-filtered kernel density estimation (GARCH-KDE) as in e.g.\ \cite{McNeil2000}. Let $(R_{t})_{t \geq 0}$ denote BTC log-returns and $(\hat\sigma_t)_{t \geq 0}$ the estimated GARCH(1,1) volatility, \citep{Bollerslev1986}. The kernel density estimation is performed on "de-garched" residuals 
\begin{equation}\label{eq:degarched_res}
    \hat z_t = \displaystyle \frac{R_t}{\hat\sigma_t}.
\end{equation}
The rationale is to capture the time-variation of volatility by the GARCH filter and perform kernel density estimation on standardised residuals. The estimated density function is 
\begin{equation}\label{eq:degarched_dist}
\widehat{f}^{h}(z)=\frac{1}{n h} \sum_{t=1}^{n} \mathcal{K}\left(\frac{\hat{z}_{t}-z}{h}\right),
\end{equation}
where $\mathcal{K} \left(\cdot \right)$ denotes the Gaussian Kernel. The resulting generated paths are 
$(S(0,i),\ldots, S(T,i))$, $i=1,\ldots, n$, with\footnote{The simulated, discretised prices are denoted by $S(t,i)$ are opposed to $S_t$, which refers to the continuous-time process.}
\begin{equation}
    S(t,i) = \displaystyle S(0) \exp\left[\sum_{k=1}^t \hat\sigma_k \hat{z}_k\right], \quad t=0,\ldots, T.
\end{equation}
Throughout the paper, the parametric and the semi-parametric method are referred to as the SVCJ and GARCH-KDE framework, respectively. 
\subsection{Valuation}\label{sec:svi_interpolation}
This section describes how option prices are derived from the implied volatility surface. As the market for CC claims, during the time period of our dataset, is still relatively immature with only a limited number of actively traded options on \href{https://www.deribit.com/}{\texttt{Deribit}} and the Chicago Mercantile Exchange, 
arbitrage-free option prices are derived through the stochastic volatility inspired (SVI) parameterization of the volatility surface of \cite{Gatheral2014}.
Let $\sigma^{\mathrm{BS}}(k, \tau)$ denote the BS IV with log-moneyness $k=\ln \left( K /S_0 \right)$ and total implied variance $w(k, \tau)=\{\sigma^{\mathrm{BS}}(k, \tau)\}^{2} \tau$. For a fixed $\tau$, the raw SVI parameterization of a total implied variance smile, as initially presented in \cite{Gatheral2004APA}, is 
\begin{equation}\label{eq:IV_params}
w(k, \chi_{R})=a+b\left\{\rho^{SVI}(k-m)+\sqrt{(k-m)^{2}+\left(\sigma^{SVI}\right)^{2}}\right\}.
\end{equation}
In the parameter set $\chi_{R}=\{a, b, \rho^{SVI}, m, \sigma^{SVI}\}$, 
$a \in \mathbb{R}$ governs the general level of variance,
$b \geq 0$ regulates the slopes of the wings, $\rho^{SVI} \in [-1, 1]$ controls the skew, $m \in \mathbb{R}$ enables horizontal smile shifts and $\sigma^{SVI}>0$ is the ATM curvature of the smile \citep{Gatheral2014}.
For each maturity, the smile is recalibrated daily. The implied volatility is obtained by a simple root-finding procedure, whereas the parameters $\chi_{R}$ are calibrated according to the optimization technique explained in Section \ref{sec:calibration}.  
In addition, the calibration is subject to non-linear constraints prescribed in \cite{Gatheral2014}. These constraints ensure convexity of the option price, which rules out butterfly arbitrage. Calendar spread arbitrage is avoided by penalizing fitted smiles, which induce a decrease in the level of the total implied variance for a given strike level. For interpolation, the at-the-money (ATM) total implied variance $\theta_{T}^{SVI}=w(0,T)$ is interpolated for $t_1 < T < t_2$ as in \cite{Gatheral2014}, where $t_1, t_2$ refer to time points at which implied volatilities are observed. The resulting option price is the convex combination 
\begin{equation}
C(T, K)=\alpha_{T} C(t_1, K)+\left(1-\alpha_{T}\right) C(t_2, K),
\end{equation}
where $\alpha_{T}=\displaystyle\frac{\sqrt{\theta_{t_{2}}^{SVI}}-\sqrt{\theta_{T}^{SVI}}}{\sqrt{\theta_{t_{2}}^{SVI}}-\sqrt{\theta_{t_{1}}^{SVI}}} \in\left[0,1\right]$.

\subsection{Hedge routine}\label{sec:models}
This section describes the models selected to hedge BTC options as well as the model parameter calibration procedure. Given these model classes, hedge strategies are chosen for the hedge routine.
\subsubsection{Hedge models}\label{sec:models_hedge}
For hedging purposes, the choice of a hedge model faces the trade-off between sufficient complexity to describe the actual market dynamics and market completeness \citep{Detering2015}. In practice, a trader may for instance  initiate hedging with an evidently wrong but simple model, such as the complete BS option pricing model. A lower number of parameters provides a parsimonious setup with potentially manageable explanatory power. In our setting, a European option is hedged employing models of increasing complexity. In the following, the model granularity is gradually extended by the addition of risk-factors such as local volatility, jumps, stochastic volatility and others. This covers the empirical finding of the previous literature on CC's, e.g. \citep{Kim2019Crix,Scaillet2017}. Accordingly, the hedge models selected encompass affine jump diffusion models and infinite activity Levy processes. 

The class of affine jump diffusion models covers well-known models nested in \eqref{eq:svcjsqe}.
Due to its popularity in the financial world, the simple but complete BS option pricing is selected as
one of the hedge models. 
The volatility is constant with $V_t=\sigma$ and there are no discontinuities from jumps $N_{t} = 0$. 
A slightly more complex model is the JD model. It assumes constant volatility with $V_{t}=\theta$, $\sigma^{v}=0$ and extends the BS model by allowing for jumps in returns. The jump size is $\operatorname{log} \xi \sim N\left(\mu^{s}, \left(\delta^{s}\right)^{2} \right)$ distributed.

Evidence for stochastic volatility motivates the choice of the SV model. The jump component is excluded with $\lambda = 0$ and $N_{t} = 0$. We also examine the SVCJ model itself as a model used for hedging. It serves as the most general model and its hedge performance provides a meaningful insight for the comparison of the SVCJ and GARCH-KDE framework, while in the SVCJ framework, it provides ``anticipated'' hedge results (cf.\ \cite{Branger2012Hedging}). Due to the jump scarcity and latent nature of the variance process $V_{t}$, we also consider the SVJ model for hedging. In difference to the SVCJ model, this model has jumps in returns but no jumps in volatility. 

In contrast to affine jump processes, there exists a well-established class of processes that do not entail a continuous martingale component. Instead, the dynamics are captured by a right-continuous pure jump process, such as the Variance Gamma (VG) model \citep{MadanVC1998}. The underlying $S_t$ evolves as 
\begin{equation}\label{eq:vg_dyn}
\begin{aligned}
d S_{t} &=r S_{t-} d t+ S_{t-} d X^{\mathrm{VG}}_{t} \\
X^{\mathrm{VG}}_{t} &=\theta^{VG} G_{t} +\sigma^{VG} W_{G_{t}},
\end{aligned}
\end{equation}
with the characteristic function of the VG-process $X^{\mathrm{VG}}_{t}$ given by  
\begin{equation}\label{eq:vgchar}
\varphi_{\mathrm{VG}}(u ; \sigma^{VG}, \nu , \theta^{VG})=\left(1-\mathrm{i} u \theta^{VG}  \nu+\frac{1}{2} \left(\sigma^{VG} \right)^{2} \nu u^{2}\right)^{-1 / \nu},
\end{equation}
where $r$ is the risk-free rate, $W_{t}$ is a Wiener process and $G_{t}$ is a Gamma process. The overall volatility level is represented by $\sigma^{VG}$; $\theta^{VG}$ governs the symmetry of the distribution and therefore controls the implied volatility skew; $\nu$ controls for tails, kurtosis and thus regulates the shape of the volatility surface. An alternative representation of the $VG$ process appealing for practical interpretation is given by the characteristic function \begin{equation}\label{eq:vgcgm}
\varphi_{\mathrm{VG}}(u ; C, G, M)=\left(\frac{G M}{G M+(M-G) \mathrm{i} u+u^{2}}\right)^{C},
\end{equation}
where $C, \ G, \ M >0$. The detailed link between \eqref{eq:vgchar} and \eqref{eq:vgcgm} is described in Appendix \ref{appendix:vgtocgm}. 
An increase in $G$ ($M$) increases the size of upward jumps (downward jumps). Accordingly, $\theta^{VG} $, $M$ and $G$ account for the skewness of the distribution. An increase in $C$ widens the L\'evy-measure. An extension of the VG model is the CGMY model by \cite{Carr2002}. On a finite time interval, the additional parameter $Y$ permits infinite variation as well as finite or infinite activity. Formally, in \eqref{eq:vg_dyn} the source of randomness is replaced by a CGMY process $X^{CGMY}_{t}$ with the characteristic function 
\begin{equation}
\varphi_{\mathrm{CGMY}}\left(u ; C, G, M,Y \right)
=\exp \left[C t \Gamma(-Y)\left\{(M-\mathrm{i} u)^{Y}-M^{Y}+(G+\mathrm{i} u)^{Y}-G^{Y}\right\}\right].
\end{equation}
The $X^{VG}_{t}$-process in the representation in Equation \eqref{eq:vgchar} is a special case of the CGMY process for $Y=1$. On a finite time interval, the behavior of the path depends on $Y$. For $Y<0$, there is a finite number of jumps, else infinite activity. In case of $Y \in (1,2]$, there is also infinite variation.
\subsubsection{Calibration routine}\label{sec:calibration}
The model parameters are calibrated following the FFT option pricing technique of \cite{CM99}. The price of a European-style option $C (T,K)$ is given by 
\begin{equation}
\begin{aligned}
C(T,K) &=\frac{1}{\pi} \operatorname{e}^{-\alpha \ln K} \int_{0}^{\infty} \operatorname{e}^{-\mathrm{i} v \ln K} \psi_T(v) d v, \quad\text{ with} \\
\psi_T(v) &=\frac{\exp^{-rT} \phi_{T}(v-(\alpha+1)\mathrm{i})}{\alpha^{2}+\alpha-v^{2}+\mathrm{i}(2 \alpha+1) v}, 
\end{aligned}
\end{equation}
where $\phi_T$ is the characteristic function of the $\alpha$-damped option price $c_{T}\left[\ln (K)\right]=e^{\alpha \ln K} C(T,K), \ \alpha >0$. The ill-posed nature of calibration can lead to extreme values of the model parameters. This is avoided by employing a Tikhonov $L_2$-regularization \citep{TikhonovLeonovYagola+2011+505+512}. At the cost of accepting some bias, this penalizes unrealistic values of the model parameters by giving preference to parameters with smaller norms. Calibration is performed by the optimizer
\begin{equation}
\begin{aligned}
\theta^* &= \underset{\theta\in\Theta}{\operatorname{argmin}} R(\theta) \\
R(\theta) &= \sqrt{\frac 1n \sum_i \{IV^{Model}(T^i, K^i, \theta) - IV^{Market}((T^i, K^i))\}^2} + \theta^{\top} \Gamma \theta,
\end{aligned}
\end{equation}
where $IV^{Model} \left(\cdot\right), IV^{Market}\left(\cdot\right)$ describe model and market implied volatilities for maturity and strike $T^i, K^i$. $\Gamma$ is a diagonal positive semi-definite matrix. It corresponds to the Tikhonov $L_2$-regularization, which gives preference to parameters with smaller norms. The entries in the matrix $\Gamma$ are chosen individually for each parameter to ensure that they maintain the same reasonable order of magnitude. 

The parameter space $\Theta \subset \mathbb R^d$ of each model in scope is subject to linear inequality constraints. Given that the objective is not necessarily convex, it may have multiple local minima. In order to explore the entire parameter space, simplex-based algorithms are more appropriate than local gradient-based techniques. In our case, we employ the Sequential Least Squares Programming optimization \citep{kraft1988software} routine. We adjust for time effects by calibrating parameters on the IV surface instead of option prices. 

We impose liquidity and moneyness cut-offs. Claims must have a positive trading volume and an absolute BS Delta in $[0.25,0.75]$.  This filters options that are close to ATM as is custom in FX trading, see \cite{clark2011foreign}.

\subsubsection{Hedging strategies}\label{sec:hedgeroutine}
Any hedging strategy's target is to protect against market movements and to minimize Profit-and-Loss (P\&L) of the hedged position. Hedges either reduce risk by eliminating market-risk-related sensitivities $\left(\Delta, \Gamma, \mathcal{V}  \right) = \displaystyle
\left( \frac{\partial C }{\partial S} , 
\frac{\partial^{2} C}{\partial^{2} S},
\frac{\partial C }{\partial \sigma} \right)$ or by minimising a risk measures, such as a hedged position's variance.
Broadly, hedging strategies are split into single- and multiple instrument hedges. 
Single instrument hedges incorporate the $\Delta$- and MV-hedging. \cite{foellmer1985hedging}'s MV hedge aims to find the strategy that minimizes the mean-squared error under $\mathbb{Q}$ 
\begin{equation}
\left(\Pi_{0}, \xi^{MV}_{t}\right) =\underset{\Pi_{0} , \xi^1_t}{\text{argmin}}\  \mbox{\sf E}_\mathbb{Q}\left[\left(C_{T} -\Pi_{0} -\int_{0}^{T} \xi^{1}_{u} d S_u\right)^{2}\right].
\end{equation}
Under the assumption of symmetric losses and gains, the minimizing strategy is denoted by $\xi^{MV}_{t}$. The $\Delta$-hedge targets to protect the position against first-order changes in the underlying $(S_t)_{t\leq T}$. 

In addition to hedging $\Delta$, multiple instrument hedges eliminate higher-order sensitivities or sensitivities of risk factors other than the underlying, e.g. $\sigma$. 
 To achieve 
$\Delta$-$\Gamma$- or $\Delta$-$\mathcal V$-neutrality, an additional liquid option $C_{2}(S(t),T,K_{1})$ with strike $K_1 \neq K$ is priced from the SVI parameterized IV surface, as explained in Section \ref{sec:svi_interpolation}. For performance comparison of linear and non-linear effects, the dynamic $\Delta$- 
and $\Delta$-$\Gamma$-hedging strategies are applied to all hedge models. The $\Delta$-$\mathcal V$-hedge is only considered for affine jump diffusion models. The technical aspects of the dynamic hedging strategies are described in Appendices \ref{sec:appendix_dynamic_delta} and \ref{sec:appendix_dynamic_delta_vega}.
The calibrated model parameters are used to compute hedging strategies $(\xi_t)_{0\leq t\leq T}$ for each model.
Table \ref{tab:Hedge model summary} summarizes the hedging strategies applied to the respective hedge models.
\begin{table}[t]
\begin{center}
\scalebox{0.8}{
\begin{tabular}{ll}
\hline
\hline
model         &  strategies applied           \\ \hline
BS & $\Delta_{BS}$,            $\Delta$-$\Gamma_{BS}$, $\Delta$-$\mathcal{V}_{BS}$\\
SV        & $\Delta$-$\mathcal{V}_{SV}$, $\Delta_{SV}$, $\Delta$-$\Gamma_{SV}$, MV \\
JD       & $\Delta_{JD}$, $\Delta$-$\Gamma_{JD}$, $\Delta$-$\mathcal{V}_{JD}$,  MV      \\
SVJ           &  $\Delta_{SVJ}$, $\Delta$-$\Gamma_{SVJ}$, $\Delta$-$\mathcal{V}_{SVJ}$, MV
\\
SVCJ& $\Delta_{SVCJ}$, $\Delta$-$\Gamma_{SVCJ}$ $\Delta$-$\mathcal{V}_{SVCJ}$, MV       \\ 
VG  &  $\Delta_{VG}$, $\Delta$-$\Gamma_{VG}$, MV     \\ CGMY         & $\Delta_{CGMY}$, $\Delta$-$\Gamma_{VG}$, MV        \\ 
\hline
\hline
\end{tabular}}
\caption{Hedge strategy summary applied to the hedge models described in Section  \ref{sec:models_hedge}.} 
\label{tab:Hedge model summary}
\end{center}
\end{table}
The methods for computing sensitivities depends on the model. Where possible, analytic formulas are used (e.g.\ BS-model). In cases where not analytic formulas are available, e.g.\ the VG-model, finite differences are applied to FFT-generated option prices. 
\subsubsection{Backtesting hedges on historical data}
In addition to evaluating the hedges in Monte Carlo simulations, the hedging strategies are backtested on the historical BTC price path. The principal idea is to write   an at-the-money option with fixed expiry (2 months in our setting) each day. Each option is hedged by a  self-financing hedging strategy with daily rebalancing. At expiry, the P\&L is recorded. This gives a sample of P\&L's on real data. Details of the self-financing strategy are given in Appendix \ref{sec:hedge_routine}. The choice of 2-month expiry allows to construct P\&L samples of size 60 for each market regime (bullish, calm, Covid). 

This setup follows the empirical study in \cite{Detering2016}. A similar type of backtest, recording daily P\&L instead of terminal P\&L is conducted in \cite{TingEwald2013}. Daily P\&L, however, depends on the option price and is therefore model-dependent.

\subsubsection{Hedge performance measures}\label{sec:performance}
Each model's hedge performance is evaluated by indicators derived from the relative P\&L
\begin{equation}\label{eq:pnl}
    \pi^{rel} = e^{-r T} \frac{\Pi_T }{C (T,K)}.
\end{equation}
In a perfect hedge in a complete market, we have $\Pi_T=0$, and therefore $\pi^{rel}=0$. However, in practice, due to model incompleteness, discretization and model uncertainty, $\pi^{rel}\neq 0$. We evaluate the hedge performance with the relative hedge error $\varepsilon^{hedge}$ as applied in e.g.\ \cite{Poulsen2009HE}, defined as 
\begin{equation}\label{eq:hedgeerrorpoulsen}
\varepsilon^{hedge} =100  \sqrt{
\operatorname{Var}\left( \pi^{rel} \right)}.
\end{equation}
The rationale behind $\varepsilon^{hedge}$ is that standard deviation represents a measure of uncertainty. A sophisticated hedge strategy reduces or ideally eliminates uncertainty \citep{Branger2012Hedging}. The tail behavior is captured by the expected shortfall
\begin{equation}
\text{ES}_{\alpha} = \mathbb E\left[\pi^{rel} \mid \pi^{rel}>F_{\pi^{rel}}^{(-1)}(\beta)\right],
\end{equation}
where $F_{\pi^{rel}}^{(-1)}(\beta)$ denotes an $\beta$-quantile. 
In the empirical part, these measures are estimated via the empirical distributions from Monte Carlo, resp.\ historical simulation.
\section{Empirical results}\label{sec:results}
\subsection{Data}
The models are calibrated on the market prices of European-style \href{https://www.deribit.com/}{\texttt{Deribit}} options written on BTC futures. 
The number of liquidly traded instruments varies significantly with maturity. Therefore, the data is filtered with liquidity cut-offs. 
\subsection{Option pricing}
\begin{figure}[ht!]
    \centering
    \begin{subfigure}[b]{0.328\textwidth}\includegraphics[width=\textwidth]{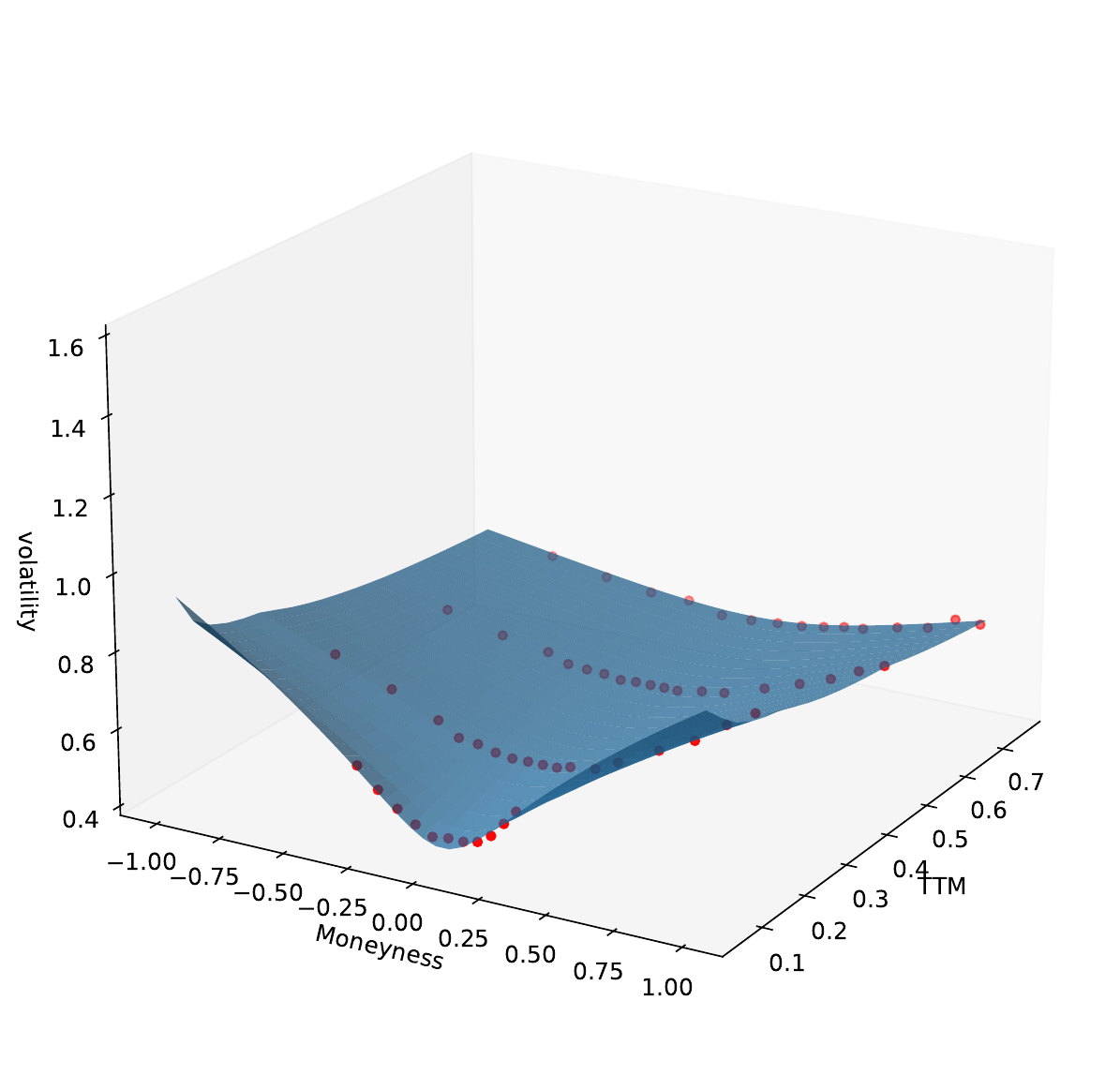}
    \caption{bullish}
    \end{subfigure}
    \begin{subfigure}[b]{0.328\textwidth}\includegraphics[width=\textwidth]{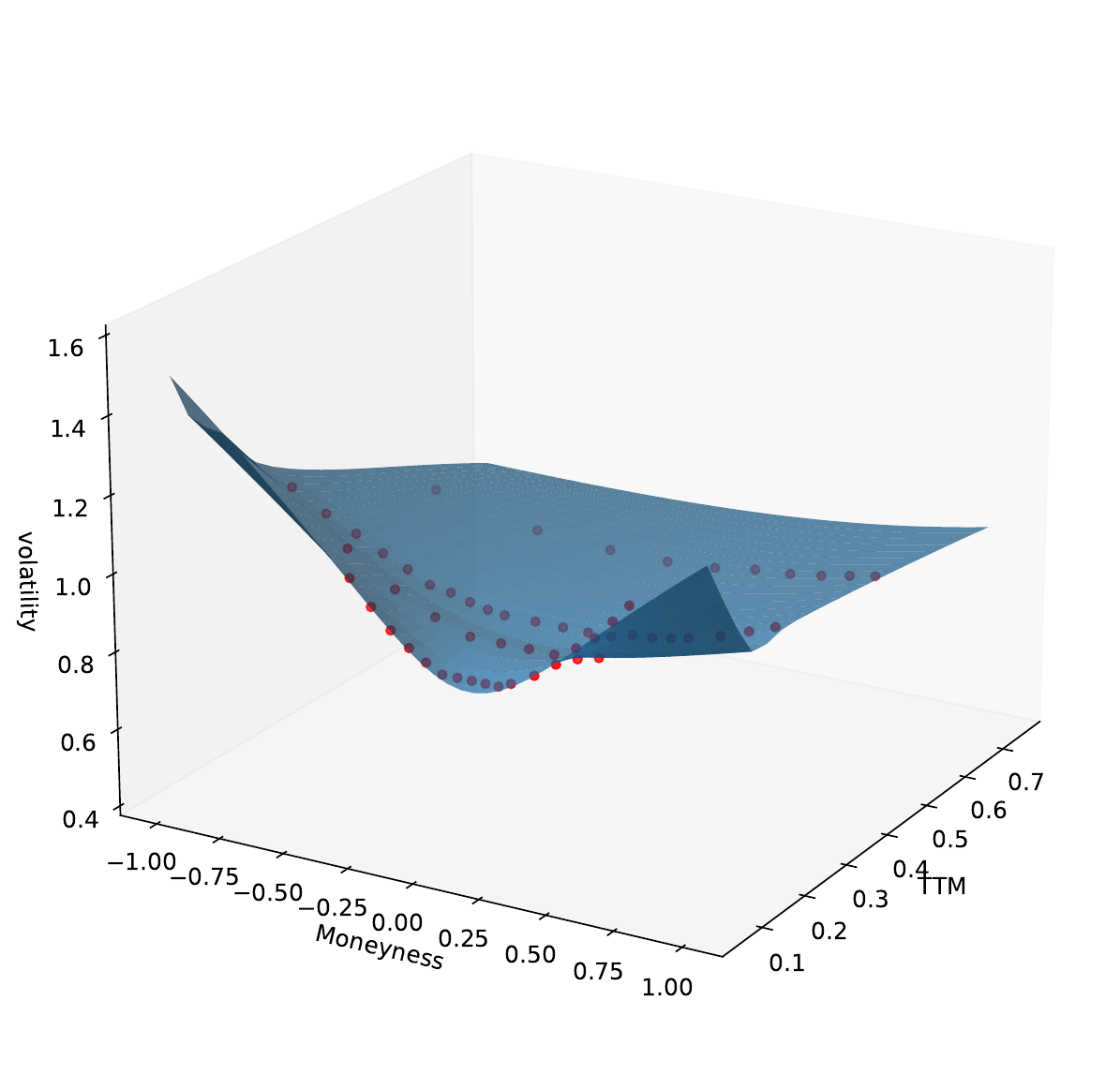}
    \caption{calm}
    \end{subfigure}
    \begin{subfigure}[b]{0.328\textwidth}\includegraphics[width=\textwidth]{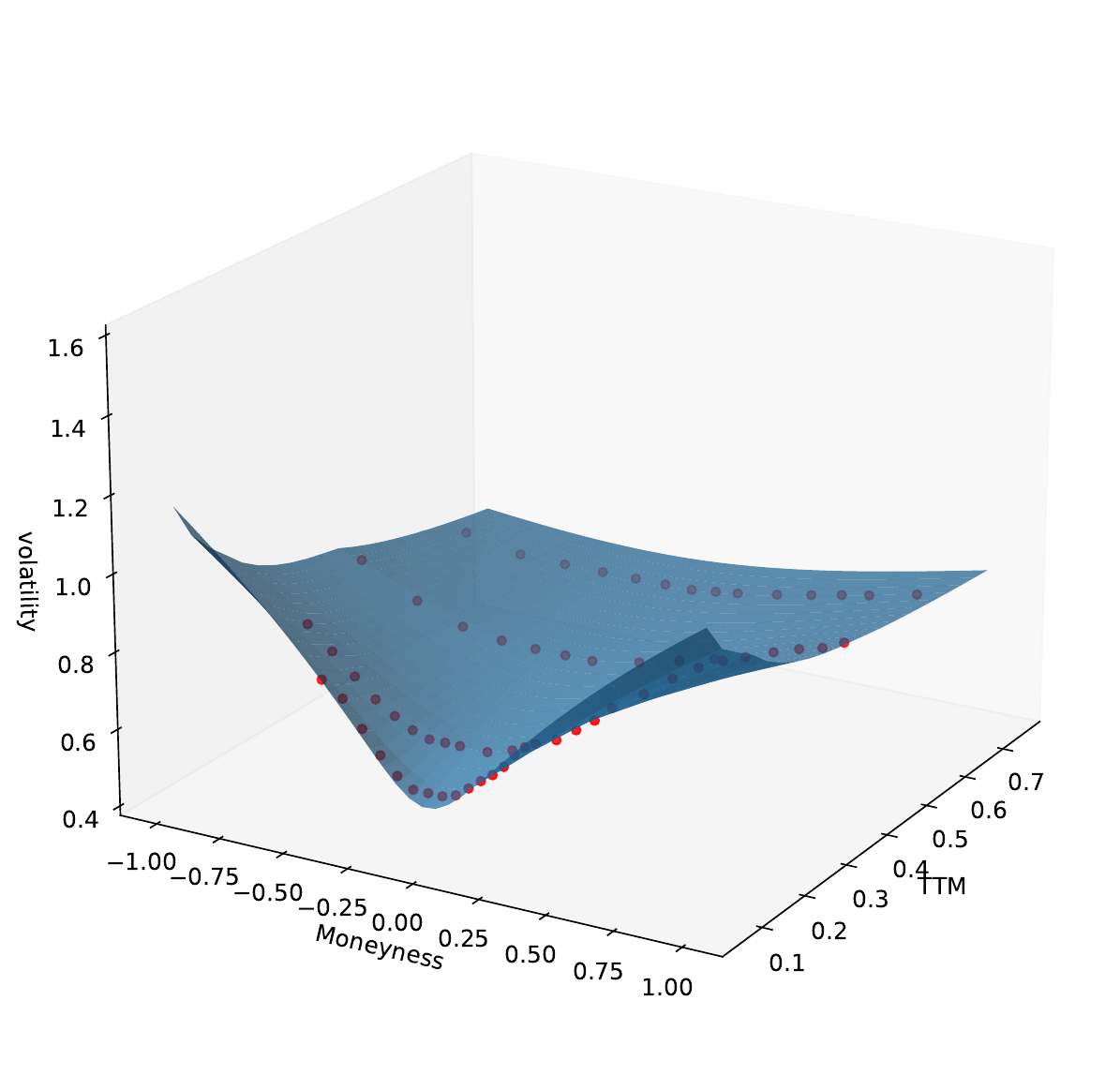}
     \caption{covid}
    \end{subfigure}
  \caption{\textcolor{red}{Market IVs} in \textcolor{red}{red} and \textcolor{blue}{interpolated IV surface} in \textcolor{blue}{blue} on (a) $1^{st}$ April 2019 (b) $1^{st}$ October 2019 (c) $1^{st}$ February 2020. Fitted smiles with very short maturities less than 1 week are excluded from plots, because they are not relevant for the hedging routine. Calibrated SVI parameters shorter maturities are given in Table \ref{tab:svi_parameters}. \includegraphics[scale=0.2]{quanlet.png}\href{https://github.com/QuantLet/hedging_cc}{ hedging\_cc}}
    \label{fig:ivsurface_interpolation}
\end{figure}

Option prices are obtained on every day of the hedging period. This is necessary for the calculation of the initial value of the hedging portfolio and to perform multi-asset dynamic hedging. Each option is priced according to the IV surface on the given day. If the option is not traded for the given strike or maturity, the SVI parametrized IV surface is interpolated in an arbitrage-free way. For illustration, we take a look at CC option prices at the beginning of each market period. 
Figure  \ref{fig:ivsurface_interpolation} displays the SVI parametrized interpolated IV surfaces for SVI parameters listed in Table \ref{tab:svi_parameters}. The resulting option prices used in the hedging routine are displayed in Table \ref{tab:svi_prices}. 
\begin{table}[t]
\centering
\scalebox{0.8}{
\begin{tabular}{lrrr}
\hline
\hline
   &  \multicolumn{1}{c}{$F_{0}$}
   & \multicolumn{1}{c}{1 M} & 
\multicolumn{1}{c}{3 M}    \\
\hline
\text{BULLISH}  &4088.16  & 206.38 &  417.87    \\
\text{CALM}    & 8367.51 & 838.01 & 1449.82 \\
\text{COVID} & 9804.85 & 610.36 & 1201.46   \\
\hline
\hline
\end{tabular}}
\caption{ Interpolated 1-month and 3-month ATM option prices. $F(0)$ denotes the price of the underlying at $t=0.$ \includegraphics[scale=0.2]{quanlet.png}\href{https://github.com/QuantLet/hedging_cc}{ hedging\_cc}
}
\label{tab:svi_prices}
\end{table}
Recall that for a given IV surface the SVI parameters related by the formula \eqref{eq:IV_params} are calibrated for each TTM. The temporal dynamics of the SVI parameters provide the following insights: parameter $a$ increases with TTM, which aligns with the increase of the ATM total variance as TTM rises. Parameter \textbf{$\sigma^{SVI}$} decreases with TTM, indicating decrease of the ATM curvature. Increasing values of parameter $b$ indicate higher slopes of the wings as TTM increases. Skewness, expressed in terms of the parameter \textbf{$\rho^{SVI}$}, varies across market segments. Usually negative values of $\rho^{SVI}$ indicate a preference for OTM puts over OTM calls. In the bullish period, skewness is close to zero across most maturities.

\subsection{Scenario generation results}\label{sec:scenario_gen}
\begin{figure}[ht!]
\begin{center}
\includegraphics[scale=0.35]{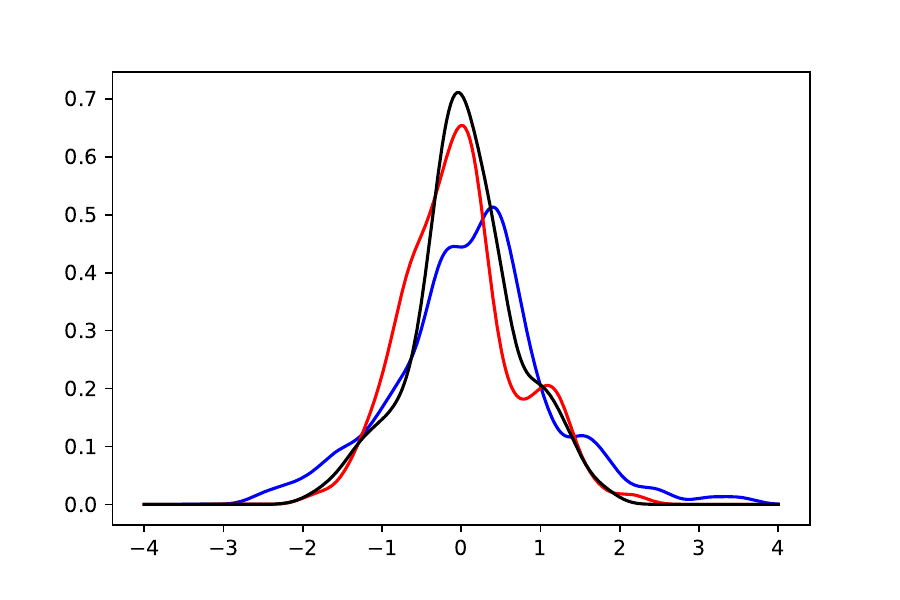}
\caption[Data]{Estimated residual density $\widehat{f}^{h}(z)$ in \eqref{eq:degarched_dist} during \textcolor{blue}{bullish market behavior}, \textcolor{black}{calm period} and the stressed scenario during 
\label{fig:densities}
\textcolor{red}{the Corona crisis} for $h=0.2$. \includegraphics[scale=0.2]{quanlet.png}\href{https://github.com/QuantLet/hedging_cc}{ hedging\_cc}
}
\end{center}
\end{figure}

\begin{figure}[ht!]
\centering
\begin{subfigure}[b]{0.328\textwidth}\includegraphics[width=\textwidth]{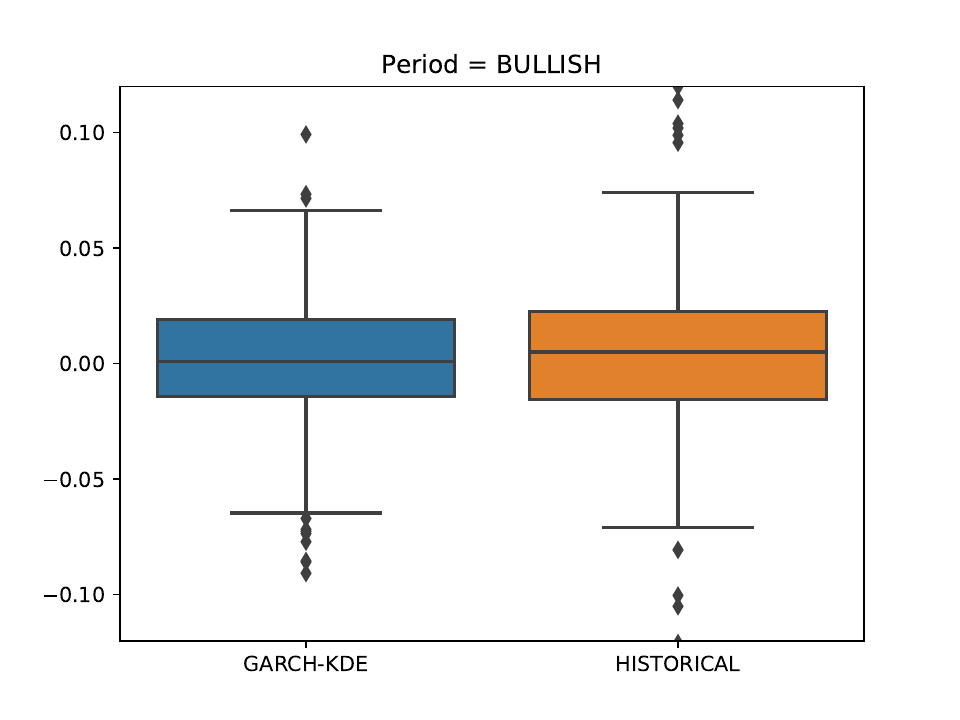}
    \caption{bullish}
\end{subfigure}
\begin{subfigure}[b]{0.328\textwidth}\includegraphics[width=\textwidth]{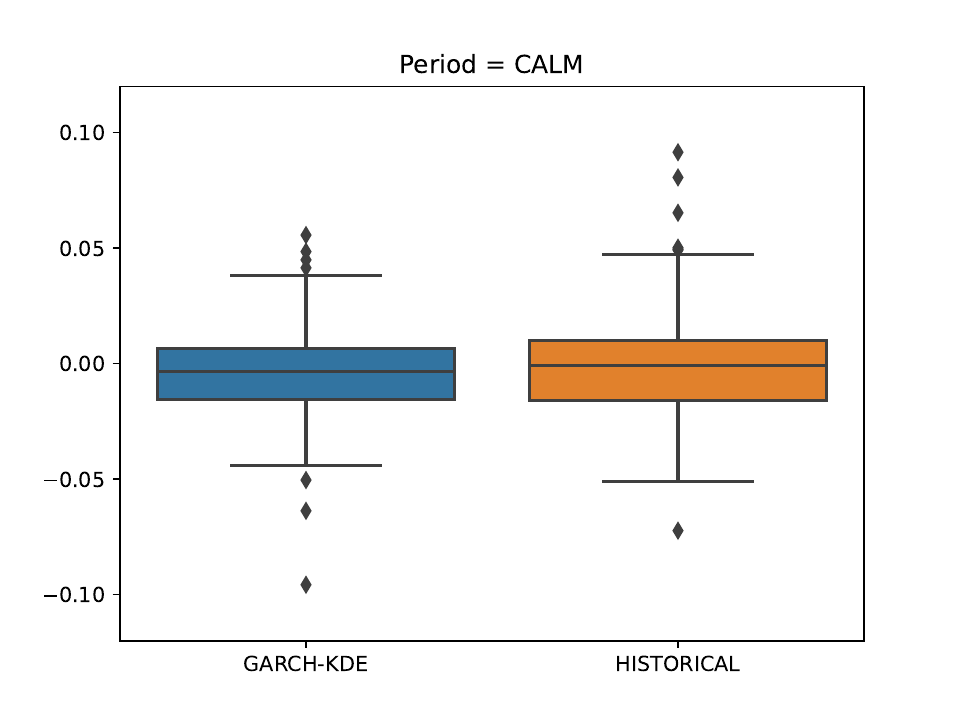}
    \caption{calm}
\end{subfigure}
\begin{subfigure}[b]{0.328\textwidth}\includegraphics[width=\textwidth]{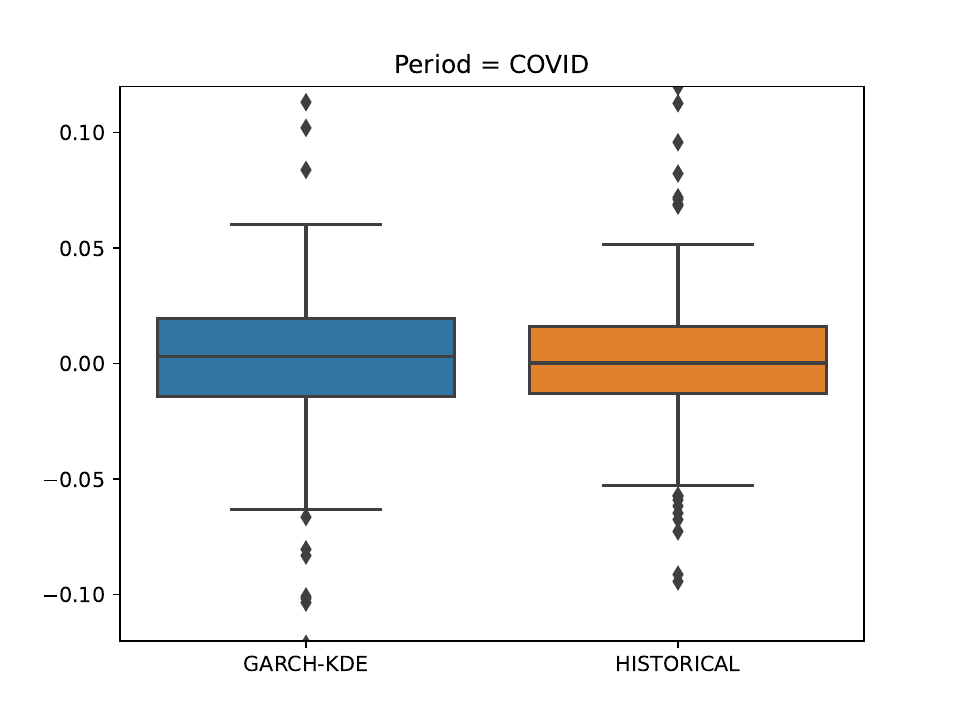}
 \caption{covid}
 \end{subfigure}
  \caption{Distributions of one sampled GARCH-KDE path and historical returns.
  \includegraphics[scale=0.2]{quanlet.png}\href{https://github.com/QuantLet/hedging_cc}{ hedging\_cc}}
    \label{fig:boxplots_oos}
\end{figure}

\begin{table}[t]
\centering
\scalebox{0.7}{
\begin{tabular}{lrrrrrrrr}
\hline
\hline
period   & \multicolumn{1}{c}{mean} & \multicolumn{1}{c}{std} & \multicolumn{1}{c}{skew} & \multicolumn{1}{c}{kurt} &
\multicolumn{1}{c}{$q^{25}$} & \multicolumn{1}{c}{$q^{50}$} & \multicolumn{1}{c}{$q^{75}$} \\
\hline
BULLISH &  0.13 &  0.99 & 0.17 & 0.87 & -0.44 &  0.15 &  0.66 \\
CALM    & -0.02 &  0.74 & 0.34 & 0.12 & -0.51 & -0.06 &  0.38  \\
COVID   &  0.05 &  0.70 & -0.04 & 0.23 & -0.34 &  0.04 &  0.47  \\
\hline
\hline
\end{tabular}}
\caption{Summary statistics of estimated historical densities $\hat{z_{t}}$ defined in 
\eqref{eq:degarched_res} for a respective scenario. \includegraphics[scale=0.2]{quanlet.png}\href{https://github.com/QuantLet/hedging_cc}{ hedging\_cc}}
\label{tab:summary stats densities}
\end{table}
For the GARCH-KDE approach, the estimated residual distributions $\widehat{f}^{h}(z)$ from \eqref{eq:degarched_dist} are displayed in Figure \ref{fig:densities}. The empirical moments and quantiles are listed in Table \ref{tab:summary stats densities}. Figure \ref{fig:garch_estimates} illustrates the GARCH(1, 1) volatility estimates of BTC returns and the 7-day historical volatility. As a consequence of \textit{de-garching}, all three distributions are roughly symmetric and mean-zero. Deviations are direct results from market moves: the upward-moving market behavior in the \textit{bullish} period leads to a left-skewed residual distribution. High drops in the \textit{stressed} period result in a negatively skewed distribution.
To demonstrate that the GARCH-KDE method is an appropriate method of sampling "close-to-actual-market" paths, the boxplots in Figure \ref{fig:boxplots_oos} illustrate the distributions of one simulated GARCH-KDE path and the corresponding historical distribution. The strength of the GARCH-KDE approach, of course, lies in the fact that through Monte Carlo simulation, the analysis is not restricted to one path.

SVCJ paths are simulated with daily re-calibrated parameters, which are summarized in Appendix Table \ref{tab:affine_calibration}. Selected statistical properties of both scenario generation approaches are given in Table \ref{tab:scenario_summary}. We observe differences in tails, extreme values and standard deviation. 
Discrepancies in $\hat{\sigma}$ are natural consequences from different methodological assumptions. The SVCJ approach assumes volatility to be stochastic, whereas GARCH-KDE models $\sigma_{t}$ with GARCH(1,1). Discrepancies in path extremes result from the SVCJ model assumptions on return jump size $\Xi$ in \eqref{eq:svcj_jumps}.
In the calibration routine, the $L_{2}$-regularization is applied to control extreme parameter values. Yet, estimated return jump sizes can be very large. Resulting Euler discretized paths contain trajectories with extreme moves of the underlying. These are e.g. extremely low and high prices during the calm and stressed scenario displayed in Table \ref{tab:scenario_summary}. The sometimes erratic BTC price evolution suggests that such price moves are entirely implausible.
\begin{table}[t]
\centering
\scalebox{0.7}{
\begin{tabular}{lrrrrrrrr}
\hline
\hline
\multicolumn{1}{c}{segment} &
 \multicolumn{1}{c}{$\hat{\mu}$} & \multicolumn{1}{c}{$\hat{\sigma}$} &
 \multicolumn{1}{c}{min} &
  \multicolumn{1}{c}{$q^{1}$} &
 \multicolumn{1}{c}{$q^{50}$} & 
\multicolumn{1}{c}{$q^{99}$} &
\multicolumn{1}{c}{max} \\
\hline
bullish & -0.03 &  0.18 & -0.39 & -0.37 & -0.00 &  0.46 &  0.61\\ calm & -0.23 &  0.24 & -0.44 & -0.43 & -0.34 &  0.53 &  0.58 \\   covid & -0.28 &  0.17 & -0.49 & -0.48 & -0.33 &  0.11 &  0.67 \\
\hline
\hline
\end{tabular}}
\caption{Summary statistics of calibrated SVCJ jump size $\Xi$ per market segment. \includegraphics[scale=0.2]{quanlet.png}\href{https://github.com/QuantLet/hedging_cc}{ hedging\_cc}}
\label{tab:svcj_jumps}
\end{table}

 \begin{table}[t]
\begin{center}
\scalebox{0.7}{
\begin{tabular}{lrrrrrrrrrrr}
\hline\hline
 period 
 & $\kappa$
 & $\rho$
 & $V_{0}$
 & $\theta$
 & $\sigma$
 & $\lambda$
 & $\mu_{y}$ 
 & $\sigma_{y}$ 
 & $\mu_v$ \\
\hline
$BS_{bullish}$ &     - &  - &  - &  - & 0.84 &  - &      - &   - & -\\
$BS_{calm}$    &   - &  - &  - &  - & 0.68 &  - &      - &   - & -\\
$BS_{covid}$    &   - &  - &  - &  - & 0.78 &  - &      - &   - & -\\
$Merton_{bullish}$ &   - &  - &  - &  - &   0.17 &  0.11 &      0.0 &   0.82 & -\\
$Merton_{calm}$    &   - &  - &  - &  - & 0.42 &  0.72 &      0.0 &   0.55 & -\\
$Merton_{covid}$   - &  - &  - &  - & &    0.48 &  0.40 &      0.0 &   0.69& -\\
$SV_{bullish}$ &  0.75 &  0.16 &  0.76 &  0.42 &  0.82 & - & - & - & - \\
$SV_{calm}$    &  1.60 &  0.17 &  0.35 &  1.10 &  0.68 & - & - & - & - \\ 
$SV_{covid}$   &  1.43 &  0.01 &  0.63 &  0.95 &  0.56 & - & - & - & - \\ 
$SVJ_{bullish}$ &  0.72 &  0.15 &  0.75 &  0.42 &  0.80 &    0.16 &  0.01 &      0.0 & - \\ 
$SVJ_{calm}$  &   1.28 &  0.18 &  0.33 &  1.05 &  0.68 &    0.37 &  0.01 &  0.0 & - \\ 
$SVJ_{covid}$   &  0.98 &  0.14 &  0.50 &  0.74 &  0.72 &    0.86 & -0.15 & 0.0 & - \\ 
$SVCJ_{bullish}$ &   0.51 &  0.14 &  0.74 &  0.09 &  0.88 &    0.31 & -0.04 &      0.0 &  0.45 \\
$SVCJ_{calm}$  &   0.75 &  0.28 &  0.30 &  0.38 &  0.83 &    0.85 & -0.30 & 0.0 &  0.99 \\
$SVCJ_{covid}$   &   0.61 &  0.22 &  0.52 &  0.18 &  0.89 &    1.04 & -0.35 &      0.0 &  0.54 \\
\hline\hline
\end{tabular}}
\caption{Average calibrated SVCJ parameters with market segments. \includegraphics[scale=0.2]{quanlet.png}\href{https://github.com/QuantLet/hedging_cc}{ hedging\_cc}
}
\label{tab:affine_calibration}
\end{center}
\end{table}
\subsection{Calibration results}
In each period, calibration is performed daily using instruments satisfying the liquidity and moneyness requirements specified in Section \ref{sec:calibration}. For an overview, average numbers of options per maturity range used for calibration are summarized in Table \ref{tab:calibrated_options}.
As a consequence of the moneyness requirement, more longer-dated options are selected. The average parameter values per period are summarized in Table \ref{tab:affine_calibration}. Section \ref{sec:affine_calibration} and 
\ref{sec:levy_calibration} 
provide a detailed perspective on the dynamics of the calibrated parameters. 
Calibration is carried out on the market's mid IVs. Of course, ignoring bid-ask spreads and the possibility of stale prices may produce arbitrage opportunities as well as spikes in parameters and calibration errors. However, this is considered a minor issue and ignored. RMSE's for the models are illustrated in Appendix \ref{sec:rmse_appedix}. Naturally, the model fit improves with increasing model complexity. Hence, the BS model has the highest RMSE values on average while the SVCJ model has the lowest.
\begin{table}[t]
\centering
\scalebox{0.7}{
\begin{tabular}{lrrrrr}
\hline
\hline 
segment / maturity     & $ \leq 1$ W  & $(1 \text{ W},  2\text{ W}]$   &  $ (2 \text{ W},  3\text{ M}]$  &  $ (3 \text{ M},  6 \text{ M}]$   & $ (6 \text{ M},  9\text{ M}]$    \\
\hline
bullish & 2.77 & 1.72 & 4.61 & 7.14 & 2.53 \\
calm    & 2.53 & 2.24 & 3.75 & 4.28 & 3.18 \\
crisis  & 3.00 & 3.03 & 4.44 & 5.58 & 5.33 \\
\hline
\hline
\end{tabular}}
\caption{Overview of average maturity counts of all options in a daily IV surface fullfiling the liquidity and moneyness requirements (Section \ref{sec:calibration}). \includegraphics[scale=0.2]{quanlet.png}\href{https://github.com/QuantLet/hedging_cc}{ hedging\_cc}}
\label{tab:calibrated_options}
\end{table}
\subsubsection{Affine jump diffusion models}\label{sec:affine_calibration}

\begin{figure}[ht!]
\begin{center}
\includegraphics[scale=0.3]{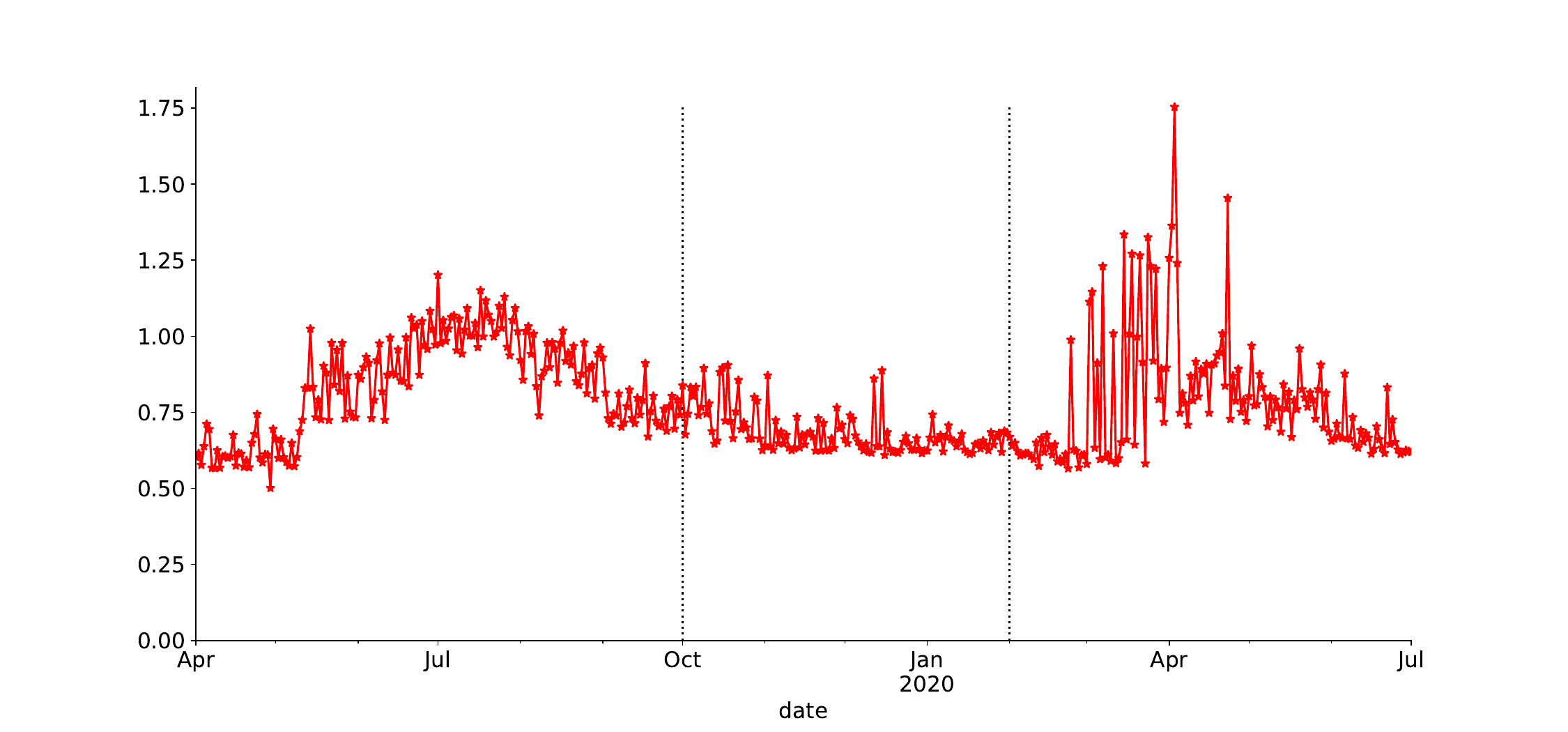}
\caption{Daily calibration \textcolor{red}{$\sigma^{BS}$} segregated by market segment in chronological order. Volatility levels are very high compared to equities or indices such as S \& P 500. \includegraphics[scale=0.2]{quanlet.png}\href{https://github.com/QuantLet/hedging_cc}{ hedging\_cc}}
\label{fig:calibration:black_scholes}
\end{center}
\end{figure}

\begin{figure}[ht!]
\begin{center}
\includegraphics[scale=0.27]{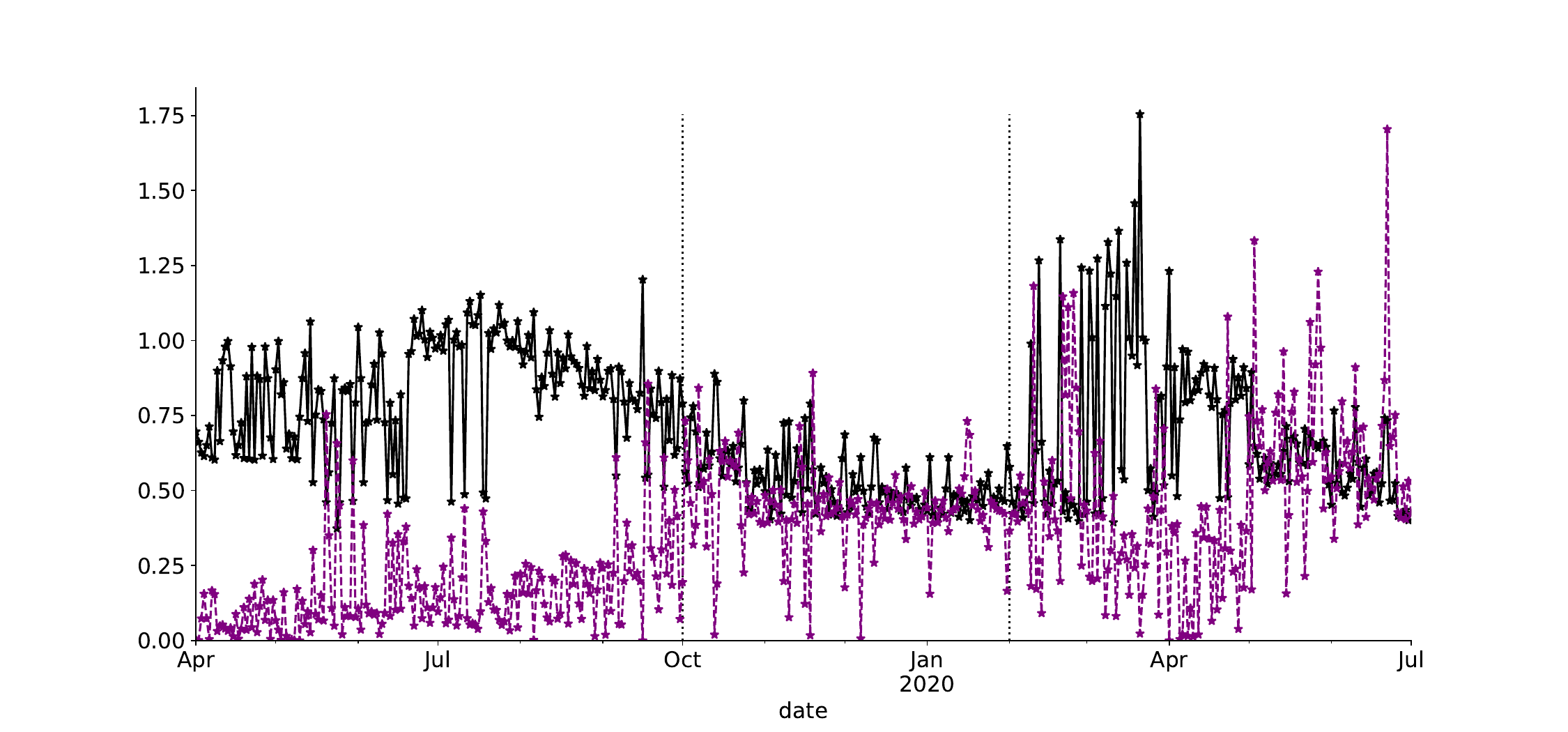}
\caption{Interplay between $\sigma^{JD}$ and \textcolor{purple}{$\lambda^{JD}$} segregated by market segment in chronological order. Mostly, for high levels of $\sigma^{JD}$ we observe low levels of \textcolor{purple}{$\lambda^{JD}$} and vice versa. \includegraphics[scale=0.2]{quanlet.png}\href{https://github.com/QuantLet/hedging_cc}{ hedging\_cc}}
\label{fig:calibration:interplay_merton}
\end{center}
\end{figure}

The calibrated parameter $\sigma^{BS}$ provides meaningful insights into market expectations. Levels vary in the range $\sigma^{BS} \in [50 \ \%,175 \ \%]$, with summary statistics for this parameter provided in Table \ref{tab:bs_calibration}. Due to the volatile nature of the CC markets, levels of $\sigma^{BS}$ are generally higher than in traditional markets \citep{Schoutens2019}. In comparison, the VIX index in the time period 1990-2021 ranges between 9.5\% and 60\%, with the 95\%-quantile at 33.5\%. Figure \ref{fig:calibration:black_scholes} shows the dynamics of $\sigma^{BS}$ over the entire time frame. In the bullish period, volatility levels rise up to $
120 \%$. In the calm period, as expected, the levels are lower than in the other two periods with $\sigma^{BS} \in [0.61,0.91]$.
\begin{table}[t]
     \centering
     \scalebox{0.7}{
      \begin{tabular}{lrrrrrrr}
      \hline 
      \hline
behavior     &     mean   &      std. dev. &      min &      $q^{25}$ &      $q^{50}$  &      $q^{75}$ &       max \\
\hline 
 bullish &  0.84 &  0.16 &  0.50 &  0.72 &  0.85 &  0.97 &  1.20 \\
 calm    &  0.68 &  0.06 &  0.61 &  0.64 &  0.66 &  0.70 &  0.89 \\
stressed   &  0.78 &  0.21 &  0.57 &  0.63 &  0.73 &  0.87 &  1.75 \\
\hline
\hline
 \end{tabular}}
     \caption{Summary statistics of daily $\sigma^{BS}$ calibration.\includegraphics[scale=0.2]{quanlet.png}\href{https://github.com/QuantLet/hedging_cc}{ hedging\_cc}}
\label{tab:bs_calibration}
 \end{table} 
Figure \ref{fig:calibration:interplay_merton} plots the calibrated parameters $\sigma^{JD}$ and $\lambda^{JD}$ of the JD model over time. In general, levels of $\sigma^{JD}$ are lower than $\sigma^{BS}$, clearly visible during the \textit{calm} and \textit{stressed} scenario. As the JD model is an extension of the BS model, higher levels of $\sigma^{BS}$ are partially compensated by the jump component. On many days, $\sigma^{JD}$ is close to $\sigma^{BS}$. The reason for this are generally low values of the 
annual jump intensity $\lambda^{JD}$ and jump size $\mu_{y}$. On average, the JD model expects less than one jump in returns per year.
The evolution of $\lambda^{JD}$ is compared to the jump intensities of extended models $\lambda^{SVJ}$ and $\lambda^{SVCJ}$ in Figure Appendix \ref{fig:calibration:lambda_jumps}. Throughout, annualised  jump intensities are low with mostly $\lambda^{SV(C)J} \leq 2.5 $. Overall, the conclusion is that jumps are infrequent. We observe contrasting levels of 
$\lambda^{SVCJ}$ and $\lambda^{JD}$. They are not directly comparable, as the jump intensity $\lambda^{SVCJ}$ contributes to simultaneous jumps in returns and stochastic volatility, while $\lambda^{JD}$ and $\lambda^{SVJ}$ corresponds solely to jumps in returns. For example, levels of $\lambda^{SVCJ}$ in the calm period are high whereas $\lambda^{SVJ}$ is close to zero. 

The plausibility of the stochastic volatility assumption is analyzed by the evolution and levels of $\sigma^{v}$. In most periods, levels of $\sigma^{v}$ are higher compared to traditional markets. In the broad picture, the evolution of $\sigma^{v}$ does not depend on model choice a shown in Figure Appendix \ref{fig:calibration:SV_volvol}. Table \ref{tab:v0_average} summarizes statistical properties of this parameter by model and market segment.  In the \textit{bullish} and \textit{calm} period, the indication for stochastic volatility is strong with vol-of-vol levels at $q^{50} \geq 80 \%$ and $q^{50} \geq 75\%$, respectively. In the \textit{stressed} period, levels of $\sigma^{v}$ in SV,SVJ,SVCJ remain high at $q^{50} \geq 73 \%$.

Empirical evidence suggests that in traditional markets the correlation parameter $\rho^{SV(CJ)}$ is usually negative. Specifically, when prices fall, volatility increases. However, across all three market segments and models, $\rho^{SV(CJ)}$ is mainly positive and close to zero as illustrated in Figure Appendix \ref{fig:calibration:SV_rho}. \cite{crixpricing2019} name this phenomenon the \textit{inverse leverage effect} in CC markets, that was previous reported on commodity markets by \cite{Schwartz2009}. 

This relationship in the CC markets is also supported by the correlation between the \href{https://thecrix.de/}{
\texttt{CRIX}} and the \href{https://thecrix.de/}{
\texttt{VCRIX}} under the physical measure $\mathbb P$. Pearson's correlation coefficient is $\rho^{pearson}=0.51$ in the \textit{bullish} and $\rho^{pearson}=0.64$ in the \textit{calm} period, respectively. In the stressed segment, correlation is negative with $\rho^{pearson}=-0.73$.
\subsubsection{VG and CGMY}\label{sec:levy_calibration}
\begin{figure}[ht!]
\begin{center}
\includegraphics[scale=0.27]{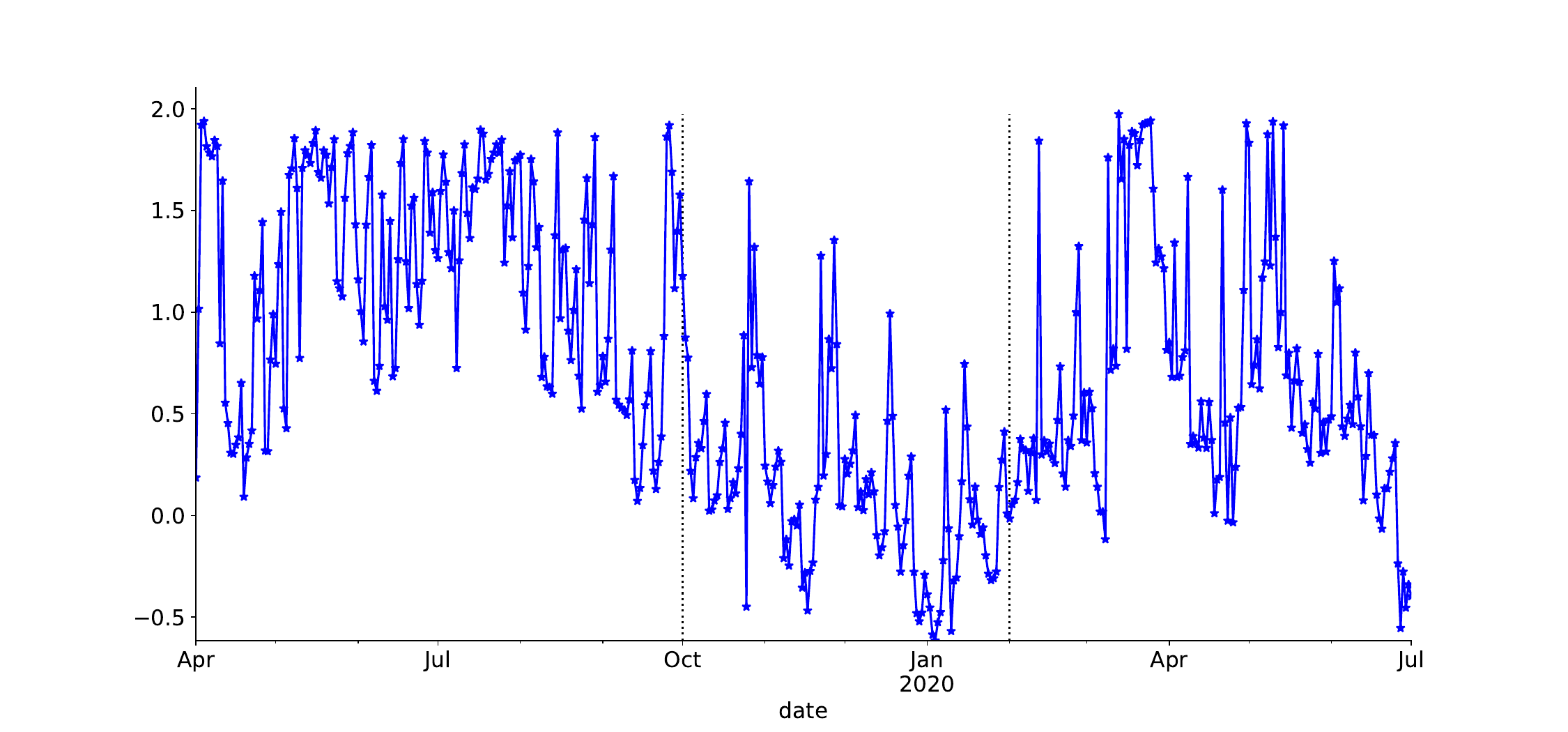}
\caption{Daily \textcolor{blue}{$Y^{CGMY}$} calibration segregated by market segment. Often, we observe $Y^{CGMY} >0$. This provides indication for infinite activity. As $Y^{CGMY} \in (1,2 ]$ in the bullish segment, there is evidence for infinite variation. \includegraphics[scale=0.2]{quanlet.png}\href{https://github.com/QuantLet/hedging_cc}{ hedging\_cc}}
\label{fig:calibration:cgmy_y}
\end{center}
\end{figure}

The prospect of infinite variation is evaluated by the calibration of the CGMY model with average calibrated parameters in Table \ref{tab:cgmy_calibration}. Precisely, we are interested in the evolution of the infinite activity parameter $Y^{CGMY}$ portrayed in Figure \ref{fig:calibration:cgmy_y}. As in each market segment we mostly have $Y^{CGMY} > 0$, there is evidence for infinite activity. In the bullish period, there is also evidence of infinite variation, as we mostly have for $Y^{CGMY} \in (1,2 ]$ \citep{Carr2002}.
\begin{table}[t]
     \centering
\scalebox{0.7}{
      \begin{tabular}{lrrrr}
      \hline 
      \hline
market segment & C &      G &      M &     Y \\
\hline 
$CGMY_{bullish}$ &   4.24 &  22.21 &  24.79 &  1.20 \\
$CGMY_{calm}$   &  10.37 &   7.67 &   9.30 &  0.14 \\
$CGMY_{covid}$  &   7.94 &  11.38 &  17.24 &  0.68 \\
\hline
\hline
 \end{tabular}}
     \caption{Average calibrated parameters of the CGMY model segregated by market segment. \includegraphics[scale=0.2]{quanlet.png}\href{https://github.com/QuantLet/hedging_cc}{ hedging\_cc}}
     \label{tab:cgmy_calibration}
 \end{table}
 The bullish period catches high magnitudes of jump size direction increase parameters $G_{CGMY}$ and $M_{CGMY}$, 
 reflecting the nature of this market segment. Similarly, the increase in decreased jump size parameter $M_{CGMY}$ is mainly higher in the stressed scenario. 
 A graphical illustration is given in Figure Appendix \ref{fig:cgmy_jumps}. 
 The \textbf{VG} is calibrated under representation \eqref{eq:vgchar}. Overall, volatility levels of $\sigma^{VG}$ are comparable to $\sigma^{BS}$, as illustrated in Figure Appendix \ref{fig:calibration:vg_sigma}. 
 
\subsection{Hedge results}\label{sec:hedge_results}
\begin{figure}[ht!]
    \centering
    \begin{subfigure}[b]{0.45\textwidth}\includegraphics[width=\textwidth]{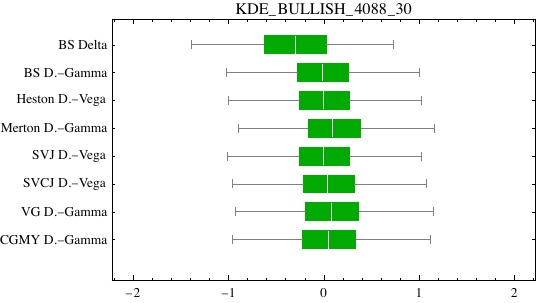}
    \caption{GARCH-KDE bullish}
    \end{subfigure}
    \begin{subfigure}[b]{0.45\textwidth}\includegraphics[width=\textwidth]{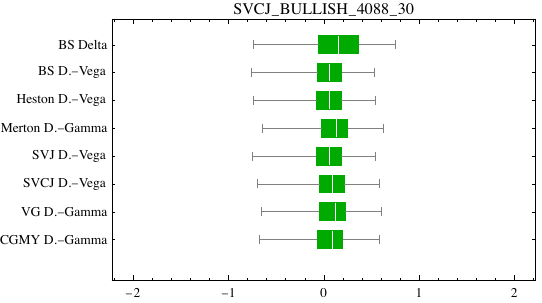}
    \caption{SVCJ bullish }
    \end{subfigure}
    \begin{subfigure}[b]{0.45\textwidth}\includegraphics[width=\textwidth]{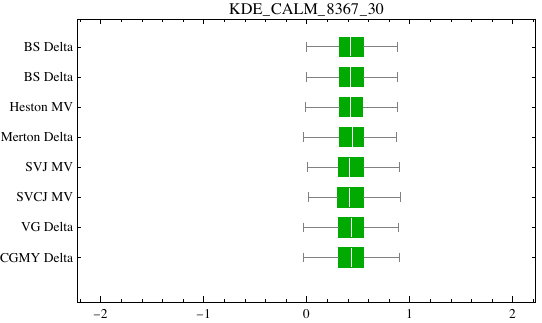}
    \caption{GARCH-KDE calm}
    \end{subfigure}
    \begin{subfigure}[b]{0.47\textwidth}\includegraphics[width=\textwidth]{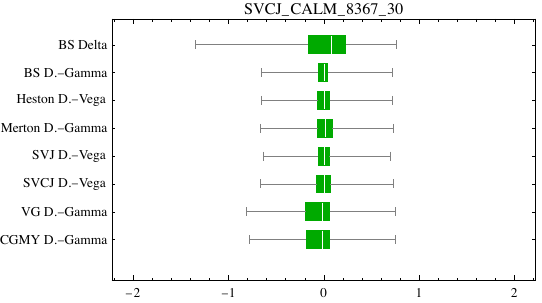}
    \caption{SVCJ calm}
    \end{subfigure}
\begin{subfigure}[b]{0.45\textwidth}\includegraphics[width=\textwidth]{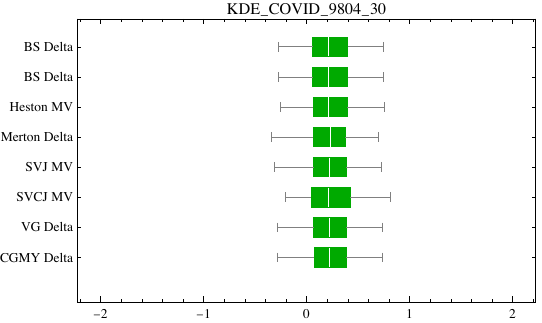}
    \caption{GARCH-KDE Covid}
    \end{subfigure}
    \begin{subfigure}[b]{0.47\textwidth}\includegraphics[width=\textwidth]{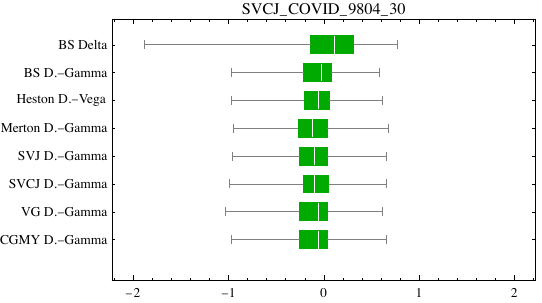}
    \caption{SVCJ Covid}
    \end{subfigure}
  \caption{1-month option hedge performance boxplots of $\pi^{rel}$ under (a) GARCH-KDE and (b) SVCJ market simulation. For illustrative purposes $\pi^{rel}$ is truncated at $q^{5}$ and $q^{95}$. The vertical axis portrays $\Delta^{BS}$ hedge results compared each model's
  best performing strategy. This best performing strategy is selected according to the minimal $\text{ES}^{5 \%}$.}
    \label{fig:pnl_boxplots_1M}
\end{figure}

\begin{figure}[ht!]
    \centering
    \begin{subfigure}[b]{0.45\textwidth}\includegraphics[width=\textwidth]{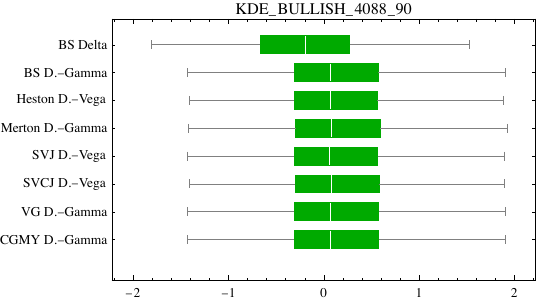}
    \caption{GARCH-KDE bullish}
    \end{subfigure}
    \begin{subfigure}[b]{0.45\textwidth}\includegraphics[width=\textwidth]{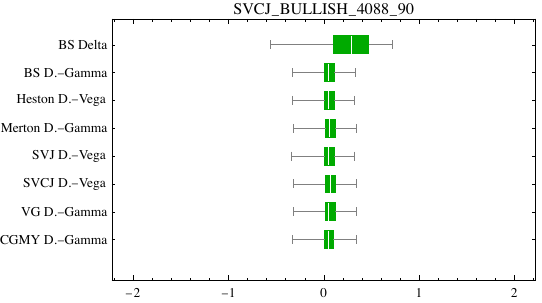}
    \caption{SVCJ bullish }
    \end{subfigure}
    \begin{subfigure}[b]{0.45\textwidth}\includegraphics[width=\textwidth]{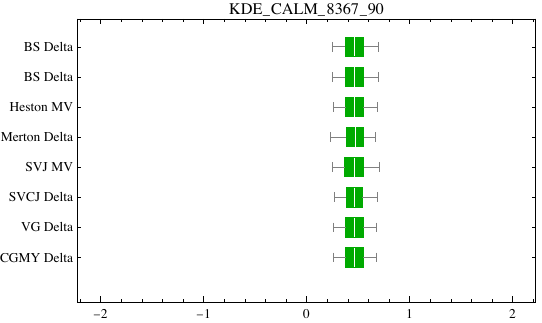}
    \caption{GARCH-KDE calm}
    \end{subfigure}
    \begin{subfigure}[b]{0.47\textwidth}\includegraphics[width=\textwidth]{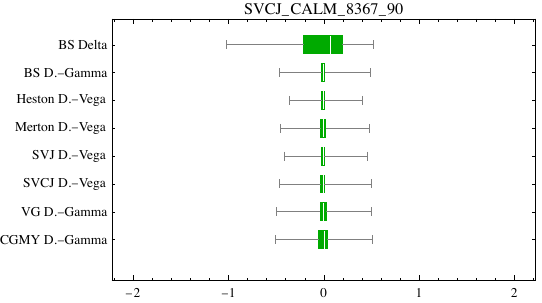}
    \caption{SVCJ calm}
    \end{subfigure}
\begin{subfigure}[b]{0.45\textwidth}\includegraphics[width=\textwidth]{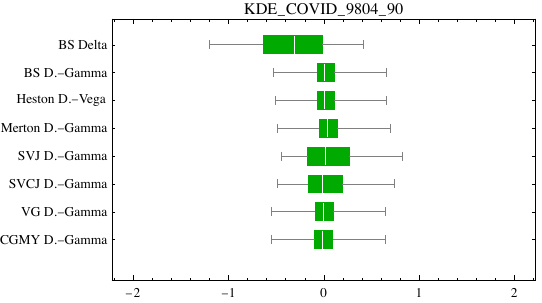}
    \caption{GARCH-KDE covid}
    \end{subfigure}
    \begin{subfigure}[b]{0.47\textwidth}\includegraphics[width=\textwidth]{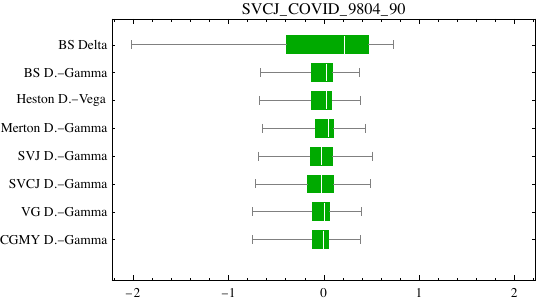}
    \caption{SVCJ covid}
    \end{subfigure}
  \caption{3-month option hedge performance boxplots of $\pi^{rel}$ under (a) GARCH-KDE and (b) SVCJ market simulation. For illustrative purposes $\pi^{rel}$ is truncated at $q^{5}$ and $q^{95}$. The vertical axis portrays $\Delta^{BS}$ hedge results compared each model's
  best performing strategy. This best performing strategy is selected according to the minimal $\text{ES}^{5 \%}$.
  \includegraphics[scale=0.2]{quanlet.png}\href{https://github.com/QuantLet/hedging_cc}{ hedging\_cc}}
\label{fig:pnl_boxplots_3M}
\end{figure}

At the beginning of each market period, we short 1- and 3-month ATM options with option premiums listed  in Table \ref{tab:svi_prices}. 
As outlined earlier, the price process is simulated in both SVCJ and the GARCH-KDE setting. The exposure in each option is dynamically hedged using the strategies summarized in Table \ref{tab:Hedge model summary}. The hedge performance is evaluated in terms of the hedge error $\varepsilon^{rel}$ and tail measures $\text{ES}^{5 \%}$ and  $\text{ES}^{95 \%}$. The hedge results are shown in Tables \ref{tab:1m_hedges} and \ref{tab:3m_hedges}.
For a concise graphical representation, the best performing hedge strategies across models are compared in boxplots displayed in Figures \ref{fig:pnl_boxplots_1M} and Figures \ref{fig:pnl_boxplots_3M}.
For each model, the best performing strategy is selected according to $\operatorname{ES}^{5 \%}$, as this provides a trade-off between an extreme, yet plausible tail summary.
%

These are the main findings: First, with some exceptions, using multiple instruments for hedging, i.e., Delta-Gamma and Delta-Vega hedges, when compared to a simple Delta-hedge lead to a substantial reduction in tail risk. Hence, whenever liquidly traded options are available for hedging, they should be used.

Exceptions are the calm and COVID periods in the GARCH-KDE approach for the short-maturity option as well as the calm period and GARCH-KDE approach for the long-date option  -- here, no significant improvement is achieved by including a second hedge instrument. In any case, no deterioration takes place when using a second security for hedging. Contrary to the SVCJ approach, which models several risk factors (jumps, stochastic volatility) explicitly, the GARCH-KDE approach, with a smooth KDE density, exhibits less sensitivity to concrete risk factors (e.g.\ Vega) in the calm period, even despite the GARCH filter, see Figure \ref{fig:garch_estimates}.

Second, for short-dated options, no substantial differences occur in the optimal hedging strategies across models. The sole exception is worse performance of the VG- and CGMY-models in the calm period when price paths are generated in the SVCJ model. 

Third, turning to the long-dated option, although not always best performing, it can be said that stochastic volatility models perform {\em consistently\/} well. Amongst the stochastic volatility models, the SV model as the simplest model, does not underperform and sometimes even is the best-performing model. For the choice of a SV hedge model, the $\Delta^{SV}$-$\mathcal{V}^{SV}$ hedge is a replicating strategy \citep{Kurpiel999Heston} and performs often better than other models under the same or different strategies. 
As calibrated jump intensities $\lambda^{SVJ}$ and $\lambda^{SVCJ}$ are low, the SVJ or SVCJ are often similar to the SV leading to comparable hedge results. 

The simulated hedge results are confirmed in the historical hedge backtest. As before, with expections (calm period), hedges involving multiple hedge instruments consistently achieve desirable variance and tail risk reduction. For example, in the bullish period, the $\Delta$-$\mathcal{V}_{SV}$ strategy strikingly outperforms other best performing strategies.
\clearpage
\begin{table}[t!]
\centering
\scalebox{0.7}{
\begin{tabular}{rrrrrrrrr}
\hline
\hline
 &  &  &  & \textbf{bullish}  &  &  &  & \\
\hline
\text{GARCH-KDE} & $\Delta_{BS}$ & $\Delta$-$\Gamma_{BS}$ & $\Delta$-$\mathcal{V}_{SV}$ & $\Delta$-$\Gamma_{JD}$ & $\Delta$-$\mathcal{V}_{SVJ}$ & $\Delta$-$\mathcal{V}_{SVCJ}$ & $\Delta$-$\Gamma_{VG}$ & $\Delta$-$\Gamma_{CGMY}$ \\
 \text{Min} & \textcolor{red}{\textbf{-3.35}} & -2.58 & -2.62 & \textcolor{green}{\textbf{-2.48}} & -2.63 & -2.59 & -2.51 & -2.53 \\
$ \text{ES}^{5 \%}$ & \textcolor{red}{\textbf{-1.75}} & -1.34 & -1.32 & \textcolor{green}{\textbf{-1.21}} & -1.32 & -1.27 & -1.24 & -1.27 \\
$ \text{ES}^{95 \%}$ & \textcolor{green}{\textbf{1.17}} & 1.49 & 1.51 & \textcolor{red}{\textbf{1.65}} & 1.5 & 1.57 & 1.64 & 1.61 \\
 \text{Max} & \textcolor{red}{\textbf{3.31}} & 5.32 & 5.29 & 5.33 & 5.28 & \textcolor{green}{\textbf{5.35}} & 4.77 & 5.05 \\
 $\varepsilon^{rel}$ & \textcolor{red}{\textbf{63.14}} & 59.55 & 59.39 & 60.43 & \textcolor{green}{\textbf{59.40}} & 59.75 & 60.97 & 60.87 \\
\hline
\text{SVCJ} & $\Delta_{BS}$ & $\Delta$-$\mathcal{V}_{BS}$ & $\Delta$-$\mathcal{V}_{SV}$ & $\Delta$-$\Gamma_{JD}$ & $\Delta$-$\mathcal{V}_{SVJ}$ & $\Delta$-$\mathcal{V}_{SVCJ}$ & $\Delta$-$\Gamma_{VG}$ & $\Delta$-$\Gamma_{CGMY}$ \\
 \text{Min} & \textcolor{red}{\textbf{-11.35}} & -9.46 & -9.65 & -9.69 & -9.65 & -9.58 & -8.13 & \textcolor{green}{\textbf{-8.07}} \\
$ \text{ES}^{5 \%}$ & \textcolor{red}{\textbf{-1.48}} & -1.16 & -1.16 & \textcolor{green}{\textbf{-1.06}} & -1.16 & -1.12 & -1.08 & -1.10 \\
$ \text{ES}^{95 \%}$ & 1.02 & \textcolor{green}{\textbf{0.98}} & \textcolor{green}{\textbf{0.98}} & 1.11 & \textcolor{green}{\textbf{0.98}} & 1.04 & \textcolor{red}{\textbf{1.12}} & 1.10 \\
 \text{Max} & \textcolor{red}{\textbf{18.69}} & 20.15 & 20.46 & 20.51 & 20.46 & 20.58 & 22.56 & \textcolor{green}{\textbf{24.47}} \\
 $\varepsilon^{rel}$ & \textcolor{red}{\textbf{56.12}} & 50.7 & 50.2 & 51.36 & \textcolor{green}{\textbf{49.86}} & 50.37 & 52.32 & 52.56 \\
\hline
 &  &  &  & \textbf{calm} &  &  &  & \\
\hline
\text{GARCH-KDE}& $\Delta_{BS}$ & $\Delta_{BS}$ &
$\text{MV}_{SV}$ & $\Delta_{JD}$ & $\text{MV}_{SVJ}$ & $\text{MV}_{SVCJ}$ & $\Delta_{VG}$ & $\Delta_{CGMY}$ \\
 \text{Min} & \textcolor{green}{\textbf{-0.94}} &\textcolor{green}{\textbf{-0.94}} & 
  -1.01 & -1.07 & -1.03 & -1.1 & -1.16 & \textcolor{red}{\textbf{-1.18}} \\
$ \text{ES}^{5 \%}$ & -0.16 & -0.16  & -0.17 & -0.19 & \textcolor{green}{\textbf{-0.15}} & -0.15 & \textcolor{red}{\textbf{-0.2}} & \textcolor{red}{\textbf{-0.2}} \\
$ \text{ES}^{95 \%}$ & 1.04 & 1.04 & 1.05 & \textcolor{green}{\textbf{1.03}} & 1.07 & \textcolor{red}{\textbf{1.09}} & 1.08 & 1.08 \\
 \text{Max} & \textcolor{red}{\textbf{1.77}} & \textcolor{red}{\textbf{1.77}} & 1.81 & 1.8 & \textcolor{green}{\textbf{1.91}} & 1.86 & 1.8 & 1.81 \\
 $\varepsilon^{rel}$ & \textcolor{green}{\textbf{25.44}} & \textcolor{green}{\textbf{25.44}} & 25.52 & 25.97 & 25.78 & 26.01 & 26.8 & \textcolor{red}{\textbf{26.87}} \\
\hline
\text{SVCJ} & $\Delta_{BS}$ & $\Delta$-$\Gamma_{BS}$ & $\Delta$-$\mathcal{V}_{SV}$ & $\Delta$-$\Gamma_{JD}$ & $\Delta$-$\mathcal{V}_{SVJ}$ & $\Delta$-$\mathcal{V}_{SVCJ}$ & $\Delta$-$\Gamma_{VG}$ & $\Delta$-$\Gamma_{CGMY}$ \\
 \text{Min} & \textcolor{red}{\textbf{-8.07}} & \textcolor{green}{\textbf{-4.45}} & \textcolor{green}{\textbf{-4.45}} & -5.07 & \textcolor{green}{\textbf{-4.45}} & -4.46 & -5.04 & -6.24 \\
$ \text{ES}^{5 \%}$ & \textcolor{red}{\textbf{-2.20}} & -1.01 & -1.00 & -1.01 & \textcolor{green}{\textbf{-0.96}} & -1.01 & -1.19 & -1.14 \\
$ \text{ES}^{95 \%}$ & 1.13 & 1.12 & 1.12 & 1.13 & \textcolor{red}{\textbf{1.09}} & 1.13 & 1.15 & \textcolor{green}{\textbf{1.17}} \\
 \text{Max} & 8.81 & 8.86 & 8.88 & \textcolor{green}{\textbf{12.07}} & 8.88 & 9.69 & \textcolor{red}{\textbf{8.73}} & 9.95 \\
 $\varepsilon^{rel}$ & \textcolor{red}{\textbf{67.72}} & 43.78 & 43.66 & 44.58 & \textcolor{green}{\textbf{42.29}} & 44.34 & 48.69 & 48.24 \\
\hline
 &  &  &  & \textbf{covid} &  &  &  & \\
\hline
\text{GARCH-KDE}& $\Delta_{BS}$ & $\Delta_{BS}$
& $\text{MV}_{SV}$ & $\Delta_{JD}$ & $\text{MV}_{SVJ}$ & $\text{MV}_{SVCJ}$ & $\Delta_{VG}$ & $\Delta_{CGMY}$ \\
 \text{Min} & \textcolor{red}{\textbf{-1.39}}
 & \textcolor{red}{\textbf{-1.39}}
 & -1.28 & -1.38 & -1.29 & \textcolor{green}{\textbf{-1.23}} & \textcolor{red}{\textbf{-1.39}} & \textcolor{red}{\textbf{-1.39}} \\
$ \text{ES}^{5 \%}$ & -0.49  & -0.49 & -0.46 & \textcolor{red}{\textbf{-0.55}} & -0.51 & \textcolor{green}{\textbf{-0.39}} & -0.48 & -0.48 \\
$ \text{ES}^{95 \%}$ & 0.88 & 0.88 & 0.89 & \textcolor{red}{\textbf{0.83}} & 0.87 & \textcolor{green}{\textbf{0.96}} & 0.88 & 0.88 \\
 \text{Max} & 1.37 & 1.37 & 1.39 & \textcolor{red}{\textbf{1.33}} & 1.38 & \textcolor{green}{\textbf{1.54}} & 1.44 & 1.43 \\
 $\varepsilon^{rel}$ & 30.21 & 30.21 & \textcolor{green}{\textbf{29.52}} & 30.3 & 30.08 & \textcolor{red}{\textbf{30.78}} & 29.62 & 29.56 \\
\hline
\text{SVCJ} & $\Delta_{BS}$ & $\Delta$-$\Gamma_{BS}$ & $\Delta$-$\mathcal{V}_{SV}$ & $\Delta$-$\Gamma_{JD}$ & $\Delta$-$\Gamma_{SVJ}$ & $\Delta$-$\Gamma_{SVCJ}$ & $\Delta$-$\Gamma_{VG}$ & $\Delta$-$\Gamma_{CGMY}$ \\
 \text{Min} & -16.51 & -10.93 & \textcolor{green}{\textbf{-10.88}} & -14.36 & -14.92 & \textcolor{red}{\textbf{-29.05}} & -24.66 & -17.07 \\
$ \text{ES}^{5 \%}$ & \textcolor{red}{\textbf{-3.13}} & \textcolor{green}{\textbf{-1.64}} & -1.72 & -1.76 & -1.76 & -1.84 & -1.85 & -1.75 \\
$ \text{ES}^{95 \%}$ & 1.08 & \textcolor{red}{\textbf{0.98}} & 1.01 & 1.09 & 1.08 & \textcolor{green}{\textbf{1.11}} & 1.00 & 1.06 \\
 \text{Max} & 7.74 & 8.92 & \textcolor{red}{\textbf{7.00}} & \textcolor{green}{\textbf{21.48}} & 14.13 & 20.24 & 11.11 & 11.54 \\
 $\varepsilon^{rel}$ & \textcolor{red}{\textbf{88.09}} & \textcolor{green}{\textbf{56.03}} & 57.62 & 60.19 & 60.53 & 63.85 & 61.3 & 58.33 \\
\hline
\hline
\end{tabular}
}
\caption{1-month ATM hedge performances with the \textcolor{green}{\textbf{best}} and  \textcolor{red}{\textbf{worst}} performing strategy. The table corresponds to results in Figure \ref{fig:pnl_boxplots_1M}.
}
\label{tab:1m_hedges}
\end{table}
\clearpage

\clearpage

\begin{table}[t!]
\centering
\scalebox{0.7}{
\begin{tabular}{rrrrrrrrr}
\hline
\hline
 &  &  &  & \textbf{bullish}  &  &  &  & \\
\hline
 \text{GARCH KDE}& $\Delta_{BS}$ & $\Delta$-$\Gamma_{BS}$ & $\Delta$-$\mathcal{V}_{SV}$ & $\Delta$-$\Gamma_{JD}$ & $\Delta$-$\mathcal{V}_{SVJ}$ & $\Delta$-$\mathcal{V}_{SVCJ}$ & $\Delta$-$\Gamma_{VG}$ & $\Delta$-$\Gamma_{CGMY}$ \\
\text{Min} & \textcolor{red}{\textbf{-6.55}} & -6.36 & -6.35 & \textcolor{green}{\textbf{-6.32}} & -6.35 & -6.34 & -6.36 & -6.37 \\
 $ \text{ES}^{5 \%}$ & \textcolor{red}{\textbf{-2.38}} & -1.99 & \textcolor{green}{\textbf{-1.95}} & -1.96 & -1.97 & \textcolor{green}{\textbf{-1.95}} & -1.98 & -1.99 \\
 $ \text{ES}^{95 \%}$ & \textcolor{green}{\textbf{2.43}} & 2.83 & 2.8 & \textcolor{red}{\textbf{2.85}} & 2.81 & 2.81 & 2.83 & 2.83 \\
 \text{Max} & 11.46 & 11.73 & 11.74 & \textcolor{red}{\textbf{11.76}} & \textcolor{green}{\textbf{11.00}} & 11.73 & 11.72 & 11.71 \\
 $\varepsilon^{rel}$ & \textcolor{red}{\textbf{101.91}} & 101.76 & \textcolor{green}{\textbf{100.30}} & 101.77 & 101.02 & 100.72 & 101.75 & 101.75 \\
\hline
\text{SVCJ} & $\Delta_{BS}$ & $\Delta$-$\Gamma_{BS}$ & $\Delta$-$\mathcal{V}_{SV}$ & $\Delta$-$\Gamma_{JD}$ & $\Delta$-$\mathcal{V}_{SVJ}$ & $\Delta$-$\mathcal{V}_{SVCJ}$ & $\Delta$-$\Gamma_{VG}$ & $\Delta$-$\Gamma_{CGMY}$ \\
 \text{Min} & \textcolor{red}{\textbf{-14.67}} & -11.58 & -11.57 & -11.51 & -11.55 & \textcolor{green}{\textbf{-9.30}} & -11.6 & -11.6 \\
 $\text{ES}^{5 \%}$ & \textcolor{red}{\textbf{-1.10}} & -0.64 & -0.63 & -0.62 & -0.63 & \textcolor{green}{\textbf{-0.62}} & -0.63 & -0.63 \\
$\text{ES}^{95 \%}$ & \textcolor{red}{\textbf{0.84}} & 0.64 & 0.62 & 0.66 & \textcolor{green}{\textbf{0.62}} & 0.64 & 0.65 & 0.65 \\
 \text{Max} & 10.14 & \textcolor{red}{\textbf{11.42}} & 11.29 & 11.34 & 11.26 & \textcolor{green}{\textbf{9.02}} & 11.27 & 11.27 \\
$\varepsilon^{rel}$ & \textcolor{red}{\textbf{44.14}} & 26.5 & 25.86 & 26.45 & 25.89 & \textcolor{green}{\textbf{25.26}} & 26.55 & 26.39 \\
\hline
 &  &  &  & \textbf{calm}  &  &  &  & \\
\hline
GARCH-KDE &  $\Delta_{BS}$ &  $\Delta_{BS}$  & $\text{MV}_{SV}$ & $\Delta_{JD}$ & $\text{MV}_{SVJ}$ & $\Delta_{SVCJ}$ & $\Delta_{VG}$ & $\Delta_{CGMY}$ \\
\text{Min} & \textcolor{red}{\textbf{-0.29}}
& \textcolor{red}{\textbf{-0.29}} 
& -0.27 & -0.28 & \textcolor{green}{\textbf{-0.25}} & \textcolor{green}{\textbf{-0.25}} & -0.28 & -0.28 \\
 $\text{ES}^{5 \%}$ & 0.18 & 0.18 & \textcolor{red}{\textbf{0.20}} & \textcolor{green}{\textbf{0.15}} & 0.19 & \textcolor{red}{\textbf{0.20}} & 0.19 & 0.19 \\
$\text{ES}^{95 \%}$ & 0.76 & 0.76 & 0.76 & \textcolor{green}{\textbf{0.73}} & \textcolor{red}{\textbf{0.77}} & 0.75 & 0.75 & 0.75 \\
 \text{Max} & \textcolor{green}{\textbf{1.04}} 
 & \textcolor{green}{\textbf{1.04}}
 & 1.06 & 1.05 & \textcolor{red}{\textbf{1.12}} & 1.07 & 1.12 & 1.12 \\
 $\varepsilon^{rel}$ & 13.59 & 13.59 & 13.11 & 13.53 & \textcolor{red}{\textbf{13.82}} & \textcolor{green}{\textbf{12.82}} & 13.18 & 13.18 \\
\hline
\text{SVCJ} & $\Delta_{BS}$ & $\Delta$-$\Gamma_{BS}$ & $\Delta$-$\mathcal{V}_{SV}$ & $\Delta$-$\mathcal{V}_{JD}$ & $\Delta$-$\mathcal{V}_{SVJ}$ & $\Delta$-$\mathcal{V}_{SVCJ}$ & $\Delta$-$\Gamma_{VG}$ & $\Delta$-$\Gamma_{CGMY}$ \\
 \text{Min} & -12.63 & -8.68 & \textcolor{red}{\textbf{-12.75}} & \textcolor{green}{\textbf{-6.32}} & -7.79 & -12.75 & -12.73 & -12.74 \\
 $\text{ES}^{5 \%}$ & \textcolor{red}{\textbf{-1.56}} & -0.85 & \textcolor{green}{\textbf{-0.71}} & -0.79 & -0.78 & -0.89 & -0.96 & -0.97 \\
$\text{ES}^{95 \%}$ & 0.88 & 0.82 & \textcolor{green}{\textbf{0.69}} & 0.77 & 0.79 & 0.88 & 0.89 & \textcolor{red}{\textbf{0.90}} \\
 \text{Max} & 7.74 & 5.19 & 7.79 & \textcolor{green}{\textbf{4.15}} & 7.78 & 8.99 & 8.97 & \textcolor{red}{\textbf{9.25}} \\
$\varepsilon^{rel}$ & \textcolor{red}{\textbf{53.39}} & 33.36 & \textcolor{green}{\textbf{28.28}} & 31.01 & 31.26 & 36.05 & 38.82 & 39.09 \\
\hline
 &  &  &  & \textbf{covid}  &  &  &  & \\
\hline
\text{GARCH-KDE} & $\Delta_{BS}$ & $\Delta$-$\Gamma_{BS}$ & $\Delta$-$\mathcal{V}_{SV}$ & $\Delta$-$\Gamma_{JD}$ & $\Delta$-$\Gamma_{SVJ}$ & $\Delta$-$\Gamma_{SVCJ}$
 & $\Delta$-$\Gamma_{VG}$ & $\Delta$-$\Gamma_{CGMY}$\\
 \text{Min} & \textcolor{red}{\textbf{-4.36}} & -2.69 & -2.64 & -2.64 & \textcolor{green}{\textbf{-2.44}} & -2.58 & -2.7 & -2.71 \\
 $\text{ES}^{5 \%}$ & \textcolor{red}{\textbf{-1.56}} & -0.8 & -0.76 & -0.77 & \textcolor{green}{\textbf{-0.70}} & -0.78 & -0.83 & -0.84 \\
$\text{ES}^{95 \%}$ & \textcolor{green}{\textbf{0.6}} & 0.93 & 0.9 & 0.97 & \textcolor{red}{\textbf{1.11}} & 1.00 & 0.91 & 0.9 \\
 \text{Max} & 3.88 & 3.33 & \textcolor{green}{\textbf{3.32}} & 4.52 & \textcolor{red}{\textbf{4.57}} & 4.45 & 4.49 & 4.55 \\
 $\varepsilon^{rel}$ & \textcolor{red}{\textbf{50.06}} & 34.48 & \textcolor{green}{\textbf{33.09}} & 34.57 & 40.02 & 37.4 & 34.63 & 34.67 \\
 \hline
\text{SVCJ} & $\Delta_{BS}$ & $\Delta$-$\Gamma_{BS}$ & $\Delta$-$\mathcal{V}_{SV}$ & $\Delta$-$\Gamma_{JD}$ & $\Delta$-$\Gamma_{SVJ}$ & $\Delta$-$\Gamma_{SVCJ}$ & $\Delta$-$\Gamma_{VG}$ & $\Delta$-$\Gamma_{CGMY}$ \\
 \text{Min} & -13.53 & \textcolor{green}{\textbf{-7.89}} & -7.9 & -14.3 & -11.76 & -11.75 & \textcolor{red}{\textbf{-20.99}} & -11.72 \\
 $\text{ES}^{5 \%}$ & \textcolor{red}{\textbf{-2.77}} & \textcolor{green}{\textbf{-1.18}} & -1.26 & -1.34 & -1.36 & -1.39 & -1.26 & -1.25 \\
$\text{ES}^{95 \%}$ & 0.87 & 0.71 & \textcolor{green}{\textbf{0.68}} & 0.78 & \textcolor{red}{\textbf{0.94}} & 0.93 & 0.73 & 0.73 \\
 \text{Max} & 13.48 & 10.78 & \textcolor{green}{\textbf{10.77}} & 13.60 & 13.66 & 13.6 & \textcolor{red}{\textbf{13.67}} & 13.65 \\
$\varepsilon^{rel}$ & \textcolor{red}{\textbf{88.42}} & \textcolor{green}{\textbf{38.24}} & 39.34 & 43.95 & 48. & 49.06 & 42.99 & 41.27 \\
\hline
\hline
\end{tabular}
}
\caption{3-month ATM hedge performance with the \textcolor{green}{\textbf{best}} and  \textcolor{red}{\textbf{worst}} performing strategy.
The table corresponds to results in Figure \ref{fig:pnl_boxplots_3M}.}
\label{tab:3m_hedges}
\end{table}
\clearpage


\begin{figure}[ht!]
    \centering
    \begin{subfigure}[b]{0.7\textwidth}\includegraphics[width=\textwidth]{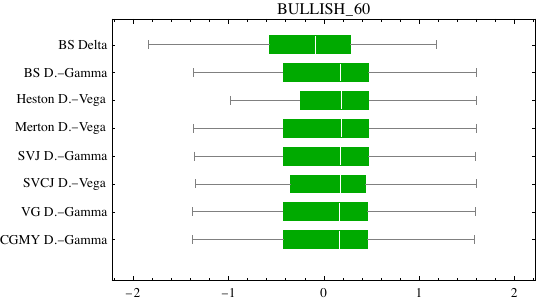}
    \caption{Bullish}
    \end{subfigure}
    \begin{subfigure}[b]{0.69\textwidth}\includegraphics[width=\textwidth]{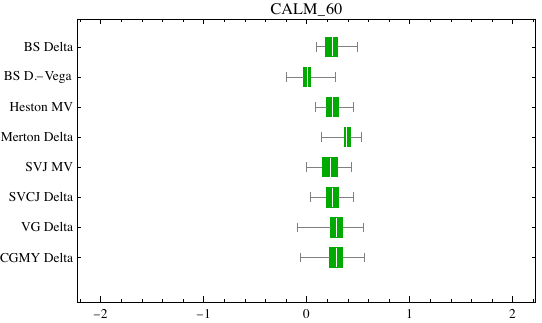}
     \caption{Calm}
    \end{subfigure}
    \begin{subfigure}[b]{0.7\textwidth}\includegraphics[width=\textwidth]{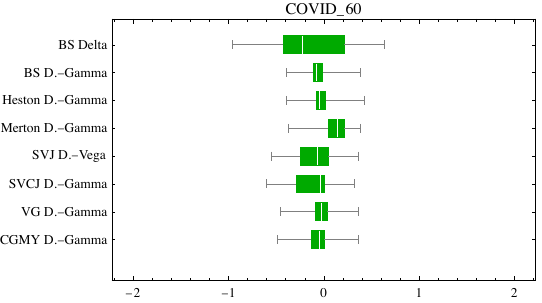}
    \caption{Covid}
    \end{subfigure}
  \caption{Historical backtest hedge performance; $\pi^{rel}$ for 2-months ATM options. For illustrative purposes $\pi^{rel}$ is truncated at $q^{5}$ and $q^{95}$. The vertical axis portrays $\Delta^{BS}$ hedge results compared each model's best performing strategy. This best performing strategy is selected according to the minimal $\text{ES}^{5 \%}$. \includegraphics[scale=0.2]{quanlet.png}\href{https://github.com/QuantLet/hedging_cc}{ hedging\_cc}}
\label{fig:hedge_backtest}
\end{figure}
\clearpage

\begin{table}[t!]
\centering
\scalebox{0.7}{
\begin{tabular}{rrrrrrrrr}
\hline
 &  &  &  & \textbf{bullish}  &  &  &  & \\
\hline
\text{Backtest} & $\Delta_{BS}$ & $\Delta$-$\Gamma_{BS}$ & $\Delta$-$\mathcal{V}_{SV}$ & $\Delta$-$\mathcal{V}_{JD}$ & $\Delta$-$\Gamma_{SVJ}$ & $\Delta$-$\mathcal{V}_{SVCJ}$ & $\Delta$-$\Gamma_{VG}$ & $\Delta$-$\Gamma_{CGMY}$ \\
\hline
\text{Min} & \textcolor{red}{\textbf{-4.34}} & -3.85  & \textcolor{green}{\textbf{-1.35}} & -3.85 & -3.85 & -2.27 & -3.86 & -3.86\\
 $\text{ES}^{5 \%}$ &  \textcolor{red}{\textbf{-4.34}} & -2.5 & \textcolor{green}{\textbf{-1.13}} & -2.5 & -2.5 & -1.66 & -2.51 & -2.51\\
 $ \text{ES}^{95 \%}$ & \textcolor{red}{\textbf{1.75}} & \textcolor{green}{\textbf{2.20}} & 2.09 & 
 \textcolor{green}{\textbf{2.20}} & 2.09 & 2.08 & 2.19 & 2.19 \\
 \text{Max} & \textcolor{red}{\textbf{2.11}} & \textcolor{green}{\textbf{2.56}} & \textcolor{green}{\textbf{2.56}}  & \textcolor{green}{\textbf{2.56}} & \textcolor{green}{\textbf{2.56}} & 2.55 & 2.55 & 2.55\\
 $\varepsilon^{rel}$ & 98.75 & 101.15 & \textcolor{green}{\textbf{73.56}} & \textcolor{red}{\textbf{101.25}} & 96.69 & 81.22 & 101.13 & 101.09 \\
\hline
 &  &  &  & \textbf{calm}  &  &  &  & \\
\hline
\text{Backtest} & $\Delta_{BS}$ & $\Delta$-$\mathcal{V}_{BS}$ & $\Delta$-$\mathcal{V}_{SV}$ & $\Delta_{JD}$ & $\text{MV}_{SVJ}$ & $\Delta_{SVCJ}$ & $\Delta_{VG}$ & $\Delta_{CGMY}$ \\
\text{Min} & -0.20 & \textcolor{red}{\textbf{-0.48}} & -0.16 & -0.15 & -0.15 & \textcolor{green}{\textbf{-0.13}} & -0.15 & -0.15\\
 $\text{ES}^{5 \%}$ & -0.05 & \textcolor{red}{\textbf{-0.36}} & -0.05 & \textcolor{green}{\textbf{-0.02}} & -0.08 & -0.06 & -0.12 & -0.1 \\
 $ \text{ES}^{95 \%}$ & 0.62 & \textcolor{red}{\textbf{0.32}} & 0.62 & 0.6 & 0.58 & 0.62 & \textcolor{green}{\textbf{0.68}} & 0.67\\
 \text{Max} & 0.82 & \textcolor{red}{\textbf{0.41}} & 0.84 & 0.75 & 0.78 & 0.82 & 0.88 & \textcolor{green}{\textbf{0.90}}\\
 $\varepsilon^{rel}$ & 14.31 & \textcolor{green}{\textbf{13.22}} & 14.37 & 13.48 & 14.43 & 14.52 & \textcolor{red}{\textbf{16.53}} & 16.30 \\
\hline
 &  &  &  & \textbf{covid}  &  &  &  & \\
\hline
\text{Backtest} & $\Delta_{BS}$ & $\Delta$-$\Gamma_{BS}$ & $\Delta$-$\Gamma_{SV}$ & $\Delta$-$\Gamma_{JD}$ & $\Delta$-$\mathcal{V}_{SVJ}$ & $\Delta$-$\Gamma_{SVCJ}$ & $\Delta$-$\Gamma_{VG}$ & $\Delta$-$\Gamma_{CGMY}$ \\
\text{Min} & \textcolor{red}{\textbf{-1.96}} & -1.33 & -1.23 & \textcolor{green}{\textbf{-1.19}} & -1.32 & -1.36 & -1.27 & -1.26 \\
 $\text{ES}^{5 \%}$ & \textcolor{red}{\textbf{-1.37}}
 & -0.77 & -0.70 & \textcolor{green}{\textbf{-0.66}} & -0.81 & -0.87 & -0.75 & -0.75 \\
 $\text{ES}^{95 \%}$ & \textcolor{green}{\textbf{0.70}}
  & 0.51 & 0.56 & 0.58 & 0.56 & \textcolor{red}{\textbf{0.50}} & 0.52 & 0.51 \\
 \text{Max} & 0.78 & \textcolor{red}{\textbf{0.69}}
  & 0.82 & \textcolor{green}{\textbf{0.91}} & 0.72 & 0.70 & 0.76 & 0.76 \\
 $\varepsilon^{rel}$ & \textcolor{red}{\textbf{49.47}}
  & 27.68 & 27.34 & 27.62 & 29.97 & 30.45 & \textcolor{green}{\textbf{27.10}} & 27.11 \\
\hline
\hline
\end{tabular}
}
\caption{Backtest of hedging 2-month ATM options with the \textcolor{green}{\textbf{best}} and  \textcolor{red}{\textbf{worst}} performing strategies.
The table corresponds to results in Figure \ref{fig:hedge_backtest}.
}
\label{tab:backtest_covid}
\end{table}
\subsubsection{Hedges with jump size correlation}\label{sec:rhojsens}
The hedges above (Section \ref{sec:hedge_results})
are performed under the assumption that the jump size correlation parameter $\rho^{j}$ is zero. This assumption is particularly well-founded on the BTC market, because jump size correlation $\rho^{j}$ is reportedly insignificant
\citep{crixpricing2019}. Nevertheless, we investigate whether $\rho^{j} \neq 0$ impacts hedging and look for differences to the \textit{main hedge results} from Section \ref{sec:hedge_results}. Therefore, the hedge routines are repeated for a daily calibrated $\rho^{j}$ and for fixed parameter values $\rho^{j} \in \{-0.5, 0.5\}$. For comparison, we deliberately look at a few examples. Note that changes to $\rho^{j} \neq 0$ impact the SVCJ Monte Carlo simulation and the SVCJ hedge strategies (Table \ref{tab:Hedge model summary}). 

\begin{figure}[ht!]
\begin{center}
\includegraphics[scale=0.22]{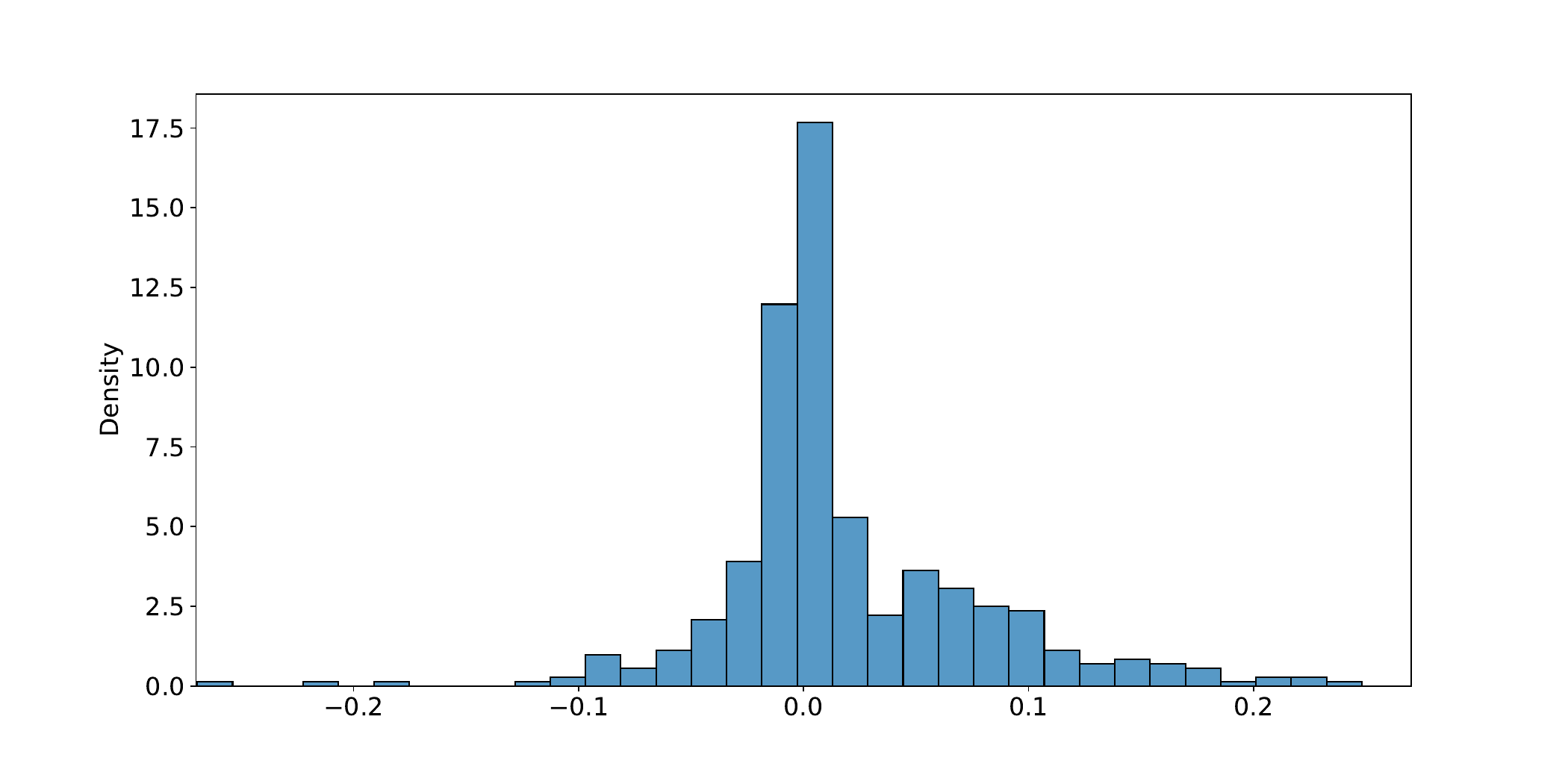}
\caption{Distribution of daily calibrated $\rho^{j}$, where most values lie close to $\rho^{j}=0$. \includegraphics[scale=0.2]{quanlet.png}\href{https://github.com/QuantLet/hedging_cc}{ hedging\_cc}}
\label{fig:calibration:rhoj}
\end{center}
\end{figure}

The calibration results from 
Figure \ref{fig:calibration:rhoj} and 
Table \ref{tab:rhoj_calibration} show that most calibrated values lie close to $\rho^{j}=0$.
As such, the SVCJ's hedge performance results are similar. An example are the SVCJ's hedge performance comparisons in the historical backtest in Table \ref{tab:hedge difference calibrated bullish} to Table \ref{tab:hedge difference calibrated covid}.
\begin{table}[t]
\centering
\scalebox{0.7}{
\begin{tabular}{lrrrrrrr}
\hline 
\hline
segment &     mean  &      std. dev. &      min &      $q^{25}$ &      $q^{50}$  &      $q^{75}$ &       max \\
\hline 
 bullish  & 0.01 & 0.07 & -0.27 & -0.01 & 0.00 & 0.02 & 0.50 \\
 calm & 0.03 & 0.07 & -0.09 & -0.01 & 0.01 & 0.06 & 0.34 \\
 covid & 0.04 &  0.13 & -0.12 & -0.01 &  0.00 &  0.07 & 0.84 \\
\hline
\hline
 \end{tabular}}
\caption{Summary statistics of daily $\rho^{j}$ calibration.\includegraphics[scale=0.2]{quanlet.png}\href{https://github.com/QuantLet/hedging_cc}{ hedging\_cc}}
\label{tab:rhoj_calibration}
\end{table} 

Appendix \ref{secB1}, table \ref{tab:hedge difference calm} shows hedge results, when bumping the correlation to $\rho^j=0.5$, resp.\ $\rho^j=-0.5$. Return jump sizes \eqref{eq:svcj_jumps} depend on $\rho^{j}$. Unsurprisingly, large correlation changes have a significant impact on the hedge performance. 

 \section{Conclusion}\label{sec:conclusion}
From a risk management perspective, CC markets are a highly interesting new asset class: on the one hand CC prices are subject to extreme moves, jumps and high volatility, while on the other hand, derivatives are actively traded -- and have been for several years -- on several exchanges. 
This paper presents an in-depth comparison of different hedging methods, providing concise answers to the trade-off between hedging in a complete, albeit oversimplified model and hedging in a more appropriate, albeit incomplete market model. 

As a central part of the methodology, we simulate price paths given the Bitcoin price history in two different ways: First, a semi-parametric approach (under the physical measures $\mathbb P$) combines GARCH volatilities with KDE estimates of the GARCH residuals. These paths are statistically close to the actual market behaviour. Second, paths are generated (under the risk-neutral measure $\mathbb Q$) in the parametric SVCJ model, where the SVCJ model parameters include valuable information on the contributing risk factors such as jumps. The time period under consideration features diverse market behaviour, and as such, lends itself to being partitioned into ``bullish'', ``calm'' and ``Covid-19'' periods. 

We hedge options with maturities of one and three months. 
If not directly quoted on the BTC market, option prices are interpolated from an arbitrage-free SVI-parametrization of the volatility surface.
The options are then hedged assuming risk managers use market models from the classes of affine jump diffusion  and infinite activity L\'evy models, which feature risk factors such as jumps and stochastic volatility. 
The calibration of these models strongly support the following risk factors: 
 stochastic volatility, infrequent jumps, some indication for infinite activity and inverse leverage effects on the market. Under GARCH-KDE and SVCJ, options are hedged with dynamic Delta, Delta-Gamma, Delta-Vega and minimum variance hedging strategies. 

For  longer-dated options, multiple-instrument hedges lead to considerable tail risk reduction. For the short-dated option, using multiple hedging instruments did not significantly outperform a single-instrument hedge. 
 This is in-line with traditional markets, where even in  highly volatile market periods, short-dated options are less sensitive to volatility or Gamma effects. For longer-dated options, multiple-instrument hedges consistently improve the hedge quality. Hence, if several liquidly traded options are available for hedging, they should be used. Among all models, persistently good hedge results are achieved by hedging with stochastic volatility models. This demonstrates that complete market models with stochastic volatility perform well, while models allowing for jump risk, although more realistic, do not produce better hedges due to the associated market incompleteness. These findings are confirmed for a historical backtest, where a 2-month option is written every day, generating a series of daily P\&L's from hedging at expiry.

\backmatter

\section*{Statements and declarations}

`Statements and Declarations': 'The code is available as quantlets, accessible through \href{https://www.quantlet.com/}{\texttt{Quantlet}} under the name \href{https://github.com/QuantLet/hedging_cc}{\includegraphics[scale=0.2]{quanlet.png}}\href{https://github.com/QuantLet/hedging_cc}{ hedging\_cc}. Quotes and BTC prices are provided by \href{https://www.https://tardis.dev/}{\texttt{Tardis.dev}} and the Blockchain Research Center  \href{https://blockchain-research-center.de/}{\texttt{BRC}}. The data is available upon request. 
\bmhead{Acknowledgments}
Financial support from the Deutsche Forschungsgemeinschaft via the IRTG 1792 “High Dimensional Non Stationary Time Series”, Humboldt-Universität zu Berlin, is gratefully acknowledged. The European Union’s Horizon 2020 research and
innovation program “FIN-TECH: A Financial supervision and Technology compliance training programme” under the grant agreement No 825215 (Topic: ICT-35-2018, Type of action: CSA), the European Cooperation in Science \& Technology COST Action grant CA19130 - Fintech and Artificial Intelligence in Finance - Towards a transparent financial industry, the Yushan Scholar Program of Taiwan and the Czech Science Foundation’s grant no. 19-28231X / CAS: XDA 23020303 are greatly acknowledged.

\begin{appendices}

\section{Hedging details}\label{secA1}
\subsection{Hedge routine}\label{sec:hedge_routine}
We illustrate the dynamic hedging routine on a single instrument self-financed hedging strategy $\xi$ and apply it analogously for all other hedging strategies considered in this study. The simulated, discretised prices are denoted by $S(t,i)$ are opposed to $S_t$, which refers to the continuous-time process.

At time $t= 0$ and for $B(0) = B_{0,i} = 1$ the value of the portfolio for the self-financed strategy $\xi$ is
\begin{equation}
\begin{aligned}
\Pi(0)&=C\left(0, S (0)\right)=\xi (0) S(0)+\left\{C \left(0, S(0)\right)-\xi (0) S(0)\right\}  B(0)\\
M(0) &=C\left(0, S(0)\right)-\xi (0) S(0) 
\end{aligned}
\end{equation}
where $B(t)$ is a risk-free asset and $M(t)$ the money market account vector. The value of the portfolio at time $t > 0$ is
\begin{equation}
\begin{aligned}
M (t) &=M(t-d t)+ \left\{\xi(t-d t) - \xi(t)\right\} \frac{S(t)}{B(t)} \\
\Pi (t) &=\xi(t-dt) S(t)+M(t-d t) B(t-d t) e^{r d t}
=\xi(t) S(t)+\underbrace{\frac{\Pi(t)-\xi (t) S(t)}{B(t)}}_{=M (t)}  B (t)
\end{aligned}
\end{equation}
At maturity $T$, the final $PnL$ distribution vector is
\begin{equation}\label{eq:finalposition}
\Pi (T)= \xi(T-d t)  S(t)+M (T-d t)B (t)
\end{equation}
\subsection{Dynamic Delta-hedging}\label{sec:appendix_dynamic_delta}
The option writer shorts the call $C(t)$, longs the underlying $S(t)$ and sends the remainder to a money market account $B(t)$ for which
\begin{equation*}
    dB (t) = r B (t) dt
\end{equation*}
At time $t$, the value of portfolio $\Pi (t)$ is 
 \begin{equation}\label{eq:selffianancedbs}
      \Pi (t) = - C(t)+ \Delta (t) S(t) + \frac{\left\{ C(t) - \Delta(t) S(t)\right\}}{B(t)} B(t)
  \end{equation}
  The changes evolve through
  \begin{equation}
   d \Pi (t)  = - dC (t) + \Delta (t)  dS(t) + \left\{C (t) - \Delta (t) S(t)\right\} r dt
 \end{equation}
 \subsection{Dynamic multiple-instrument-hedging
 }\label{sec:appendix_dynamic_delta_vega}
 We will explain the $\Delta-\mathcal{V}$ hedge in detail. The $\Delta-\Gamma$- hedge is performed accordingly. 
This strategy eliminates the sensitivity to changes in the underlying and changes in volatility.
The option writer shorts the call option $C$, takes the position $\Delta$ in the asset and $\Lambda$ in the second contingent claim. At time $t$, the value of the portfolio is \begin{equation}
\Pi ( t ) = - C ( t ) + \Lambda C_{ 1 } ( t ) + \Delta  S ( t )
\end{equation}
with the change in the portfolio $\Pi (t)$
\begin{equation}
d \Pi (t)= \Delta (t) d S+\left\{C(t)-\Delta S(t)-\Lambda C_{2}(t)\right\} r d t-d C(t)+\Lambda d C_{2}(t)
\end{equation}
That is
\begin{equation}\label{deltavega}
\begin{aligned} d \Pi (t)&=\left(C(S, V, t)-\Delta S(t)-\Lambda C_{2}(S, V, t)\right) r d t \\ 
&-\left(\frac{\partial C}{\partial t}+\frac{1}{2} \frac{\partial^{2} C}{\partial S^{2}} V S^{2}+\frac{1}{2} \frac{\partial^{2} C}{\partial V^{2}} ^{2} V+\frac{\partial^{2} C}{\partial V \partial S} \rho V  S\right) d t \\ 
&+\Lambda\left(\frac{\partial C_{2}}{\partial t}+\frac{1}{2} \frac{\partial^{2} C_{2}}{\partial S^{2}} V S^{2}+\frac{1}{2} \frac{\partial^{2} C_{2}}{\partial V^{2}} ^{2} V+\frac{\partial^{2} C_{2}}{\partial V \partial S} \rho V  S\right) d t \\ &+\left(\Lambda \frac{\partial C_{2}}{\partial S}-\frac{\partial C}{\partial S}+\Delta\right) d S+\left(\Lambda \frac{\partial C_{2}}{\partial V}-\frac{\partial C}{\partial V}\right) d V \end{aligned}
\end{equation}
For the choice of 
\begin{equation*}
    \begin{aligned}
    \Delta &=\frac{\partial C}{\partial S}-\Lambda \frac{\partial C_{2}}{\partial S} \\
    \Lambda &=\frac{\partial C / \partial v}{\partial C_{2} / \partial v}
    \end{aligned}
\end{equation*}
the portfolio is $\Delta - \mathcal{V}$ hedged. Analogously, for the choice of
\begin{equation*}
    \begin{aligned}
    \Delta &=\frac{\partial C}{\partial S}-\Lambda \frac{\partial C_{2}}{\partial S} \\
    \Lambda &= \frac{\partial^{2} C}{\partial^{2} S}
    \end{aligned}
\end{equation*}
this is a $\Delta-\Gamma$ hedge. For comparison, these hedges are applied to all models in the class of affine jump diffusion models. 
 %
 %
\subsection{Alternative representation of the \textbf{VG} process}\label{appendix:vgtocgm}
The alternative representation of the $\textbf{VG}$ process has the characteristic function \begin{equation}
\varphi_{\mathrm{VG}}(u ; C, G, M)=\left(\frac{G M}{G M+(M-G) \mathrm{i} u+u^{2}}\right)^{C}
\end{equation}
where $C,G,M >0$ with
\begin{equation}
\begin{aligned}
C &=1 / \nu \\
G &=\left(\sqrt{\frac{1}{4} \left(\theta^{VG} \right)^{2} \nu^{2}+\frac{1}{2} \left(\sigma^{VG}\right)^{2}  \nu}-\frac{1}{2} \theta^{VG}  \nu \right)^{-1} \\
M &=\left(\sqrt{\frac{1}{4}  \left(\theta^{VG} \right)^{2} v^{2}+\frac{1}{2} \left(\sigma^{VG}\right)^{2}  \nu}+\frac{1}{2} \theta^{VG}  \nu\right)^{-1}
\end{aligned}
\end{equation}
An increase in $G$ increases the size of upward jumps, while an increase in $M$ increases the size of downward jumps. Accordingly, $\theta^{VG}$, $M$ and $G$ account for the skewness of the distribution. $C$ governs the Levy-measure by widening it with its increase and narrowing it with its decrease.
\section{Tables}\label{secB1}
\begin{table}[ht!]
    \centering
\scalebox{0.7}{
\begin{tabular}{l|rrrrrrr}
framework &      $\hat{\mu}$ &      $\hat{\sigma}$ &      min &       $q^{1}$ &       $q^{50}$ &       $q^{99}$ &        max \\
\hline
\hline
$SVCJ_{{\text{BULLISH}}_{30}}$ &      4087.32 &          343.05 &  1352.90 &  3411.83 &   4065.04 &   5177.02 &   15819.48 \\     
 $SVCJ_{\text{CALM}_{30}}$ &      8369.33 &         1650.21 &   646.68 &  3475.29 &   8367.51 &  13092.95 &   26271.20 \\    
 $SVCJ_{\text{COVID}_{30}}$ &      9800.32 &         1269.66 &  1435.49 &  5406.93 &   9804.85 &  13341.72 &   41464.61 \\   
 $KDE_{{BULLISH}_{30}}$ &      4393.95 &          606.01 &  2089.55 &  3237.62 &   4277.65 &   6248.48 &   10209.30 \\      
 $KDE_{\text{CALM}_{30}}$ &      8359.21 &          746.38 &  4545.46 &  6608.25 &   8349.06 &  10524.45 &   15611.32 \\     
 $KDE_{\text{COVID}_{30}}$ &      9933.81 &          836.48 &  5579.96 &  8007.32 &   9848.62 &  12365.51 &   16863.17 \\  
 \hline
 $SVCJ_{\text{BULLISH}_{90}}$ &      4087.50 &          657.29 &   419.77 &  2961.11 &   4001.56 &   6336.31 &   56189.20 \\  
 $SVCJ_{\text{CALM}_{90}}$ &      8367.54 &         2982.34 &    37.40 &  2488.41 &   8124.74 &  18415.20 &  118249.15 \\  
 $SVCJ_{\text{COVID}_{90}}$ &      9796.71 &         2456.05 &   119.85 &  3620.50 &   9682.93 &  17545.53 &  115020.35 \\   
 $KDE_{\text{BULLISH}_{90}}$ &      5116.43 &         1419.86 &  1325.30 &  3038.41 &   4762.11 &   9988.82 &   28593.53 \\   
 $KDE_{\text{CALM}_{90}}$ &      8345.58 &         1407.72 &  3034.41 &  5341.07 &   8274.30 &  12590.88 &   22406.78 \\     
 $KDE_{\text{COVID}_{90}}$ &     10718.15 &         3457.73 &  1560.16 &  4729.19 &  10007.73 &  23519.87 &   81081.55\\
 \hline
 \hline
 \end{tabular}
 }
\caption{Summary statistics of scenario generations framework per market segment and maturity \includegraphics[scale=0.2]{quanlet.png}\href{https://github.com/QuantLet/hedging_cc}{ hedging\_cc}}
  \label{tab:scenario_summary}
\end{table}

\begin{table}[h]
\begin{center}
\scalebox{0.7}{
\begin{tabular}{l|rrrrrr}
\hline
\hline
  \textbf{TTM} &         a &         b &       $\rho^{SVI}$ &         m &     $\sigma^{SVI}$ &    penalty \\
\hline
0.01 &  0.17 &  0.10 &  0.00 &  0.00 &   1.00 &    24.53 \\
0.03 &  0.003 &  0.01 &  0.15 &  0.01 &   0.17 &     0.00001 \\
0.07 &  0.01 &  0.04 &  0.00 & -0.01 &   0.08 &     0.000004\\
0.24 &  0.02 &  0.10 & -0.11 & -0.01 &   0.45 &     0.001 \\
0.49 &  0.01 &  0.17 & -0.02 &  0.04 &   0.77 &     0.002 \\
0.74 &  0.14 &  0.09 &  0.00 &  0.01 &   0.93 &     0.03 \\
\hline
0.01 &  0.001 &  0.05 & -0.13 &  0.02 &   0.08 &     0.09 \\
0.03 &  0.01 &  0.05 & -0.39 &  0.01 &   0.16 &     0.01 \\
0.07 &  0.01 &  0.10 & -0.02 &  0.12 &   0.32 &     0.02 \\
0.16 &  0.06 &  0.15 & -0.50 & -0.17 &   0.54 &     0.01 \\
0.24 &  0.04 &  0.19 & -0.27 & -0.10 &   0.76 &     0.03 \\
0.49 &  0.18 &  0.21 &  0.23 &  0.38 &   1.00 &     0.01 \\
\hline
0.02 &  0.004 &  0.02 &  0.50 &  0.02 &   0.01 &     0.03 \\
 0.04 &  0.003 &  0.05 & -0.07 & -0.03 &   0.11 &     0.01 \\
 0.07 &  0.01 &  0.08 & -0.09 & -0.05 &   0.15 &     0.02 \\
 0.15 &  0.02 &  0.13 &  0.19 &  0.07 &   0.29 &     0.04 \\
 0.40 &  0.06 &  0.20 & -0.15 & -0.21 &   0.56 &     0.01 \\
 0.65 &  0.14 &  0.18 &  0.16 & -0.12 &   0.88 &     0.02 \\
\hline
\hline
\end{tabular}}
\caption{Calibrated SVI parameters at the beginning of the bullish, calm and stressed segment. \includegraphics[scale=0.2]{quanlet.png}\href{https://github.com/QuantLet/hedging_cc}{ hedging\_cc}}
\label{tab:svi_parameters}
\end{center}
\end{table}

\begin{table}[t]
\begin{center}
\scalebox{0.7}{
\begin{tabular}{l|rrr}
\hline
\hline
&  SV &     SVJ &    SVCJ \\
\hline
$\hat{\mu}$    &    0.82 &    0.78 &    0.87 \\
$\hat{\sigma}$ &    0.32 &    0.33 &    0.35 \\ 
$\operatorname{min}$            &    0.00 &    0.00 &    0.00 \\ $q^{25}$       &    0.62 &    0.62 &    0.69 \\ 
$q^{50}$       &    0.84 &    0.81 &    0.92 \\ 
$q^{75}$       &    1.04 &    0.99 &    1.06 \\ 
$\operatorname{max}$            &    1.49 &    1.57 &    2.43 \\
\hline
$\hat{\mu}$    &    0.68 &    0.72 &    0.90 \\
$\hat{\sigma}$ &    0.30 &    0.36 &    0.37 \\ 
$\operatorname{min}$            &    0.00 &    0.00 &    0.00 \\ $q^{25}$       &    0.50 &    0.56 &    0.70 \\ 
$q^{50}$       &    0.75 &    0.79 &    1.02 \\ 
$q^{75}$       &    0.90 &    0.95 &    1.19 \\ 
$\operatorname{max}$            &    1.43 &    1.40 &    1.44 \\
\hline
$\hat{\mu}$    &    0.56 &    0.72 &    0.84 \\
$\hat{\sigma}$ &    0.49 &    0.66 &    0.45 \\ 
$\operatorname{min}$            &    0.00 &    0.00 &    0.00 \\ $q^{25}$       &    0.27 &    0.29 &    0.61 \\ 
$q^{50}$       &    0.50 &    0.73 &    0.88 \\ 
$q^{75}$       &    0.78 &    1.01 &    1.04 \\ 
$\operatorname{max}$            &    3.83 &    6.33 &    3.83 \\
\hline
\hline
\end{tabular}}
\caption{Summary statistics of $\sigma^{v}$ for all 3 market segments and models. \includegraphics[scale=0.2]{quanlet.png}\href{https://github.com/QuantLet/hedging_cc}{ hedging\_cc}}
\label{tab:v0_average}
\end{center}
\end{table}
\clearpage
\begin{table}[t!]
\centering
\scalebox{0.7}{
\begin{tabular}{rrrrrr}
\hline
\hline
$\rho^{j}$ &Measure & $\Delta$ & $\Delta$-$\Gamma$ & $\Delta$-$\mathcal{V}$ & MV  \\
\hline
0 & $\text{ES}^{5 \%}$ & -2.97 &  -2.50 &                 -1.66 &  -2.99 \\
market & $\text{ES}^{5 \%}$ & -2.99 &   -2.49 &      -2.49 &  -2.97 \\
0 & $\varepsilon^{rel}$ & 95.68 & 100.77 &  64.79 &  95.96 \\
market & $\varepsilon^{rel}$ &  97.07 & 101.51 & 102.65 &  96.93 \\
\hline
\hline
\end{tabular}
}
\caption{Historical hedge backtest performance comparison for the SVCJ hedge strategies (Table \ref{tab:Hedge model summary}) when the jump size correlation parameter $\rho^{j}$ is calibrated to the market $\rho^{j} =\text{'market'}$ vs. $\rho^j=0$ during the bullish period.}
\label{tab:hedge difference calibrated bullish}
\end{table}

\begin{table}[t!]
\centering
\scalebox{0.7}{
\begin{tabular}{rrrrrr}
\hline
\hline
$\rho^{j}$ &Measure & $\Delta$ & $\Delta$-$\Gamma$ & $\Delta$-$\mathcal{V}$ & MV  \\
\hline
0 & $\text{ES}^{5 \%}$ &    -0.06 &    -0.33 &     -0.35 & -0.07 \\
market & $\text{ES}^{5 \%}$ & -0.07 &      -0.33 &         -0.32 & -0.06 \\ 
0 & $\varepsilon^{rel}$ &  2.07 &     1.63 &  1.91 &  2.36 \\
market & $\varepsilon^{rel}$ & 2.03 &  1.79 &  1.75 &  2.02 \\ 
\hline
\hline
\end{tabular}
}
\caption{Historical hedge backtest performance comparison for the SVCJ hedge strategies (Table \ref{tab:Hedge model summary}) when the jump size correlation parameter $\rho^{j}$ is calibrated to the market $\rho^{j} =\text{'market'}$ vs. $\rho^j=0$ during the calm period.}
\label{tab:hedge difference calibrated calm}
\end{table}

\begin{table}[t!]
\centering
\scalebox{0.7}{
\begin{tabular}{rrrrrr}
\hline
\hline
$\rho^{j}$ &Measure & $\Delta$ & $\Delta$-$\Gamma$ & $\Delta$-$\mathcal{V}$ & MV  \\
\hline
0 & $\text{ES}^{5 \%}$ &    -1.38 & -0.87 & -0.89 &  -1.52 \\
 market & $\text{ES}^{5 \%}$ &    -1.41 &   -0.69 & -0.73 &  -1.41 \\
 0 & $\varepsilon^{rel}$ &    25.66 &   9.12 &   9.43 &  37.59 \\
  market & $\varepsilon^{rel}$ &   25.60 &   7.83 &   8.17 &  25.51 \\
\hline
\hline
\end{tabular}
}
\caption{Historical hedge backtest performance comparison for the SVCJ hedge strategies (Table \ref{tab:Hedge model summary}) when the jump size correlation parameter $\rho^{j}$ is calibrated to the market $\rho^{j} =\text{'market'}$ vs. $\rho^j=0$ during the covid period.}
\label{tab:hedge difference calibrated covid}
\end{table}

\clearpage
\begin{table}[t!]
\centering
\scalebox{0.7}{
\begin{tabular}{rrrrrrrr}
\hline
\hline
 &  &  &  \textbf{calm} &   &  &   \\
\hline
Approach & Maturity & $\rho^{j}$ & Measure & $\Delta$ & $\Delta$-$\Gamma$ & $\Delta$-$\mathcal{V}$ & MV  \\
\hline
SVCJ & 1\text{M} & 0 & $\text{ES}^{5 \%}$ & 
-2.31 & -1.01 & -1.01 & -2.37 \\
SVCJ & 1\text{M} & 0.5 & $\text{ES}^{5 \%}$ & -2.50 & -1.09 & -1.09 & -2.50 \\
SVCJ & 1\text{M} & -0.5 &  $\text{ES}^{5 \%}$ &  -6.33 & -1.17 & -1.17 &   -6.46 \\
SVCJ & 1 \text{M} & 0 & $\varepsilon^{rel}$  & 49.52 & 19.72 & 19.66 & 51.48 \\
SVCJ & 1 \text{M} & 0.5 & $\varepsilon^{rel}$ & 129.27 & 85.27 & 83.75 & 128.77 \\
SVCJ & 1 \text{M} & -0.5 & $\varepsilon^{rel}$ &
399.69 & 23.99 &23.50 &  408.91 \\
\hline 
SVCJ & 3 \text{M} & 0 & $\text{ES}^{5 \%}$ &
 -1.59 & -0.89 & -0.89 & -1.61 \\
 SVCJ & 3 \text{M} &  0.5 & $\text{ES}^{5 \%}$ &  
 -3.77 & -1.35 & -1.21 & -3.69\\
 SVCJ & 3 \text{M} &  -0.5 & $\text{ES}^{5 \%}$ &  -6.71 &  -0.97 &  -0.96 &   -6.85 \\
  SVCJ & 3 \text{M} & 0 & $\varepsilon^{rel}$ &
 29.86 & 13.32 & 12.99 & 31.20  \\
SVCJ & 3 \text{M} & 0.5 & $\varepsilon^{rel}$ & 
 307.47 & 284.2 & 119.3 & 303.13 \\
SVCJ & 3 \text{M} & -0.5 & $\varepsilon^{rel}$ &
445.28 &    16.03 & 16.04 &  457.92 \\
\hline
GARCH-KDE & 1 \text{M} & 0 & $\text{ES}^{5 \%}$ & 
-0.16 & -0.49 & -0.48 & -0.15\\
GARCH-KDE & 1 \text{M} & 0.5 &  $\text{ES}^{5 \%}$ & -0.25 & -0.49 & -0.49 & -0.24\\
GARCH-KDE & 1 \text{M} & -0.5&& $\text{ES}^{5 \%}$ -0.15 &  -0.48 &  -0.48 & -0.14 \\
GARCH-KDE & 1 \text{M} & 0 & $\varepsilon^{rel}$ &
6.6 & 5.32 & 5.3 & 6.77\\
GARCH-KDE & 1 \text{M} & 0.5 & $\varepsilon^{rel}$ &
7.84 & 5.36 & 5.34 & 7.82\\
GARCH-KDE & 1 \text{M} & -0.5 & $\varepsilon^{rel}$ &
 6.90 & 5.32 &  5.31 &  7.18 \\
\hline
GARCH-KDE & 3 \text{M} & 0 & $\text{ES}^{5 \%}$ &
 0.20 & -0.20 & -0.20 & 0.20 
\\
GARCH-KDE & 3 \text{M} & 0.50 & $\text{ES}^{5 \%}$ & -0.28 & -0.31 & -0.31 & -0.27 \\
GARCH-KDE & 3 \text{M} & -0.5 & $\text{ES}^{5 \%}$ &0.13 &  -0.18 & -0.18 &  0.09 \\
 
GARCH-KDE & 3 \text{M} & 0 & $\varepsilon^{rel}$ &
 1.64 & 0.79 & 0.79 & 2.0
 \\
GARCH-KDE & 3 \text{M} & 0.5 & $\varepsilon^{rel}$ &
 12.20 & 1.88 & 1.86 & 12.17 \\
 GARCH-KDE & 3 \text{M} & -0.5 & $\varepsilon^{rel}$
 & 2.81 & 0.81 &   0.81 &  3.46 \\
 \hline
Backtest & 2 \text{M} & 0 & $\text{ES}^{5 \%}$
& -0.06 & -0.33 & -0.35 & -0.07 \\
Backtest & 2 \text{M} & 0.5 & $\text{ES}^{5 \%}$
& -0.83 &  -1.47 & -1.32 & -0.87\\
Backtest & 2 \text{M} & 0.5 & $\text{ES}^{5 \%}$ & -1.53 &-0.95& -0.97& -1.68 \\
Backtest & 2 \text{M} & 0 & $\varepsilon^{rel}$ & 
2.07 & 1.63 & 1.91 & 2.36\\
Backtest & 2 \text{M} & 0.5 & $\varepsilon^{rel}$ & 61.87 & 54.43 & 51.02 & 64.12\\
Backtest & 2 \text{M} & -0.5 & $\varepsilon^{rel}$ & 
32.61 &10.06&	10.51&	46.97\\
\hline
\hline
\end{tabular}
}
\caption{Comparison of SVCJ hedge performance for different values of $\rho^{j}$ during the calm period. We observe consistently worse hedge performances for $\rho^{j} \in \{-0.5,0.5\}$. }
\label{tab:hedge difference calm}
\end{table}

\section{Additional plots}\label{C1}
\subsection{GARCH(1,1) model}
\begin{figure}[ht!]
\begin{center}
\includegraphics[scale=0.35]{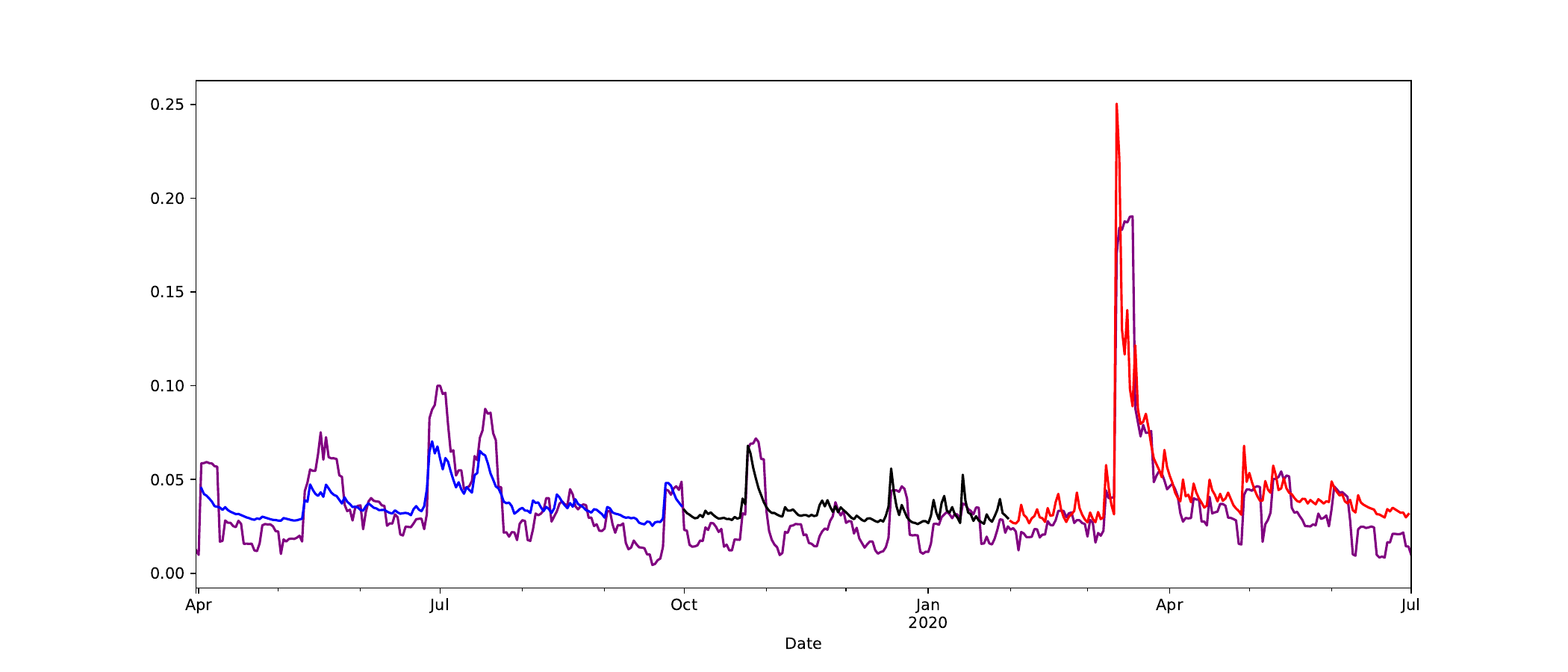}
\caption[Data]{Estimated GARCH(1,1) volatility $\hat \sigma_{t}$ during \textcolor{blue}{bullish market behavior},  \textcolor{black}{calm period} and  \textcolor{red}{stressed scenario} and the \textcolor{purple}{7-day close-to-close historical volatility}.
\includegraphics[scale=0.22]{quanlet.png}\href{htps://github.com/QuantLet/hedging_cc}{ hedging\_cc}}\label{fig:garch_estimates}
\end{center}
\end{figure}
\clearpage

\subsection{Calibration}
\begin{figure}[ht!]
\begin{center}
\includegraphics[trim={3.5cm 0 3.5cm 0},clip,scale=0.28]{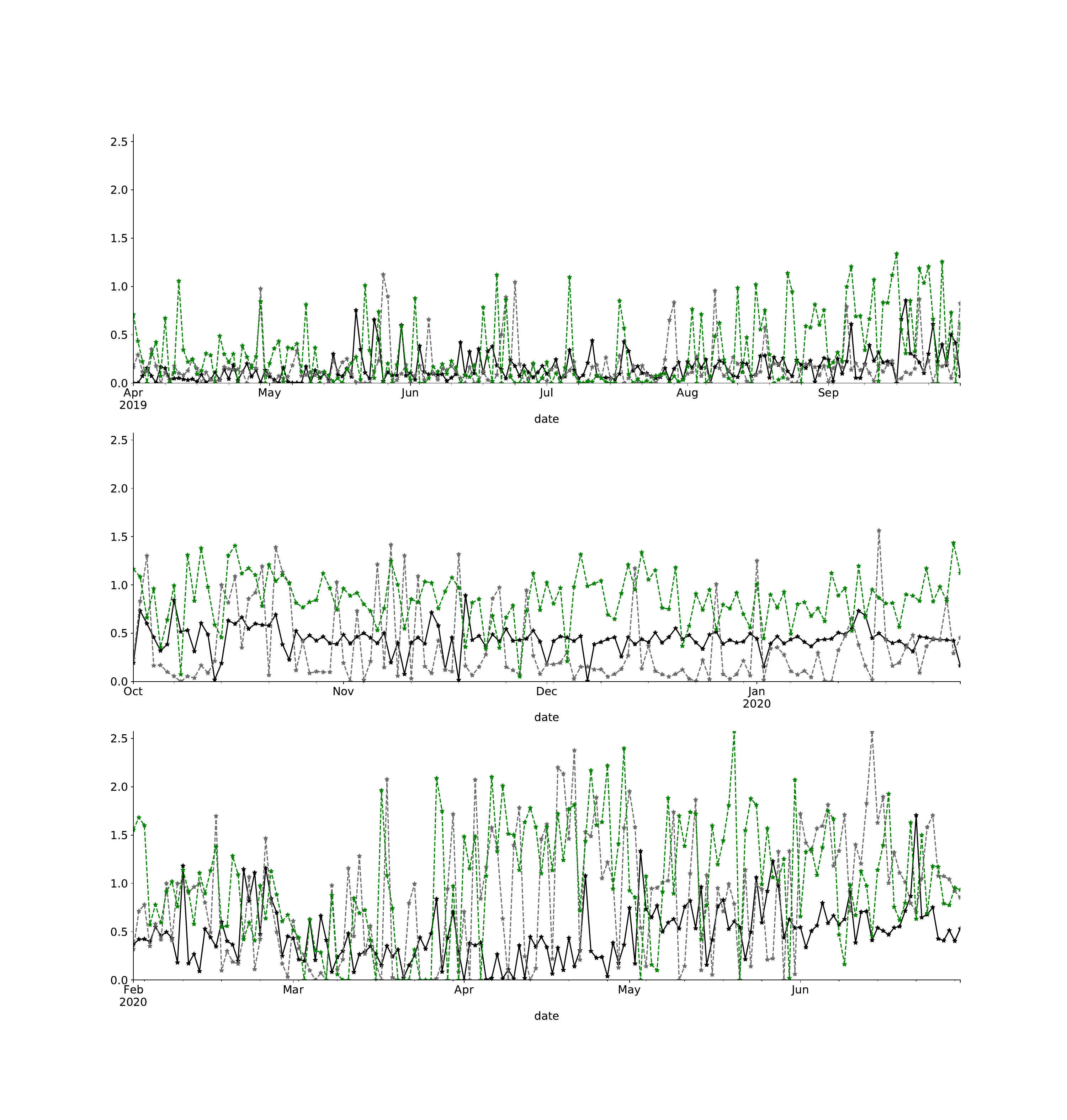}
\caption{Daily calibrated jump intensity $\lambda^{JD}$, \textcolor{gray}{$\lambda^{SVJ}$} and \textcolor{green}{$\lambda^{SVCJ}$} segregated chronologically by market segment. In each market segments, annual jump intensity is generally $\lambda  \leq 2$. \includegraphics[scale=0.2]{quanlet.png}\href{https://github.com/QuantLet/hedging_cc}{ hedging\_cc}}
\label{fig:calibration:lambda_jumps}
\end{center}
\end{figure}

\begin{figure}[ht!]
\begin{center}
\includegraphics[trim={3.5cm 0 3.5cm 0},clip,scale=0.28]
{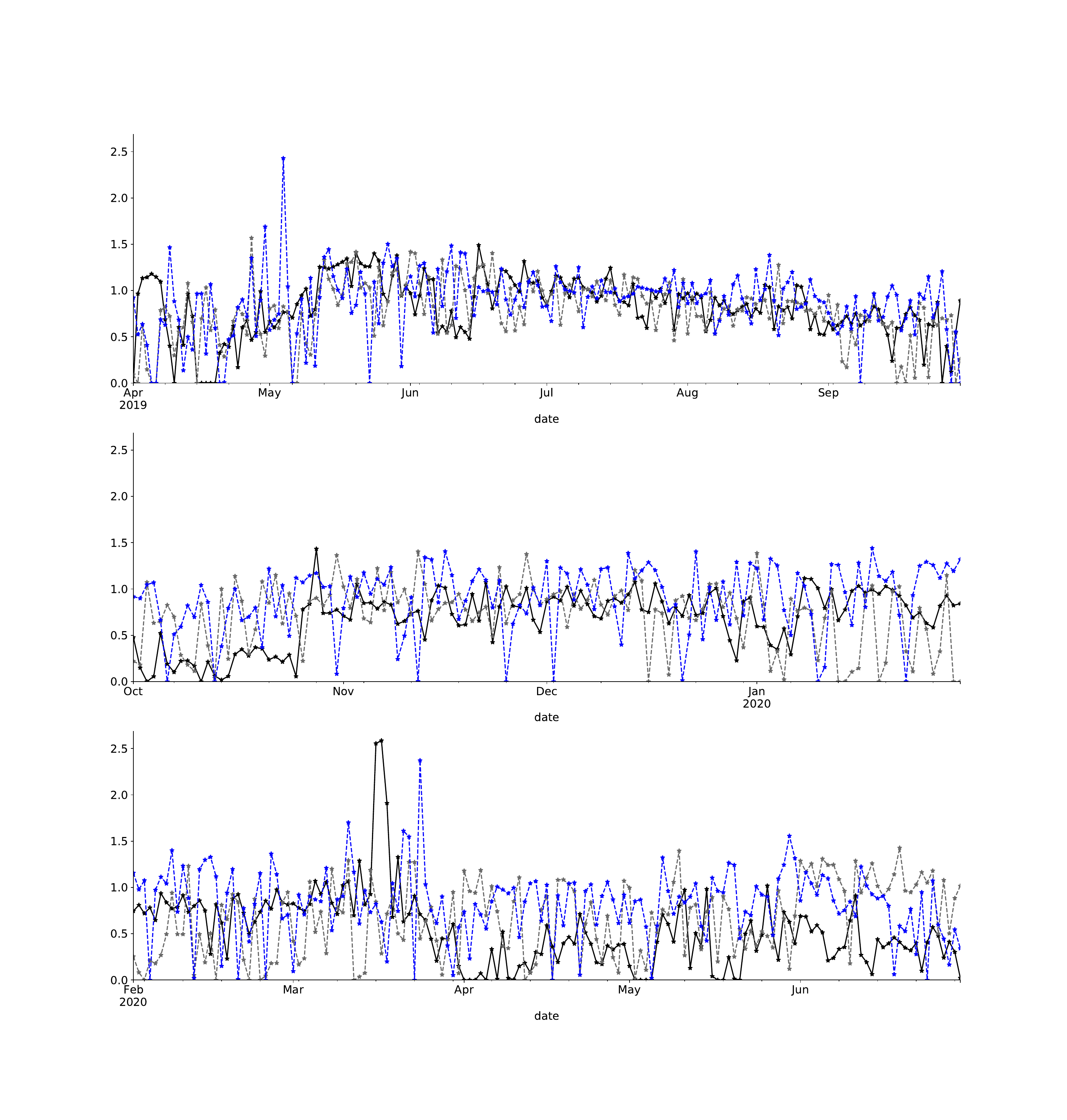}
\caption{Daily calibrated volatility of volatility $\sigma^{v}$ (SV), \textcolor{darkgray}{$\sigma^{v}$ (SVJ)} and \textcolor{blue}{$\sigma^{v}$ (SVCJ)} plotted in chronological order by market segment.
For illustrative purposes, extremes are disregarded. Information about extremes is provided in Table \ref{tab:v0_average}. Regardless of the model choice, levels of $\sigma^{v}$ are high. This provides strong indication for stochastic volatility. \includegraphics[scale=0.2]{quanlet.png}\href{https://github.com/QuantLet/hedging_cc}{ hedging\_cc}}
\label{fig:calibration:SV_volvol}
\end{center}
\end{figure}

\begin{figure}[ht!]
\begin{center}
\includegraphics[trim={3.5cm 0 3.5cm 0},clip,scale=0.28]
{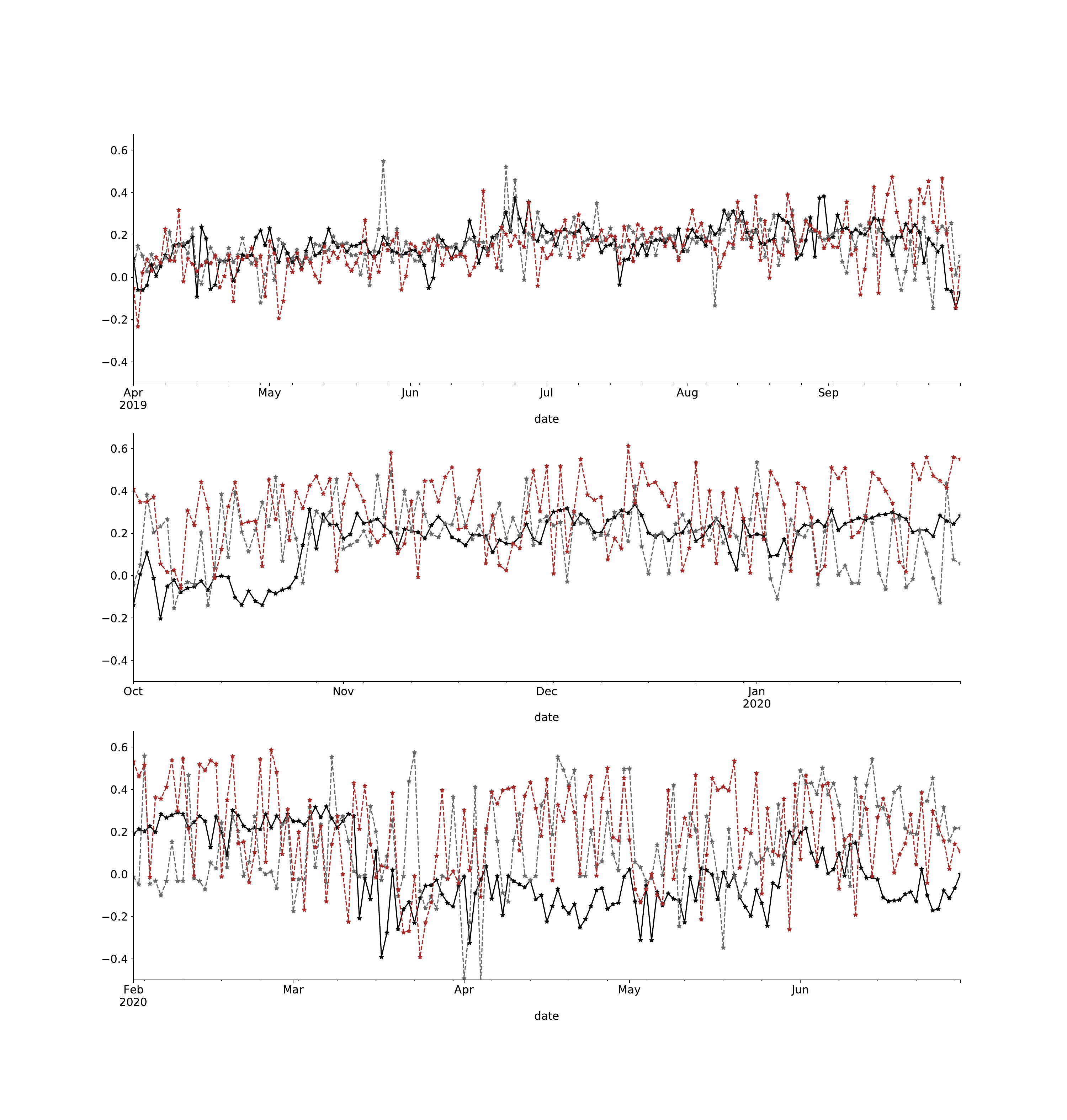}
\caption{Daily calibrated correlation parameter $\rho^{SV}$, \textcolor{darkgray}{$\rho^{SVJ}$} and \textcolor{brown}{$\rho^{SVCJ}$} plotted in chronological order by market segment.
For illustrative purposes, extremes are disregarded. As generally $\rho>0$, there is an indication for an inverse leverage effect as reported in \cite{crixpricing2019}. \includegraphics[scale=0.2]{quanlet.png}\href{https://github.com/QuantLet/hedging_cc}{ hedging\_cc}}
\label{fig:calibration:SV_rho}
\end{center}
\end{figure}

\begin{figure}[ht!]
\begin{center}
\includegraphics[scale=0.3]{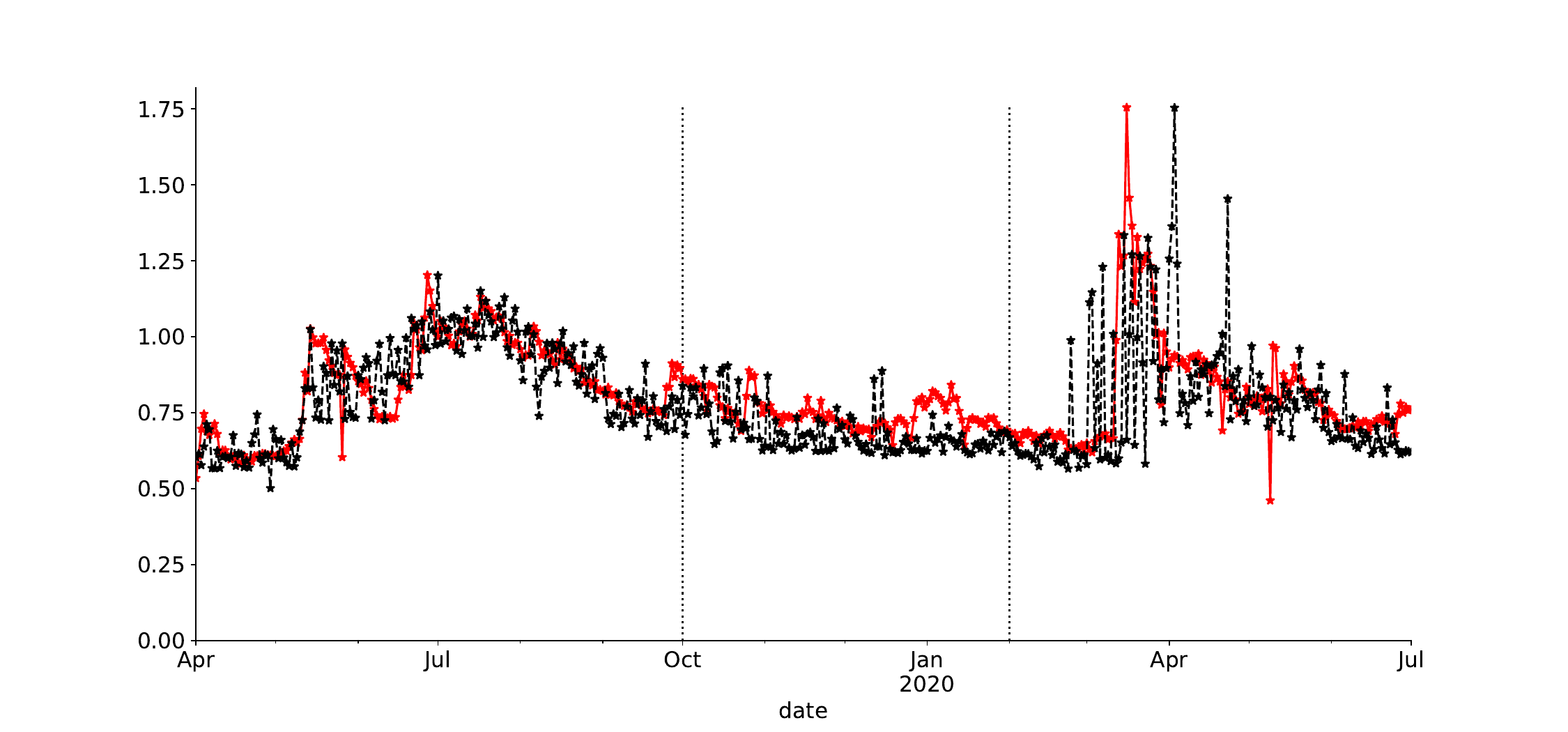}
\caption{Daily calibration of  \textcolor{black}{$\sigma^{VG}$} plotted against \textcolor{red}{$\sigma^{BS}$}. Both models capture comparable volatility levels.}
\label{fig:calibration:vg_sigma}
\end{center}
\end{figure}
\begin{figure}[ht!]
    \centering
    \begin{subfigure}[b]{0.75\textwidth}\includegraphics[width=\textwidth]{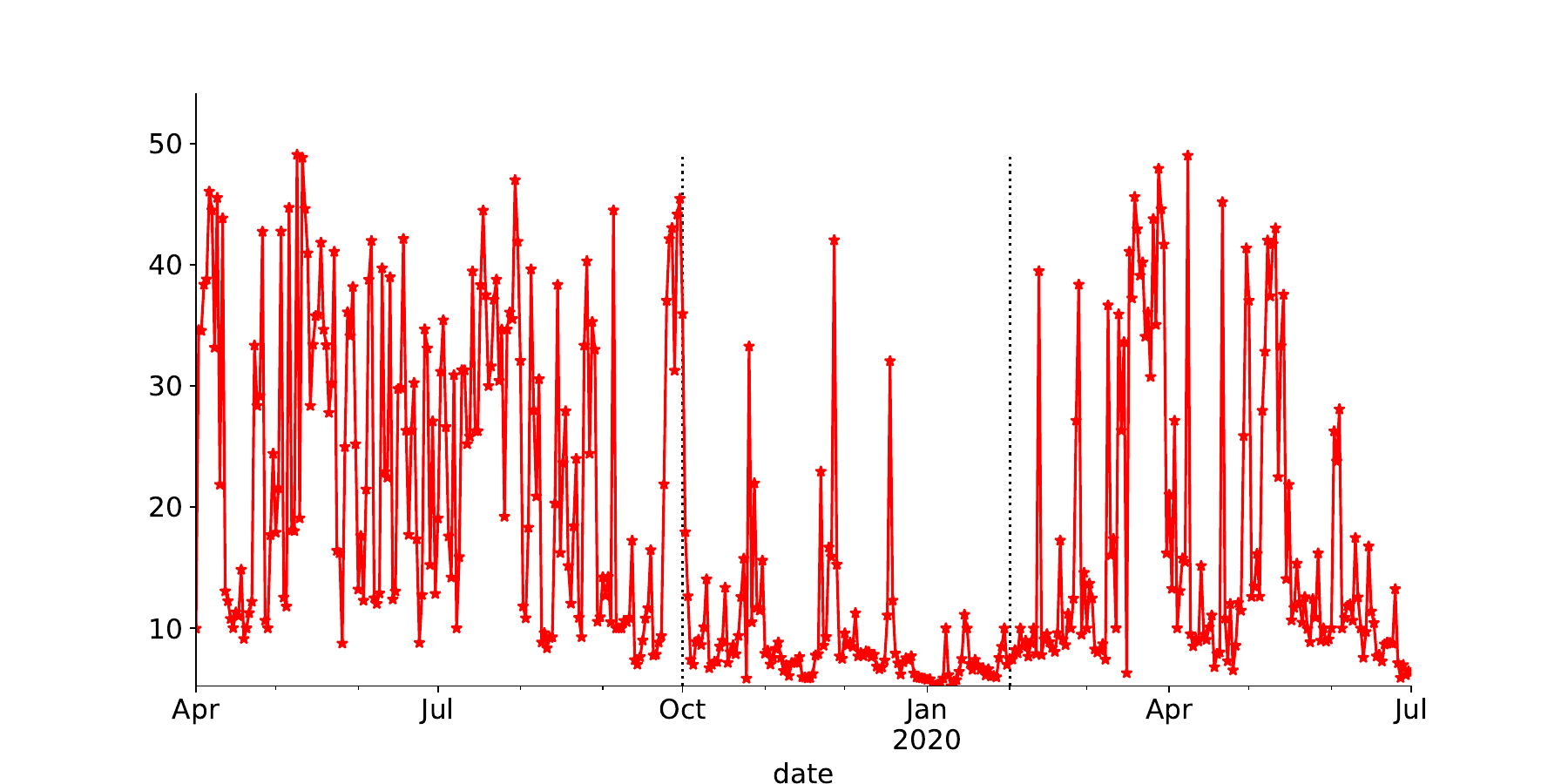}
    \caption{downward jumps M}
    \end{subfigure}
    \begin{subfigure}[b]{0.75\textwidth}\includegraphics[width=\textwidth]{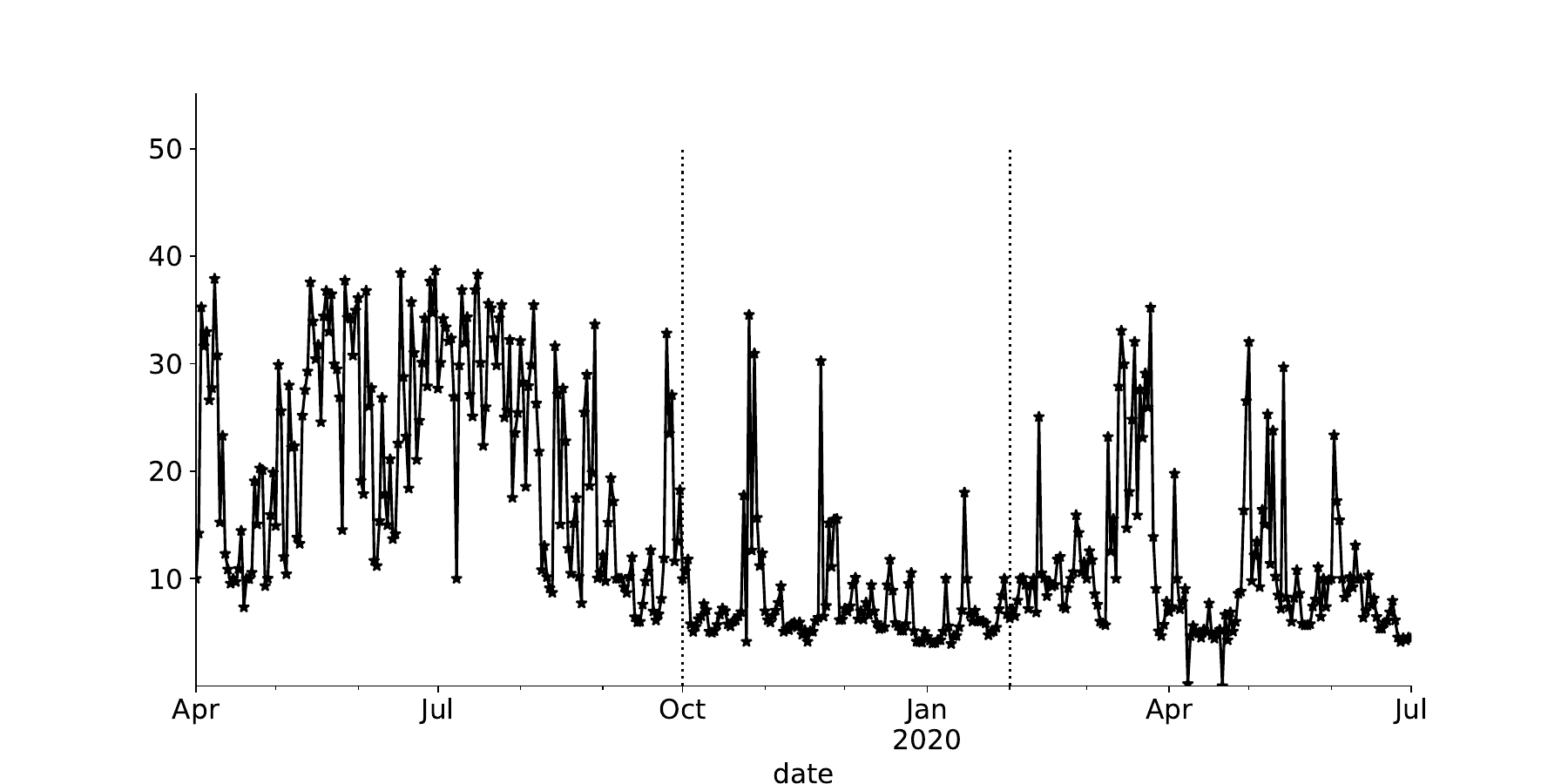}
    \caption{upward jumps G}
    \end{subfigure}
  \caption{(a) Evolution of $G_{CGMY}$ and (b) \textcolor{red}{$M_{CGMY}$} segregated by market segment. High magnitudes for both parameter values are observed during the \textit{bullish} and \textit{stressed} scenario. For illustrative purposes, extremes are excluded from this graph.}
    \label{fig:cgmy_jumps}
\end{figure}

\clearpage
\subsection{RMSE}\label{sec:rmse_appedix}
\vspace{-2em}
\begin{figure}[ht!]
    \begin{subfigure}[b]{\textwidth}
    \includegraphics[trim={3.5cm 0 3.5cm 0},clip,width=\textwidth]{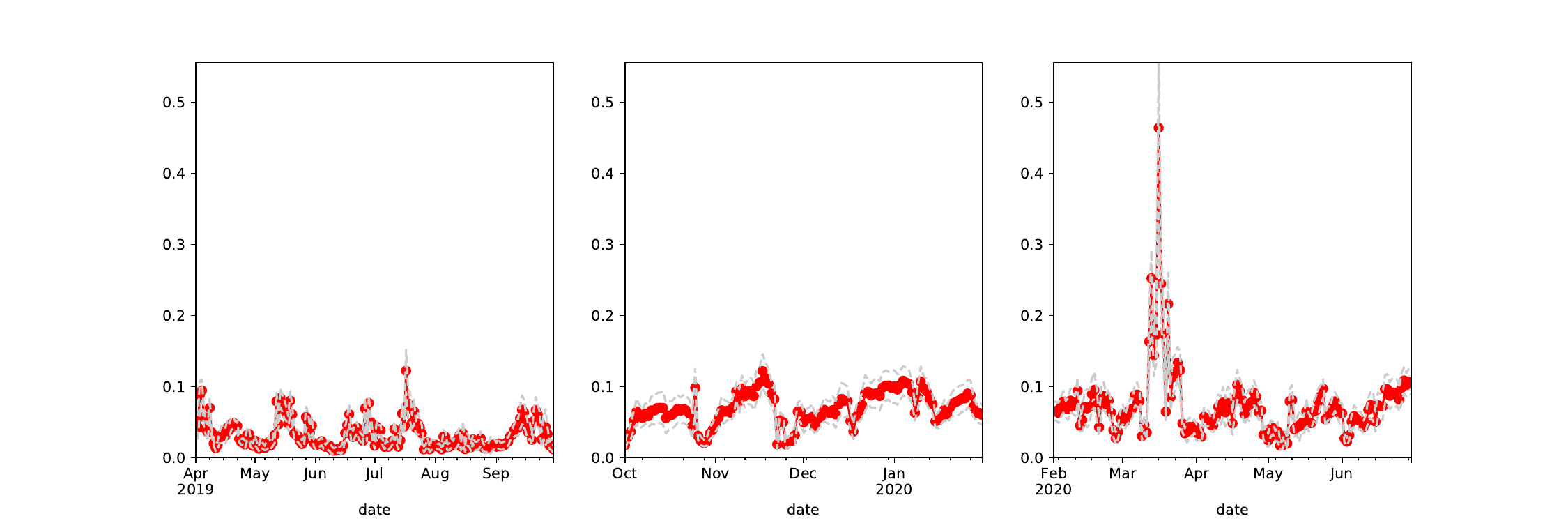}
    \caption{BS}
    \end{subfigure}
    \hspace{3cm}
    \begin{subfigure}[b]{\textwidth}\includegraphics[trim={3.5cm 0 3.5cm 0},clip,width=\textwidth]{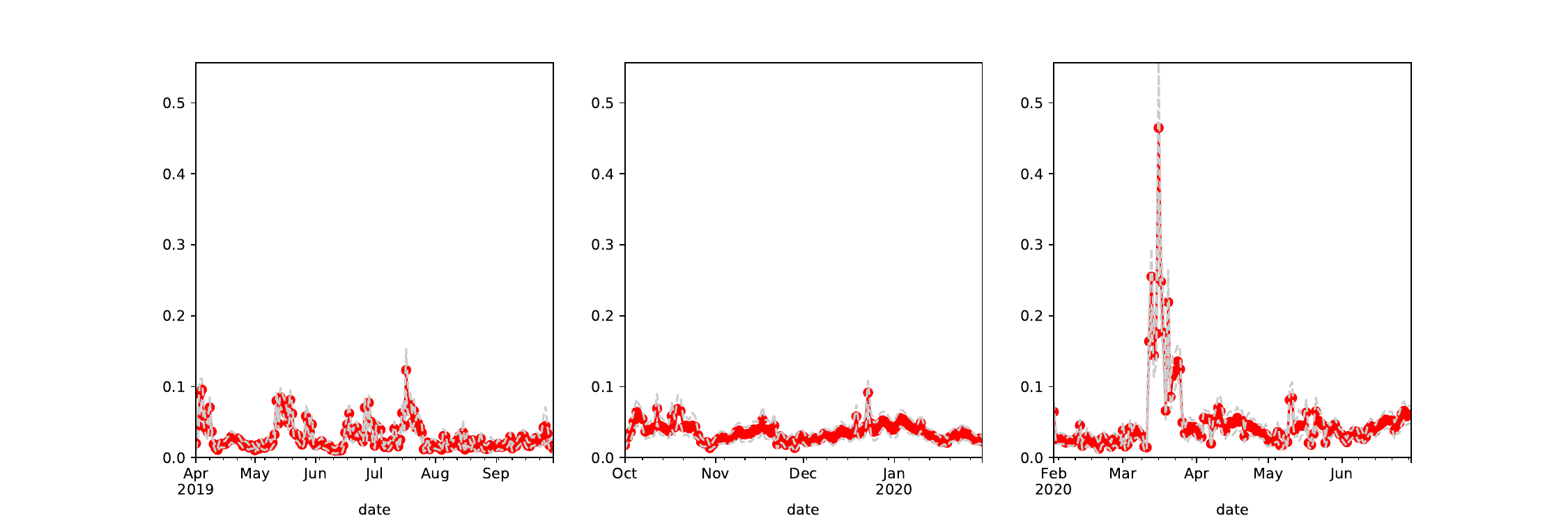}
    \caption{JD}
    \end{subfigure}
    \begin{subfigure}[b]{\textwidth}\includegraphics[trim={3.5cm 0 3.5cm 0},clip,width=\textwidth]{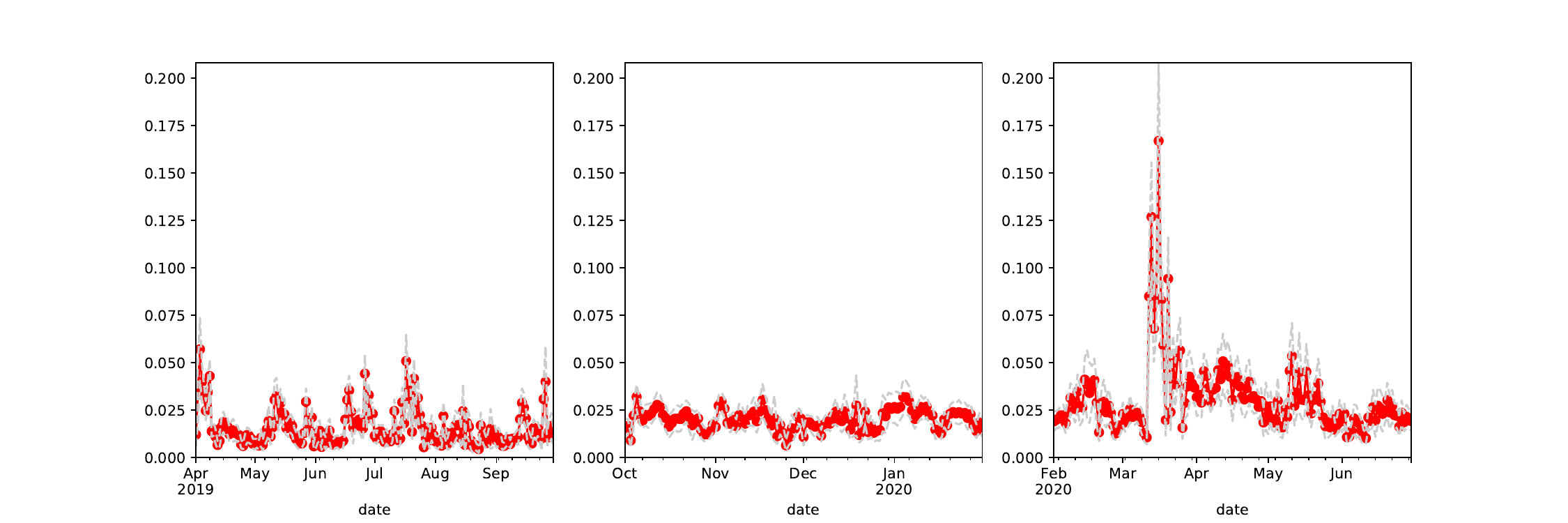}
    \caption{SV}
    \end{subfigure}
    \caption{RMSE with 95 \%-confidence interval of the (a) BS, (b) JD and (c) SV model. \includegraphics[scale=0.2]{quanlet.png}\href{https://github.com/QuantLet/hedging_cc}{ hedging\_cc}}
    \label{fig:rmse_fig1}
\end{figure}
\clearpage
\begin{figure}[ht!]
    \centering
    \begin{subfigure}[b]{\textwidth}\includegraphics[trim={3.5cm 0 3.5cm 0},clip,width=\textwidth]{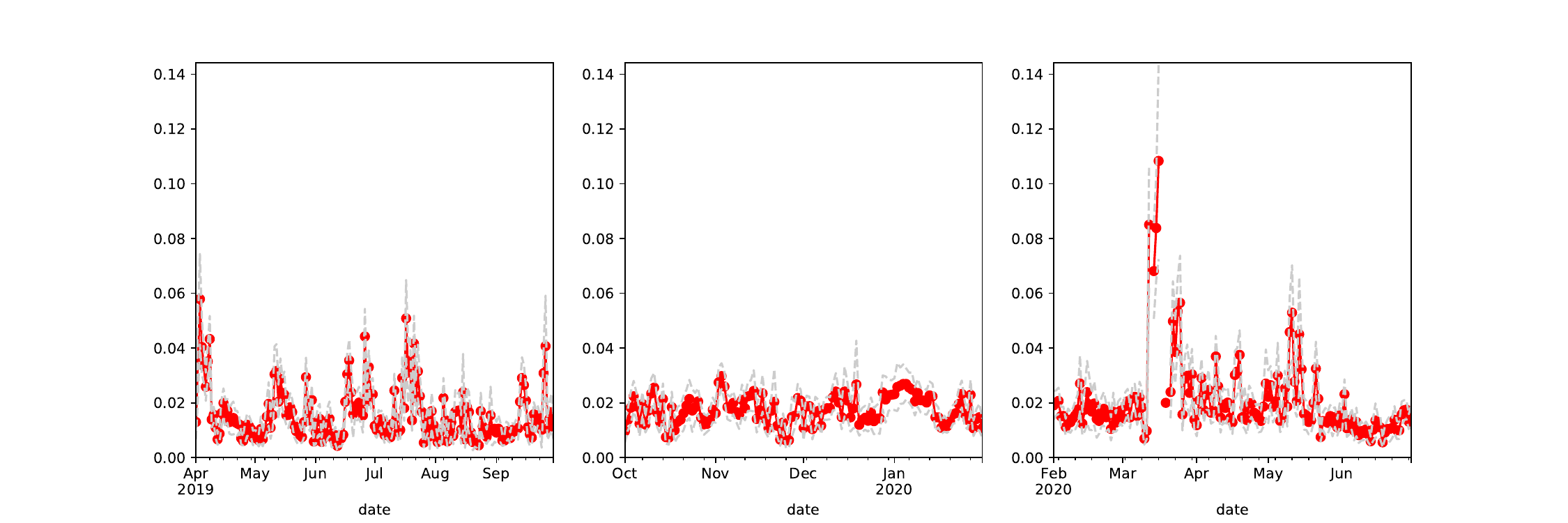}
    \caption{SVJ}
    \end{subfigure}
    \begin{subfigure}[b]{\textwidth}\includegraphics[trim={3.5cm 0 3.5cm 0},clip,width=\textwidth]{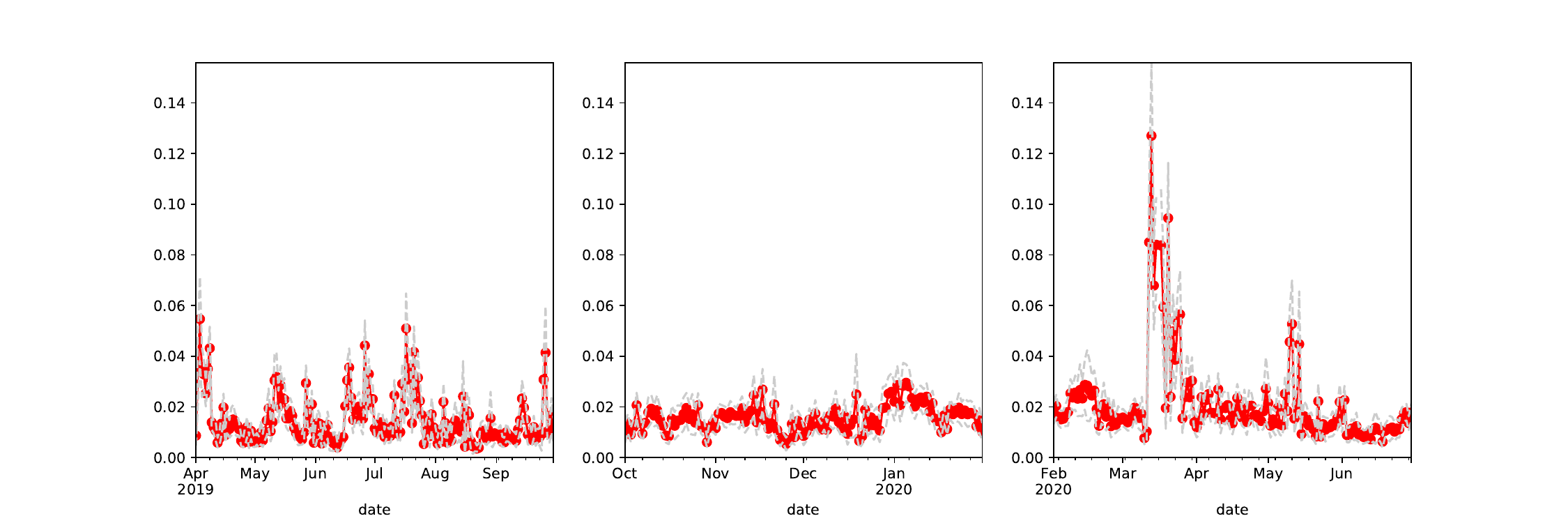}
    \caption{SVCJ}
    \end{subfigure}
  \caption{RMSE with 95 \%-confidence interval of the (a) SVJ, (b) SVCJ, (c) VG and (d) CGMY model. \includegraphics[scale=0.2]{quanlet.png}\href{https://github.com/QuantLet/hedging_cc}{ hedging\_cc}}
    \label{fig:rmse_fig2}
\end{figure}

\clearpage

\begin{figure}[ht!]
    \centering
    \begin{subfigure}[b]{\textwidth}\includegraphics[trim={3.5cm 0 3.5cm 0},clip,width=\textwidth]{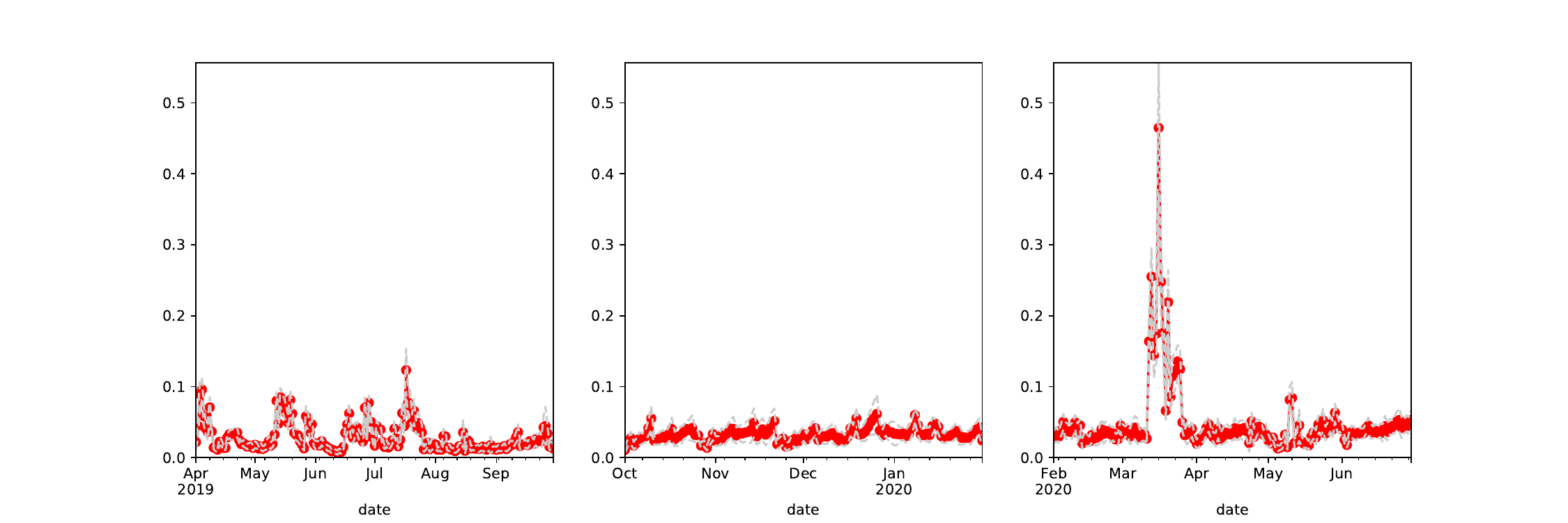}
    \caption{VG}
    \end{subfigure}
    \begin{subfigure}[b]{\textwidth}\includegraphics[trim={3.5cm 0 3.5cm 0},clip,width=\textwidth]{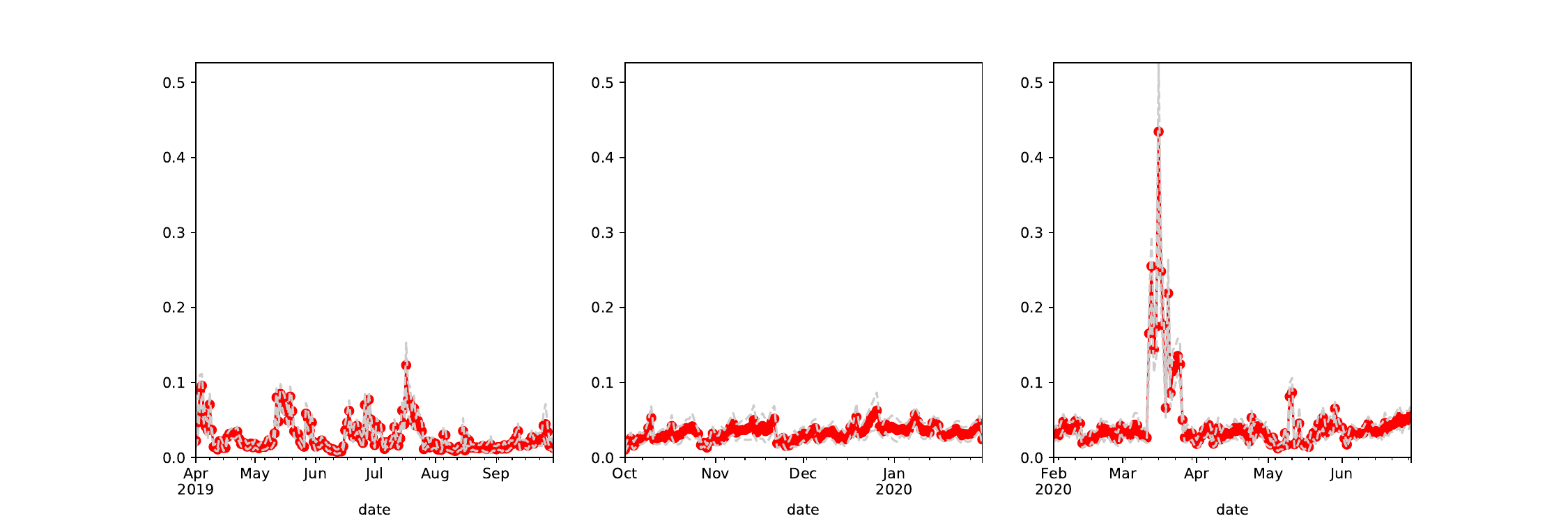}
    \caption{CGMY}
    \end{subfigure}
  \caption{RMSE with 95 \%-confidence interval of the (a) VG and (b) CGMY model. \includegraphics[scale=0.2]{quanlet.png}\href{https://github.com/QuantLet/hedging_cc}{ hedging\_cc}}
    \label{fig:rmse_fig3}
\end{figure}



\clearpage
\end{appendices}


\bibliography{sn-bibliography}


\end{document}